\documentclass[onecolumn]{elsart3p}
\usepackage{bm}
\usepackage{amssymb}

\usepackage[english,francais]{babel}
\usepackage{graphicx}
\usepackage{subfigure}

\usepackage{amsmath} 
\usepackage{amssymb} 
\usepackage{amsfonts}

\usepackage{subfigure}
\usepackage{pgfplots}

\usepackage[T1]{fontenc}

\usetikzlibrary{arrows,shapes,backgrounds,calc,
    positioning,
    intersections}

\newtheorem{e-proposition}[theorem]{Proposition}

\newtheorem{e-definition}[theorem]{Definition\rm}

\renewcommand{\theequation}{\arabic{equation}}
\def\og{\leavevmode\raise.3ex\hbox{$\scriptscriptstyle\langle\!\langle$~}}
\def\fg{\leavevmode\raise.3ex\hbox{~$\!\scriptscriptstyle\,\rangle\!\rangle$}}

\def\bs#1{\boldsymbol{#1}}

\begin{document}
%  You can place here the title of the dossier, if you know it,
%     firstly in English, then in French
\centerline{Quantum simulation / Simulation quantique}
\begin{frontmatter}

\vspace{-2.5cm}


% \author{Name\thanksref{label2}}
\selectlanguage{english}
\title{Artificial gauge fields in materials and engineered systems}

% If all authors are at the same address, the [label1] can be suppressed

\selectlanguage{english}
\author[authorlabel1,authorlabel12,authorlabel13]{M. Aidelsburger},
\ead{monika.aidelsburger@physik.uni-muenchen.de}
\author[authorlabel2]{S. Nascimbene} and
\ead{sylvain.nascimbene@lkb.ens.fr}
\author[authorlabel3]{N. Goldman}
\ead{ngoldman@ulb.ac.be}

\address[authorlabel1]{Laboratoire Kastler Brossel, College de France, CNRS, ENS-PSL Research University, UPMC-Sorbonne Universites, 11 place Marcelin-Berthelot, 75005 Paris, France}

\address[authorlabel12]{Fakult\"at f\"ur Physik, Ludwig-Maximilians-Universit\"at M\"unchen, Schellingstr. 4, 80799 Munich, Germany}
\address[authorlabel13]{Max-Planck-Institut f\"ur Quantenoptik, Hans-Kopfermann-Str. 1, 85748 Garching, Germany}
\address[authorlabel2]{Laboratoire Kastler Brossel, Coll\`ege de France, CNRS, ENS-PSL Research University, UPMC-Sorbonne Universit\'es, 11 place Marcelin-Berthelot, 75005 Paris, France}
\address[authorlabel3]{Center for Nonlinear Phenomena and Complex Systems,
Universite Libre de Bruxelles, CP 231, Campus Plaine, B-1050 Brussels, Belgium}

\begin{abstract}

Artificial gauge fields are currently realized in a wide range of physical settings. This includes solid-state devices but also engineered systems, such as photonic crystals, ultracold gases and mechanical setups. It is the aim of this review to offer, for the first time, a unified view on these various forms of artificial electromagnetic fields and spin-orbit couplings for matter and light. This topical review provides a general introduction to the universal concept of engineered gauge fields, in a form which is accessible to young researchers entering the field. Moreover, this work aims to connect different communities, by revealing explicit links between the diverse forms and realizations of artificial gauge fields.

%{\it To cite this article: A. Name1, A. Name2, C. R. Physique 6 (2005).}

%\vskip 0.5\baselineskip

%\selectlanguage{francais}
%\noindent{\bf R\'esum\'e}
%\vskip 0.5\baselineskip
%\noindent
%{\bf Here is the French title. }
%Your r\'esum\'e in French here.
%{\it Pour citer cet article~: A. Name1, A. Name2, C. R.
%Physique 6 (2005).}

%Now keywords/mots-cl�s
\keyword{Gauge fields; Quantum simulation; Condensed matter } \vskip 0.5\baselineskip
\noindent{\small{\it Mots-cl\'es~:} Champs de jauge~; Simulation quantique~;
Mati\`ere condens\'ee}}
\end{abstract}
\end{frontmatter}

\selectlanguage{english}
% main text

\tableofcontents

\section{Introduction}
\label{}

Gauge fields represent one of the most ubiquitous concepts in physics, offering surprising links between distinct research fields, ranging from high-energy physics and cosmology to quantum optics and condensed-matter physics. While gauge potentials were initially introduced as a mathematical object in the context of classical electrodynamics, their essence was fully revealed through the development of quantum mechanics and quantum field theory~\cite{aharonov:1959PR,Wu:1975PRD}. Exploring the physical consequences of gauge fields, as first evidenced by the seminal Aharonov-Bohm effect~\cite{aharonov:1959PR}, has led to remarkable theoretical and conceptual progress, which has culminated in our modern formulation of gauge theories in terms of fibre bundles~\cite{Wu:1975PRD,Simon:1983PRL,nakahara:2003book}. Interestingly, this mathematical notion of fibre bundles also played a central role in non-relativistic quantum mechanics, where it constitutes the roots of the geometric (Berry) phase~\cite{Pancharatnam:1956PIAS,Mead:1979JCP,Berry:1984PRS} and topological states of matter~\cite{Simon:1983PRL,Thouless:1982PRL,Kohmoto:1989PRB,Bernevig:2013book}. 

While gauge potentials naturally emerge from ``real" electromagnetic fields, a wide family of gauge fields can also be created and finely tuned by acting on a physical system in a proper manner. This can be realized in solid-state devises, but also in engineered systems, e.g.~ultracold atomic gases~\cite{Dalibard:2011RMP,Goldman:2014RPP,Dalibard:2015arXiv,Zhai:2015RPP,Goldman:2016fa} and photonic crystals~\cite{Lu:2014NP,Hafezi:2014JMB}. It is the aim of this review to depict an overall view on this quantum-simulation approach to gauge fields, as we now explain in the next paragraph. Before doing so, let us emphasize that interesting gauge structures can also be emulated in genuinely classical settings~\cite{Huber:2016tg}, e.g.~arrays of coupled pendula, as will be also clarified and illustrated in this review.  \\

This review aims to cover a wide and fast-growing field in a rather concise manner. We therefore invite the interested reader to consult the following reviews and progress articles for more detailed discussions:~on artificial gauge fields in solids~\cite{Vozmediano:2010dl}, ultracold atoms~\cite{Dalibard:2011RMP,Goldman:2014RPP,Dalibard:2015arXiv,Zhai:2015RPP,Goldman:2016fa}, photonics~\cite{Lu:2014NP,Hafezi:2014JMB}, and mechanical systems~\cite{Huber:2016tg}.

\subsection{Scope of the review}\label{scope_section}

In this review, we discuss how a variety of physical platforms can realize \emph{artificial gauge fields}. Specifically, we are interested in situations where the spectral properties and dynamics of a physical system are well captured by a \emph{designed Hamiltonian} of the form
\begin{equation}
 H ({\bs{p}} - \bs{A}) , \qquad \bs{A}=\bs{A}({\bs{x}}, \sigma, t).\label{Ham_general}
\end{equation}
%Here, $\hat H(\hat{\bs{p}})$ denotes the Hamiltonian in the absence of the gauge field,  
where ${\bs{p}}$ is the momentum operator, and where $\bs{A}({\bs{x}}, \sigma, t)$ represents a general gauge potential; the latter can potentially depend on the position operator ${\bs{x}}$, but also on a ``spin" (internal) degree of freedom $\sigma$, or even on time $t$. Formally, the Hamiltonian in Eq.~\eqref{Ham_general} describes a physical system that is associated with a Hamiltonian $H ({\bs{p}})$, and which is coupled to a gauge field $\bs{A}$ through the minimal coupling prescription.

Importantly, in the following, the engineered gauge potential $\bs{A}$ will always be treated as a classical and external field:~in particular, this gauge field will be non-dynamical, in the sense that the motion of particles will not affect the gauge field in return; in other words, the schemes described in this review do not reproduce a complete gauge theory (e.g.~Maxwell's equations of electromagnetism). The quantum simulation of dynamical gauge fields is discussed in the reviews~\cite{Wiese:2013Ann,Zohar:2015Rep,Dalmonte:2016hd}. 

The engineered Hamiltonians presented in this review [Eq.~\eqref{Ham_general}] are often inspired by solid-state physics. A concrete example of such target Hamiltonians is that describing the motion of a charged particle (e.g.~an electron) in an external, static and uniform magnetic field $\bs B$,
\begin{equation}
 H=\frac{1}{2M} \left ({\bs{p}} - q \bs{A}_{B} ({\bs{x}}) \right )^2 , \qquad \bs B=\bs{\nabla} \times \bs{A}_B ({\bs{x}}) ,\label{B_field_Ham}
\end{equation}
where $q$ and $M$ denote the particle's charge and mass, respectively. Another type of model Hamiltonian, which also plays an important role in condensed-matter physics, is that describing the motion of a spin-1/2 particle subjected to spin-orbit coupling~\cite{Sarma:2004RMP,Nagaosa:2010RMP,Qi:2011RMP}
\begin{equation}
H=\frac{1}{2M} \left ({\bs{p}} - {\bs{A}}_{\text{SOC}} \right )^2 , \qquad  {\bs{A}}_{\text{SOC}}=(\alpha_x  \sigma_x,\alpha_y  \sigma_y,\alpha_z   \sigma_z) , \label{SOC_Ham}
\end{equation}
where ${\sigma}_{x,y,z}$ represent the Pauli matrices. In two-dimensions, and when $\alpha_{x}\!=\!-\alpha_y\!=\!\alpha$ is constant, the Hamiltonian in Eq.~\eqref{SOC_Ham} features a non-trivial term of the form $\alpha\!\left (p_x   \sigma_x -  p_y  \sigma_y  \right )$, which corresponds to the so-called Rashba spin-orbit-coupling effect~\cite{Sarma:2004RMP,Nagaosa:2010RMP,Qi:2011RMP}. While the Hamiltonian in Eq.~\eqref{B_field_Ham} plays a central role in the physics of the quantum Hall effects~\cite{Klitzing:1986RMP,Hasan:2010RMP,Qi:2011RMP}, systems exhibiting spin-orbit coupling [Eq.~\eqref{SOC_Ham}] have been recently investigated in the context of spintronics~\cite{Sarma:2004RMP}, the anomalous Hall effect~\cite{Nagaosa:2010RMP}, topological insulators and superconductors~\cite{Qi:2011RMP}. 

Besides, Hamiltonians of the form \eqref{Ham_general} with a time-dependent gauge potential $\bs{A}\!=\!\bs{A}(t)$ emulate the effects of an external electric field. An intriguing example of such model Hamiltonians, which will be explicitly considered below, is that describing electronic systems subjected to circularly polarized light~\cite{Bennett:1965PR,Oka:09PRB}, where $ \bs{A}(t)\!=\! \lambda \left (\sin (\omega t) \bs{1}_x+ \cos (\omega t) \bs{1}_y \right )$, and where $\omega$ denotes the frequency of the applied radiation. Such configurations currently play an important role in the context of Floquet engineering (e.g.~Floquet topological insulators~\cite{Oka:09PRB,Kitagawa:2011PRB,Lindner:2011NatPhys,Cayssol:2013PS,Rechtsman:2013Nature,Jotzu:2014Nat}), and were also recently envisaged as probes for topological matter~\cite{Juan:2016arXiv,Tran:2017SciAdv}. \\

The main sections of this review article (Sections~\ref{sect:continuum}-\ref{sect:SOC}) will be dedicated to a wide variety of schemes that realize Hamiltonians of the form discussed above [Eqs.~\eqref{Ham_general}-\eqref{SOC_Ham}], both in the continuum and in lattice configurations. While many schemes have been proposed in the literature,  the focus will be set on those schemes that were actually implemented in experiments. Before presenting these different methods and physical settings, we first present a series of ``universal theoretical notions" (Section~\ref{sect:universal}), which play a central role in the realization of effective Hamiltonians displaying synthetic gauge fields.\\

\section{Universal theoretical notions}\label{sect:universal}

Artificial gauge fields generally appear in the effective dynamics of engineered quantum systems~\cite{Mead:1992RMP,Dalibard:2011RMP,Goldman:2014RPP,Dalibard:2015arXiv}. In this Section, we review general theoretical concepts that are strongly connected to the origin and the description of these artificial gauge structures. We first introduce the geometric (Berry) phase and recall how it relates to the notion of gauge fields. We then discuss how these concepts generalize to lattice systems. We conclude this section by analyzing another promising route towards the realization of gauge fields, namely methods based on shaking (Floquet engineering). Alternative strategies based on rotation or strain are introduced in Sects~\ref{sect:rotation} and \ref{sect:strain}, respectively.

\subsection{The geometric phase and its gauge structure}
\label{sect:IntroGeoPhase}

The discovery of the \emph{geometric phase}, through the successive works of Pancharatnam~\cite{Pancharatnam:1956PIAS}, Aharonov-Bohm~\cite{aharonov:1959PR}, Wu-Yang~\cite{Wu:1975PRD}, Mead-Truhlar~\cite{Mead:1979JCP} and Berry~\cite{Berry:1984PRS}, revealed a deep connection between gauge structures and geometry. Today, this connection is largely exploited to engineer gauge structures in quantum systems, as we further describe below. For a review on the effects of the geometric phase in condensed-matter physics, see Ref.~\cite{Xiao:2010RMP}.

\subsubsection{General introduction to the geometric phase}\label{gen_intro_berry}

Let us start by considering a general quantum systems, which is described by the time-dependent Hamiltonian,
\begin{equation}
H = H (\bs R), \qquad \bs R=\bs R (t),\label{H_R_berry}
\end{equation}
where $\bs R$ describes a set of parameters that depend on time (e.g.~some control parameters that are varied externally). In this general setting, geometric effects potentially emerge when the state of the system performs an adiabatic evolution under a slow change of the parameters $\bs R (t)$. To see this, one introduces the instantaneous eigenstates and eigenenergies
\begin{equation}
 H (\bs R) \vert n (\bs R)\rangle = \varepsilon_n (\bs R)\vert n (\bs R)\rangle .\label{instant_states}
\end{equation}
If the system is initially prepared in the $m$th eigenstates, e.g.~$\vert \psi (t\!=\!0) \rangle\!=\!\vert m [\bs R(t\!=\!0)]\rangle$, and if the energy manifold $\varepsilon_m (\bs R)$ is well isolated by a gap along the entire path $\bs R (t)$, then the adiabatic theorem guarantees that the system will remain in the instantaneous eigenstate $\vert m [\bs R (t)]\rangle$ at all times. Importantly, the states $\{\vert n (\bs R)\rangle \}$ in Eq.~\eqref{instant_states} are defined up to a complex phase factor, i.e.~the quantum system possesses a U(1), or gauge, degree of freedom. Formally, a geometric structure is then introduced by assuming that these phases are smoothly and uniquely defined over the path; mathematically, this corresponds to building a U(1) fibre bundle on top of the parameter space~\cite{Wu:1975PRD,Simon:1983PRL,nakahara:2003book}. Based on these few assumptions (i.e.~the validity of the adiabatic approximation and the unicity of the phase associated with the instantaneous eigenstates), one solves the Schr\"odinger equation
\begin{equation}
i \partial_t \vert \psi (t) \rangle =  H [\bs R (t)] \vert \psi (t) \rangle, \label{evolution_berry}
\end{equation}
by imposing that the state remains in the $m$th manifold at all times, namely, by writing the solution as
\begin{equation}
\vert \psi (t) \rangle= \phi_m (t) \vert m [\bs R (t)]\rangle = e^{i \gamma_m (t)} e^{-i \int_0^t \varepsilon_m [\bs R (\tau)] \text{d}\tau} \vert m [\bs R (t)]\rangle ,\label{state_explicit}
\end{equation}
where we explicitly factorized the phase factor of the evolving state in terms of two fundamentally different contributions. While the phase involving the instantaneous energies $\varepsilon_m [\bs R (t)]$ is directly analogous to the standard dynamical phase of static systems, the phase $ \gamma_m (t)$ is found to only depend on the path $\bs R (t)$ performed by the system in parameter space. Indeed, inserting Eq.~\eqref{state_explicit} into Eq.~\eqref{evolution_berry}, yields the explicit form of the \emph{geometric phase}~\cite{Mead:1992RMP,Xiao:2010RMP}
\begin{equation}
\gamma_m = \int_{\text{path}} \text{d}\bs{R} \cdot \bs{A}_{\text{geom}}, \qquad \bs{A}_{\text{geom}}=i \langle m (\bs R)\vert \partial_{\bs{R}} \vert m (\bs R)\rangle.\label{Berry_phase}
\end{equation}

Interestingly, the geometric phase acquired by the evolving state is associated with a \emph{gauge structure}. Indeed, under a local gauge transformation, the result in Eq.~\eqref{Berry_phase} imposes that
\begin{equation}
\vert m (\bs R)\rangle \rightarrow e^{i \chi (\bs R)}\vert m (\bs R)\rangle, \qquad \bs{A}_{\text{geom}} \rightarrow \bs{A}_{\text{geom}} - \partial_{\bs R} \chi (\bs R),
\end{equation}
which reveals that the \emph{Berry connection} $\bs{A}_{\text{geom}}$ transforms like the usual gauge vector of electromagnetism under a gauge transformation. Importantly,  the geometric phase can be related to a \emph{gauge-invariant quantity}, whenever the path forms a closed loop in parameters space~\cite{aharonov:1959PR,Wu:1975PRD,Berry:1984PRS}, i.e.~when $\bs R (t_{\text{final}})\!=\!\bs R (0)$. In that case, the geometric phase can be expressed as
\begin{equation}
\gamma_m = \oint_{\text{loop}} \text{d}\bs{R} \cdot \bs{A}_{\text{geom}} = \int_{\Sigma} \text{d}\bs{S} \cdot \bs{\Omega}_{\text{geom}}, \qquad \bs{\Omega}_{\text{geom}}=\bs{\nabla}_{\bs R} \times \bs{A}_{\text{geom}},\label{Berry_curvature}
\end{equation}
where $\Sigma$ denotes the surface enclosed by the loop in parameters space, and where we introduced the \emph{Berry curvature} $\bs{\Omega}_{\text{geom}}$. As apparent from Eq.~\eqref{Berry_curvature}, the Berry curvature is gauge invariant, and it takes the form of a fictitious ``magnetic field" in parameters space. In fact, the geometric (Berry) phase in Eq.~\eqref{Berry_curvature} generalizes the Aharonov-Bohm phase~\cite{aharonov:1959PR}, which is acquired by a charged particle moving in (real) space and encircling a region penetrated by a non-zero magnetic flux. The general and important result in Eq.~\eqref{Berry_curvature} indicates how artificial gauge structures (e.g.~synthetic magnetic fields) can be engineered in quantum systems, through the design of non-trivial Berry curvature in parameters space. 

\subsubsection{Non-Abelian structures}\label{non_Abelian_section}

The discussion above can be straightforwardly generalized to the case where the system populates a family of instantaneous eigenstates during the evolution~\cite{Wilczek:1984PRL,Mead:1992RMP}, i.e.~ if one replaces the unique manifold $m\rightarrow \{m_1, m_2, \dots , m_q \}$ in the formalism above; this can occur if the corresponding energies $\{\varepsilon_{m_1} [\bs R], \dots, \varepsilon_{m_q} [\bs R]\}$ undergo gap-closing events at singular points of the path $\bs R(t)$. In this case, the Berry phase is replaced by a $q\times q$ matrix acting on the relevant family of states; the corresponding Berry connection is then defined as the $q\times q$ matrix $\bs{A}_{\text{geom}}^{ab}=i \langle m_a (\bs R)\vert \partial_{\bs{R}} \vert m_b (\bs R)\rangle$, where $a,b\!=\!1, \dots, q$. This matrix-valued Berry connection acts as an effective ``spin-orbit coupling" in parameters space, i.e.~a non-Abelian gauge potential~\cite{Mead:1992RMP,Xiao:2010RMP}. This constitutes a simple route for realizing synthetic spin-orbit coupling in quantum-engineered systems~\cite{Goldman:2014RPP,Zhai:2015RPP}.

\subsubsection{The Berry connection and the effective-Hamiltonian approach}\label{berry_effective}

In the previous paragraph~\ref{gen_intro_berry}, we obtained that the Berry connection $\bs{A}_{\text{geom}}$ acts as an effective gauge potential in the adiabatic evolution of a quantum system; in particular, its associated gauge-invariant quantity, the Berry curvature $\bs{\Omega}_{\text{geom}}$, can lead to observable effects~\cite{Mead:1992RMP,Xiao:2010RMP}. We now discuss how the Berry connection explicitly appears in an effective-Hamiltonian analysis of the adiabatic evolution~\cite{Mead:1979JCP,Mead:1992RMP,Goldman:2014RPP,Dalibard:2015arXiv}. To do so, let us analyse a slightly more concrete situation: the motion of a particle of mass $M$ whose internal degrees of freedom are addressed by an external field~\cite{Dalibard:2011RMP,Goldman:2014RPP,Dalibard:2015arXiv}. We now write the Hamiltonian in Eq.~\eqref{H_R_berry} in the more specific form
\begin{equation}
H = \frac{{\bs p}^2}{2M} +  V_{\text{ext}} ({\bs{x}}, t) = \frac{{\bs p}^2}{2M} + \sum_{m,n=1}^N \vert n \rangle V_{nm} ({\bs{x}}, t) \langle m \vert,\label{H_coupl}
\end{equation}
where the first term describes the kinetic energy, and where the operator $ V_{\text{ext}}$ captures the effects of the external field on the internal degrees of freedom of the particle; the corresponding internal states are denoted $\{ \vert n \rangle \}$ in Eq.~\eqref{H_coupl}, with $n\!=\!1, \dots, N$. A concrete example, which will be further discussed below, is that of an atom immersed in a laser field, which resonantly couples a set of atomic internal states~\cite{Dalibard:2011RMP,Goldman:2014RPP,Dalibard:2015arXiv}. In direct analogy with the treatment above, let us introduce the instantaneous eigenstates (or ``dressed states"~\cite{Dalibard:2011RMP,Goldman:2014RPP}) associated with the coupling term,
\begin{equation}
 V_{\text{ext}} ({\bs{x}}, t) \vert \eta_m ({\bs{x}}, t) \rangle = \varepsilon_m ({\bs{x}}, t) \vert \eta_m ({\bs{x}}, t) \rangle . \label{dressed_states}
\end{equation}
We are interested in studying the adiabatic evolution of the driven particle, when the latter is initially prepared in a given dressed state $\vert \eta_m ({\bs{x}}, t\!=\!0) \rangle $; thus, the initial state is written as~\cite{Dalibard:2015arXiv} $$\vert \Psi ({\bs{x}}, t\!=\!0) \rangle = \phi_m ({\bs{x}}, t\!=\!0) \vert \eta_m ({\bs{x}}, t\!=\!0) \rangle,$$
where $\phi_m ({\bs{x}}, t\!=\!0)$ reflects the probability amplitude to initially detect the particle at the point ${\bs{x}}$, being in the internal state $\vert \eta_m ({\bs{x}}, t\!=\!0) \rangle $. We suppose that the velocity of the particle is sufficiently weak and that the ``band" $\varepsilon_m ({\bs{x}}, t)$ is well isolated by a gap along the path undergone by the particle, such that the adiabatic approximation indeed holds during the time-evolution. Under this assumption, one can solve for the Schr\"odinger equation~\cite{Goldman:2014RPP,Dalibard:2015arXiv}
\begin{equation}
i \partial_t \vert \Psi ({\bs{x}}, t) \rangle = H \vert \Psi ({\bs{x}}, t) \rangle , \quad \text{ with } \vert \Psi ({\bs{x}}, t) \rangle\!=\! \phi_m ({\bs{x}}, t) \vert \eta_m ({\bs{x}}, t) \rangle ,\label{troncation}
\end{equation}
where the simplified form of the solution $\vert \Psi ({\bs{x}}, t) \rangle$ reflects the adiabatic approximation~\cite{Mead:1992RMP,Goldman:2014RPP,Dalibard:2015arXiv}. Importantly, this projection of the dynamics unto the subspace spanned by the unique (non-degenerate) dressed state $\vert \eta_m ({\bs{x}}, t) \rangle$ yields an effective Schr\"odinger equation for the unknown coefficient $\phi_m ({\bs{x}}, t)$. Indeed, using Eqs.~\eqref{H_coupl}, \eqref{dressed_states} and \eqref{troncation} yields the effective Schr\"odinger equation~\cite{Mead:1979JCP,Mead:1992RMP,Goldman:2014RPP,Dalibard:2015arXiv}
\begin{equation}
i \partial_t \phi_m ({\bs{x}}, t) =  H_{\text{eff}} \, \phi_m ({\bs{x}}, t), \qquad    H_{\text{eff}} = \frac{1}{2M} \left ({\bs p} - \bs{A}_{\text{geom}}  \right )^2 + V_{\text{eff}},\label{effective_ham_berry}
\end{equation}
where the Berry connection $\bs{A}_{\text{geom}}(\bs x,t)\!=\! i \langle \eta_m ({\bs{x}}, t)\vert \bs{\nabla} \eta_m ({\bs{x}}, t) \rangle$ explicitly appears as a gauge potential. In addition to this effective gauge field, the particle also feels an effective scalar potential $V_{\text{eff}}\!=\!\varepsilon_m - i \langle \eta_m \vert \partial_t \eta_m \rangle + (1/2M) \sum_{n \ne m} \vert \langle \bs{\nabla} \eta_m \vert \eta_n \rangle \vert^2$. From this approach, one directly obtains that the gauge structure associated with the geometric phase naturally emerges through a projection procedure~\cite{Mead:1992RMP}, which is motivated and justified by the adiabatic approximation [Eq.~\eqref{troncation}]. In particular, let us point out that the effective Hamiltonian in Eq.~\eqref{effective_ham_berry} indeed has the appealing form discussed in Section~\ref{scope_section} [see  Eqs.~\eqref{Ham_general} and \eqref{B_field_Ham}].\\

The same analysis can be performed in the non-Abelian case [Section~\ref{non_Abelian_section}]. When the particle occupies a family of dressed-states during its time-evolution,  the truncation in Eq.~\eqref{troncation} should be relaxed according to $\vert \Psi ({\bs{x}}, t) \rangle\!=\! \sum_{a=1}^q \phi_{m_a} ({\bs{x}}, t) \vert \eta_{m_a} ({\bs{x}}, t) \rangle$, where $\{ \vert \eta_{m_a} \rangle \}$ denotes the family of $q$ relevant dressed states~\cite{Goldman:2014RPP,Dalibard:2015arXiv}. The resulting equations of motion take the form of Eq.~\eqref{effective_ham_berry}, but now with the non-Abelian Berry connection $\bs{A}_{\text{geom}}^{ab}\!=\! i \langle \eta_{m_{a}} ({\bs{x}}, t)\vert \bs{\nabla} \eta_{m_{b}} ({\bs{x}}, t) \rangle$, as already introduced in Section~\ref{non_Abelian_section}. Specifically, the effective Hamiltonian is now of the form introduced in Eq.~\eqref{SOC_Ham}, indicating that the non-Abelian Berry connection can be designed so as to engineer artificial spin-orbit coupling~\cite{Goldman:2014RPP,Dalibard:2015arXiv,Zhai:2015RPP}.

\subsection{Gauge potentials in lattices}
\label{sect:hofstadter}

The Hamiltonians in Eqs.~\eqref{B_field_Ham} and~\eqref{SOC_Ham} capture the effects of an external gauge field on a particle that moves in the \emph{continuum} (i.e. $\bs x$ is a continuous space variable). In both cases, these Hamiltonians are obtained through the ``minimal coupling" between matter and the gauge field, which is formally described by the substitution $ H ({\bs{p}})\rightarrow H ({\bs{p}} - \bs{A})$ leading to Eq.~\eqref{Ham_general}, where $ H ({\bs{p}})$ is a Hamiltonian in the absence of the gauge field. In fact, this substitution also appears in lattice systems (i.e.~in a discretized space), as we now discuss in this Section. Let us emphasize that the interplay between lattice structures and external gauge fields plays an important role in solid-state physics~\cite{Luttinger:1951PR,Hofstadter:1976PRB,Sarma:2004RMP,Xiao:2010RMP,Qi:2011RMP,Hasan:2010RMP}, but also in cold atoms~\cite{Bloch:2008RMP,Dalibard:2011RMP,Goldman:2014RPP} and photonics~\cite{Lu:2014NP,Lu:2016ib}, where both the lattice structure and the gauge fields can be engineered. From a theoretical point of view, the introduction of gauge fields within lattice structures has been formalized in the context of lattice gauge theories, with direct implications in high-energy physics~\cite{Kogut:1979RMP,Kogut:1983RMP}. 

Let us start by considering the motion of a particle moving in a 2D space-periodic potential,
\begin{equation}
H =\frac{{\bs{p}}^2}{2M}  +  V_{\text{lattice}} (\bs x), \qquad  V_{\text{lattice}} (\bs x+a_x\bs 1_x)= V_{\text{lattice}} (\bs x) = V_{\text{lattice}} (\bs x+a_y\bs 1_y) ,
\end{equation}
where $a_{x,y}$ denote the lattice spacings. For deep enough lattice potentials, one can solve the Schr\"odinger equation using the so-called tight-binding approximation~\cite{Jaksch:1998PRL,Simon:2013oxford}, 
\begin{equation}
 H \vert \Psi \rangle \!=\!E\vert \Psi \rangle, \qquad \vert \Psi \rangle= \sum_{n,m} \phi (n,m) \vert n,m \rangle ,
\end{equation}
where we have expanded the solution state $\vert \Psi \rangle$ over the lowest-energy (vibrational) eigenstates $\{ \vert n,m \rangle \}$ associated with the potential wells located at $\bs x\!=\!(na_x,ma_y)$, and ($n,m$) are integers. Considering the simplest case of a square lattice ($a_x\!=\!a_y\!=\!a$), the coefficients $\phi (n,m)$ satisfy an effective Schr\"odinger equation~\cite{Simon:2013oxford}
\begin{align}
E \phi (n,m) = - J \left [ \phi (n+1,m) + \phi (n-1,m)+ \phi (n,m+1)+ \phi (n,m-1) \right ] + \varepsilon  \phi (n,m),\label{tight-binding}
\end{align}
where $J$ describes hopping between nearest-neighboring sites, where $\varepsilon$ is an onsite energy, and where we neglected higher-order hopping processes (e.g.~next-nearest-neighbor tunneling). One can rewrite Eq.~\eqref{tight-binding} in terms of an effective Hamiltonian $H_{\text{eff}}\phi (n,m)\!=\!(E-\varepsilon)\phi (n,m)$, 
\begin{align}
H_{\text{eff}}=-J \left [ T_x +  T_x^{\dagger} + T_y +  T_y^{\dagger} \right ], \quad  T_x=e^{-i a  p_x} , \quad T_y=e^{-i a  p_y},\label{TB_effective}
\end{align}
which explicitly involves translation operators $T_{x,y}$ over the lattice. 

In this tight-binding picture, where the motion translates into hopping between discrete lattice sites, the effects of a gauge field $\bs{A}=(A_x,A_y)$ can be directly incorporated at the level of the effective Hamiltonian in Eq.~\eqref{TB_effective},
\begin{align}
 H_{\text{eff}}=-J \left [  T_x e^{i a q A_x} +  T_x^{\dagger} e^{-i a q A_x} + T_y e^{i a q A_y}+  T_y^{\dagger} e^{-i a q A_y} \right ].\label{TB_Peierls}
\end{align}
Here, we performed the substitution $ H_{\text{eff}} ({\bs{p}})\rightarrow  H_{\text{eff}} ({\bs{p}} - q\bs{A})$, which is generally called \emph{Peierls substitution} in this lattice context;  $q$ denotes a general coupling charge. In the context of lattice gauge theories~\cite{Kogut:1979RMP}, the quantities acting on each link, $U_{\mu}=e^{i a q A_{\mu}}$, are called link variables. These gauge-dependent quantities are related to a gauge-invariant quantity (the field-strength tensor) through \emph{loop} products, namely, by evaluating the cumulative action of the link variables upon performing a closed loop on the lattice~\cite{Kogut:1979RMP}; see below for an illustration.

\begin{figure}
\includegraphics[width=\textwidth]{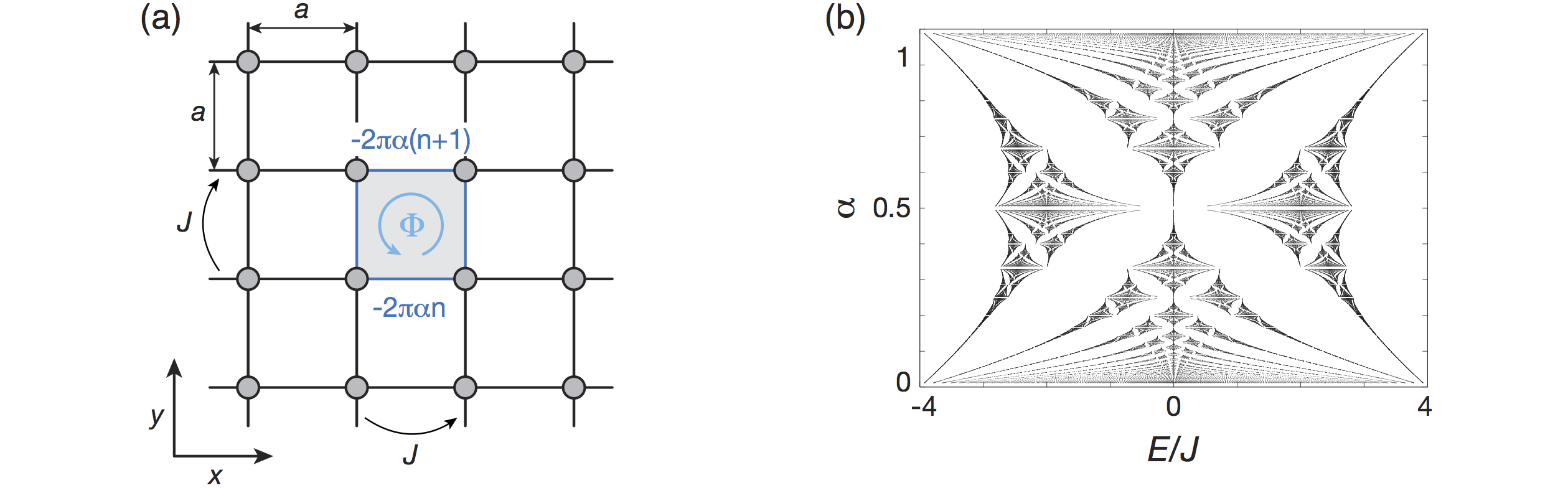}
\centering
%\vspace{0.4cm} 
\caption{\textit{\textsc{Harper-Hofstadter Hamiltonian}
(a) Schematic drawing of an effective 2D-lattice model of a charged particle in the presence of a magnetic flux $\Phi$ per lattice unit cell; $a$ is the lattice constant and $|J|$ the nearest-neighbor coupling strength. In the Landau gauge hopping along $x$ is accompanied by the Peierls phases $2\pi\alpha n$, with $\alpha = \Phi/2\pi$ and $n$ the site-index along $y$. 
(b) Single-particle energy spectrum of the Harper-Hofstadter Hamiltonian as a function of the dimensionless magnetic flux per lattice unit cell $\alpha$, also known as Hofstadter butterfly~\cite{Hofstadter:1976PRB}.}
\label{Fig:Butterfly}}
\end{figure}

A simple example of this Peierls substitution can be found in the study of a charged particle moving on a 2D square lattice, and which is subjected to a uniform magnetic field $B$ perpendicular to the plane; see Refs.~\cite{Luttinger:1951PR,Hofstadter:1976PRB}. This problem can be treated using the Landau gauge, $A_x\!=\!0$ and $A_y\!=\! B x\!=\!\Phi n/a$, where we introduced the magnetic flux $\Phi\!=\!Ba^2$ penetrating each cell of the 2D lattice, as illustrated in Fig.~\ref{Fig:Butterfly}a. In this gauge, the effective Hamiltonian in Eq.~\eqref{TB_Peierls} takes the form
\begin{align}
 H_{\text{eff}}&=-J \left [  T_x + T_x^{\dagger}  + T_y e^{i 2 \pi \alpha n}+  T_y^{\dagger} e^{-i 2 \pi \alpha n} \right ], \notag \\
&= -J \sum_{n,m} \vert n,m \rangle \langle n+1,m \vert + \vert n,m \rangle \langle n,m+1 \vert e^{i 2 \pi \alpha n} + \text{h.c.} ,\label{TB_Hofstadter}
\end{align}

\noindent where we introduced the number of magnetic flux quanta per plaquette $\alpha\!=\!\Phi/\Phi_0$, and where $\Phi_0\!=\!2 \pi$ is the quantum of flux (in units where $\hbar\!=\!e\!=\!1$). The Hamiltonian in Eq.~\eqref{TB_Hofstadter} is the emblematic Harper-Hofstadter Hamiltonian, whose spectrum displays a fractal structure (Fig.~\ref{Fig:Butterfly}b) known as the \emph{Hofstadter butterfly}~\cite{Hofstadter:1976PRB}. This model  plays an important role in the physics of topological states of 
matter~\cite{Hasan:2010RMP,Qi:2011RMP}, as it exhibits a wide family of quantum Hall states~\cite{Thouless:1982PRL}, and offers a natural platform for the realization of fractional Chern insulators~\cite{Parameswaran:2013fractional}; see Section~\ref{Sect:Hof} for more details. We point out that while the \emph{Peierls phase factors}, i.e.~the phase factors $e^{\pm i 2 \pi \alpha n}$  in Eq.~\eqref{TB_Hofstadter}, are gauge dependent, their product around a closed loop is gauge invariant; for instance, the product of Peierls phase factors around a unit cell of the lattice yields $e^{i \Phi}$, which is the Aharonov-Bohm phase~\cite{aharonov:1959PR} acquired by a particle encircling a region penetrated by the flux $\Phi$ (i.e. a gauge-invariant quantity). 

Finally, we point out that the general effective Hamiltonian in Eq.~\eqref{TB_Peierls} can also be used to describe the effects of spin-orbit coupling on a lattice, in which case the gauge potential  $\bs{A}={\bs{A}}_{\text{SOC}}=(\alpha_x  \sigma_x,\alpha_y   \sigma_y,\alpha_z   \sigma_z)$ is matrix valued with respect to spin-space; see Eq.~\eqref{SOC_Ham}.

\subsection{Floquet engineering}
\label{section_floquet}

An alternative route towards artificial gauge fields is offered by Floquet engineering, where effective Hamiltonians can be finely built through proper time-periodic modulations~\cite{Oka:09PRB,Kitagawa:2010PRB,Lindner:2011NatPhys,Kitagawa:2011PRB,Cayssol:2013PS,Goldman:PRX2014,Bukov:2014AP,eckardt2017colloquium}. To understand this strategy, let us introduce a static Hamiltonian $H_0$, which describes a quantum system in the absence of external drive. When subjecting this system to a time-periodic modulation,  $ V (t+T)\!=\!V(t)$, the time-evolution operator $ U(t,t_0)$ typically takes a complicated form as soon as $[ H_0, V(t)]\!\ne\!0$; formally, this can be written as a time-ordered integral, $U(t,t_0)\!=\! \mathcal{T} \mathrm{exp}\left(-i\int^t_{t_0}\mathrm{d}\tau H(\tau)\right)$, where $H(t)\!=\! H_0\!+\! V(t)$ denotes the total time-dependent Hamiltonian. However, by exploiting the time-periodicity $ H(t)\!=\! H (t+T)$, the time-evolution operator $U(t,t_0)$ can be rewritten in a more suggestive form~\cite{Rahav:2003PRA,Goldman:PRX2014}
		\begin{align}
			\label{eq:Floquet_thm}
			U(t,0) =\mathrm{e}^{-i  K_\mathrm{kick}(t)}\mathrm{e}^{-i t  H_\mathrm{eff}}\mathrm{e}^{i  K_\mathrm{kick}(0)},
		\end{align}  
where $H_\mathrm{eff}$ is a time-independent operator, where $ K_\mathrm{kick}(t+T)\!=\! K_\mathrm{kick}(t)$ has zero average over a period of the drive, and where we set the initial time $t_0\!=\!0$. This observation [Eq.~\eqref{eq:Floquet_thm}] indicates that the long-time-evolution of a periodically-driven system is essentially governed by an effective Hamiltonian, i.e.~$H_\mathrm{eff}$. This is even more apparent when probing the system at stroboscopic times, $t_N\!=\!NT$, where $N$ is an integer:~in this case, the stroboscopic time-evolution is generated by the operator $\left [{U}(T) \right ]^N\!=\!\mathrm{e}^{-i t_N H_\mathrm{eff}}$, up to a gauge transformation~\cite{Goldman:PRX2014} involving the operator $K_\mathrm{kick}(0)$. This highlights the fact that a driven system behaves as  an effective static system (described by $H_\mathrm{eff}$), as far as stroboscopic evolution is concerned. At arbitrary times, i.e.~within each period of the drive $T$, the system undergoes a micro-motion, which is captured by the operator $ K_\mathrm{kick}(t)$ in Eq.~\eqref{eq:Floquet_thm}.

The effective Hamiltonian $H_\mathrm{eff}$ can be computed from the operators $H_0$ and $ V(t)$, using an inverse-frequency perturbative expansion~\cite{Goldman:PRX2014,Goldman:2015kc,Bukov:2014AP,Eckardt:2015NJP,Mikami:2015PRB}, namely, assuming a high-frequency limit $\omega=2 \pi/T\rightarrow \infty$.  Up to second order $\mathcal{O} (1/\omega^2)$, this reads~\cite{Goldman:PRX2014,Eckardt:2015NJP,Mikami:2015PRB}
\begin{align}
 H_{\text{eff}}=  H_0 + \frac{1}{ \omega} \sum_{j=1}^{\infty} \frac{1}{j} [{V}^{(j)}, V^{(-j)} ] &+ \frac{1}{2  \omega^2} \sum_{j=1}^{\infty} \frac{1}{j^2} \left ( [[ V^{(j)}, H_0], V^{(-j)} ] + \text{H.c.} \right ) \label{eq_expansion} \\
&+ \frac{1}{3  \omega^2} \sum_{j,l=1}^{\infty} \frac{1}{j l} \left ( [ V^{(j)},[ V^{(l)}, V^{(-j-l)} ]] -  [V^{(j)},[ V^{(-l)}, V^{(l-j)} ]] + \text{H.c.} \right ),  \notag
\end{align}
where we set $ V(t)\!=\!  \sum_{j=1}^{\infty} V^{(j)} e^{i j \omega t} \!+\!  V^{(-j)} e^{-i j \omega t}$. This shows how the effective Hamiltonian originates from a rich interplay between the static Hamiltonian $ H_0$ and the drive $ V(t)$. 

Importantly, by exploiting the non-trivial terms that appear in the expression \eqref{eq_expansion} for the effective Hamiltonian $ H_{\text{eff}}$, one can engineer artificial gauge potentials that are not present in the initial static Hamiltonian $H_0$; this has been considered in view of generating artificial magnetic fields~\cite{Sorensen:2005PRL,Kolovsky:2011EPL,Bermudez:2011PRL,Miyake:2013PRL,Aidelsburger:2013PRL,Jotzu:2014Nat,Aidelsburger:2014NatPhys,Creffield:2016NJP,Tai:2016arxiv} and spin-orbit coupling~\cite{Hauke:2012PRL,Xu:2013PRA,Anderson:2013PRL,Goldman:PRX2014}.  We point out that this Floquet-engineering approach is particularly useful when working on a lattice, where both the static Hamiltonian $ H_0$ and the drive $ V(t)$ can be finely tuned. Examples of how Floquet engineering can lead to non-trivial gauge structures on a lattice [e.g.~as in the Harper-Hofstadter Hamiltonian displayed in Eq.~\eqref{TB_Hofstadter}] will be presented in the following sections.

\section{Artificial magnetism in the continuum}\label{sect:continuum}
%A magnetic field acting on the quantum motion of a charged particle leads to rich physical behavior, from the Aharonov-Bohm effect to the Landau quantization of cyclotron orbits, at the heart of the quantum Hall effect. However, most quantum simulators addressing orbital quantum magnetism effects consist of systems of charge neutral particles, such as atoms, excitons or photons. The effect of an orbital magnetic field on the particle motion thus needs to be artificially engineered. We describe below the most advanced schemes developed over the last years in the fields of ultracold atoms and photonics.
We first consider the simulation of orbital magnetic fields in continuous systems. The production of effective magnetic fields for ultracold atoms quickly followed the first realizations of Bose-Einstein condensates, by setting the gases into rotation or dressing the atoms in laser fields. Recent works also demonstrated the simulation of magnetic fields in systems of cavity photons and polaritons, as we now discuss in this Section.

\subsection{Rotating atomic gases}
\label{sect:rotation}
By setting atomic gases in rotation, an artificial magnetic field is induced in the rotating frame of reference, based on the similitude of the mathematical structures of the  Lorentz and Coriolis forces \cite{fetter_vortices_2001,fetter_rotating_2009}. Let us consider an atom held in an isotropic harmonic trap in the $xy$ plane, of oscillation frequency $\omega$. The Hamiltonian describing the particle motion in the frame rotating around the $z$ axis at the angular frequency $\Omega$ reads
\begin{eqnarray}
H&=&\frac{\mathbf{p}^2}{2m}+\frac{1}{2}m\omega^2\mathbf{r}^2-\Omega L_z,\nonumber\\
&=&\frac{(\mathbf{p}-\mathbf{A})^2}{2m}+\frac{1}{2}m(\omega^2-\Omega^2)\mathbf{r}^2,\label{H_rot}
\end{eqnarray}
where we introduced the angular momentum projection $ L_z=m\,(\mathbf{r}\times\mathbf{p}) \cdot \hat{\mathbf{z}}$ and the effective gauge field $\mathbf{A}=m\Omega\,\hat{\mathbf{z}}\times\mathbf{r}$; see Eq.~\eqref{B_field_Ham}. The Hamiltonian (\ref{H_rot}) is formally equivalent to the one of a particle of charge 1, placed in a uniform magnetic field $\mathbf{B}=2m\Omega\,\hat{\mathbf{z}}$ and confined in a harmonic trap of frequency $\sqrt{\omega^2-\Omega^2}$. Note that when  the centrifugal force exceeds the bare trapping force, i.e. $\Omega>\omega$, the gas becomes unstable.

\begin{figure}
\begin{tikzpicture}
\node  at (0,0) {\includegraphics[width=0.42\linewidth]{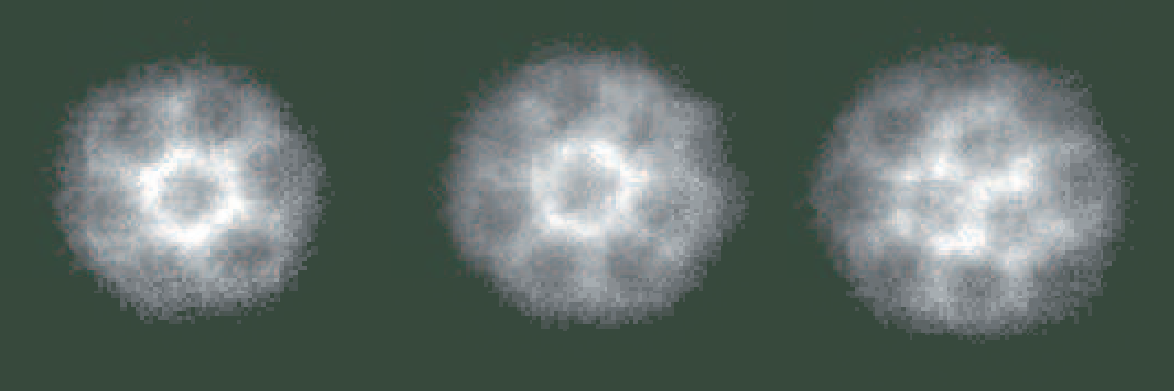}};
\node  at (6.3,0) {\includegraphics[width=0.3\linewidth]{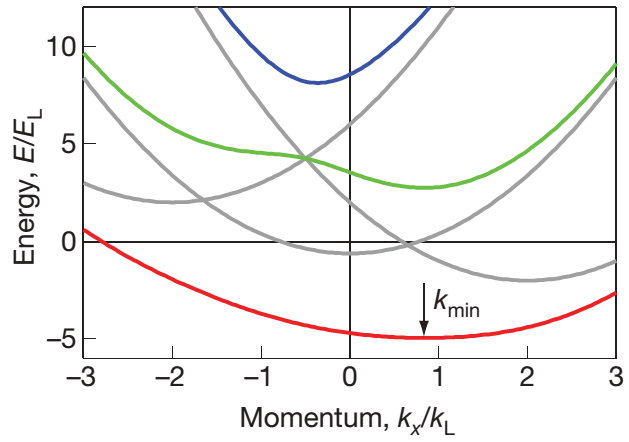}};
\node  at (11.,0) {\includegraphics[width=0.25\linewidth]{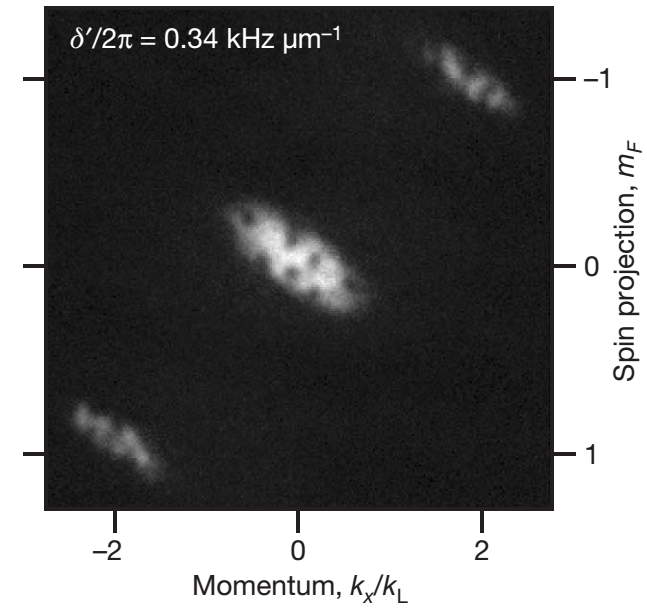}};
\node  at (-3.3,1.7) {(a)};
\node  at (3.9,1.7) {(b)};
\end{tikzpicture}
\caption{\textit{\textsc{Rotation-induced versus laser-induced magnetic fields}} (a) Abrikosov vortex lattices formed in rotating Bose-Einstein condensates (from \cite{madison_vortices_2000}). (b) Dispersion relation induced by the optical dressing used in reference \cite{lin_synthetic_2009}, and observation of quantized vortices in Bose-Einstein condensates subjected to this optical dressing and an additional magnetic field gradient (from \cite{lin_synthetic_2009}). Here $k_L$ and $E_L$ denote the wave-vector and the recoil energy associated with the Raman beams~\cite{lin_synthetic_2009}. \label{Fig_vortices}}
\end{figure}

By analogy with superfluid liquid He \cite{donnelly_quantized_1991} or type-II superconductors \cite{tilley_superfluidity_1990}, rotating Bose-Einstein condensates accomodate the effective magnetic field via the formation of quantized vortices, arranged in an Abrikosov lattice \cite{Madison:2000fg,AboShaeer:2001go} (see figure \ref{Fig_vortices}a). This behavior, considered as a smoking gun of superfluidity, also occurs in strongly-interacting Fermi gases in the superfluid phase, i.e. a condensed phase of atomic fermion pairs \cite{zwierlein_vortices_2005}. Complex vortex structures are expected to occur  in the presence of long-range electric or magnetic dipole forces \cite{baranov_theoretical_2008,lahaye_physics_2009} or in the case of a spinor internal structure \cite{kawaguchi_spinor_2012,stamper-kurn_spinor_2013}, for which unconventional Abrikosov lattices were observed \cite{schweikhard_vortex-lattice_2004}. 

In the rapid rotation limit $\Omega-\omega\ll\Omega$, the vortex density becomes high enough for the vortex cores to significantly overlap \cite{fischer_vortex_2003,cooper_rapidly_2008}. In this regime -- which was  reached in experiments  \cite{Schweikhard:2004hk,Bretin:2004jj} -- the gas effectively occupies a subset of quantum states formally equivalent to the lowest Landau level. For vortex numbers comparable to the number of trapped atoms, one expects the atomic gases to form strongly correlated states \cite{cooper_quantum_2001,cooper_rapidly_2008}, analogs to the ones observed in the fractional quantum Hall effect \cite{prange_quantum_1990}. This regime has not yet been reached in experiments, due to the very strong requirements on the trapping potential control required close to the centrifugal limit \cite{Bloch:2008RMP}.

\subsection{Berry phase induced by laser dressing\label{section_laser_dressing}}
As an alternative to rotation, another class of protocols were developped to produce orbital magnetic fields for atomic gases, based on the dressing of atoms with suitable laser fields \cite{Dalibard:2011RMP,Goldman:2014RPP}. As we described in Section~\ref{berry_effective}, by coupling the atom's motion to its internal spin, one induces position-dependent energy landscapes. An atom that adiabatically moves along a closed contour then acquires a geometrical Berry's phase \cite{Berry:1984PRS}, which can formally be viewed as the Aharonov-Bohm phase acquired by a charged particle evolving in the presence of a magnetic field \cite{aharonov:1959PR}; see Section~\ref{sect:universal}.

The physical mechanism for generating light-induced gauge fields can be illustrated assuming an atomic internal spin $F=1/2$. Two-photon optical transitions couple the quantum states $\left|m_F=-1/2,\mathbf{k}-\mathbf{K}\right\}$ and $\left|m_F=1/2,\mathbf{k}+\mathbf{K}\right\}$, where $2\mathbf{K}$ is the momentum imparted to the atoms upon a laser-induced spin-flip. Here, $m_F$ denotes the spin projection on the $z$ axis and $\mathbf{k}$ is the atom momentum. The quasi-momentum $\mathbf{q}=\mathbf{k}+\mathbf{K}\sigma_z$ is a conserved quantity, which allows writing the single-particle Hamiltonian  as
\begin{eqnarray}
H&=&\sum_{\mathbf{q}}H_{\mathbf{q}},\label{eq_laser_dressinga}\\
H_{\mathbf{q}}&=&\frac{\hbar^2(\mathbf{q}-\mathbf{K}\sigma_z)^2}{2m}+\frac{1}{2}\hbar\Omega_R\sigma_x+\frac{1}{2}\hbar\delta\sigma_z,\label{eq_laser_dressingb}
\end{eqnarray}
where  $\Omega_R$ is the two-photon Rabi frequency and $\delta$ is the detuning from resonance. The ground state dispersion relation features a minimum at $\mathbf{q}=\mathbf{q}_{\mathrm{min}}$, which can be written as 
\[
\mathbf{q}_{\mathrm{min}}\simeq\frac{\delta}{\sqrt{\delta^2+\Omega_R^2}} \mathbf{K},
\]
in the large coupling regime $\hbar\Omega_R\gg\hbar^2K^2/2m$. Around the minimum, the dispersion relation expands as $E_q=E_{q_{\mathrm{min}}}+\hbar^2(\mathbf{q}-\mathbf{q}_{\mathrm{min}})^2/2m$, allowing one to interpret $\mathbf{q}_{\mathrm{min}}$ as an effective vector potential $\mathbf{A}/\hbar$ (see figure \ref{Fig_vortices}b).

Using a gradient of the Raman laser intensity \cite{juzeliunas_slow_2004}, or by applying an additional magnetic field gradient \cite{spielman_raman_2009}, the vector potential can be made position-dependent, hence inducing an effective magnetic field $\mathbf{B}=\nabla\times\mathbf{A}$. The generation of such a light-induced magnetic field was experimentally demonstrated in reference \cite{lin_synthetic_2009}, via the observation of quantized vortices piercing the Bose-Einstein condensate (see figure \ref{Fig_vortices}b). 

For optical fields of wavelength $\lambda$, the expected order of magnitude for the vector potential is $A\sim h/\lambda$. Hence, the magnetic flux $N_{\phi}$ experienced by an ultracold gas of extent $L$, given by the contour integral around the area of interest $N_\phi=\oint \mathbf{A}\cdot d\mathbf{l}$, scales as $L/\lambda$. This moderate value limits the potential of this technique to reach the fractional quantum Hall regime, corresponding to an atom number $N\simeq
N_\phi$. A promising approach to increase the magnetic flux density was proposed in reference \cite{cooper_optical_2011}, which introduces the concept of optical flux lattices, in which spin-dependent optical lattices are used to imprint a periodic magnetic flux density of large mean density $n_\phi\sim\lambda^{-2}$.

\subsection{Synthetic magnetic fields for photons}
Recent progress on the engineering of continuum fluids of light~\cite{carusotto:2013} allowed for the simulation of orbital magnetic fields, thus opening new perspectives in the photonic framework.

\begin{figure}
\includegraphics[width=\linewidth]{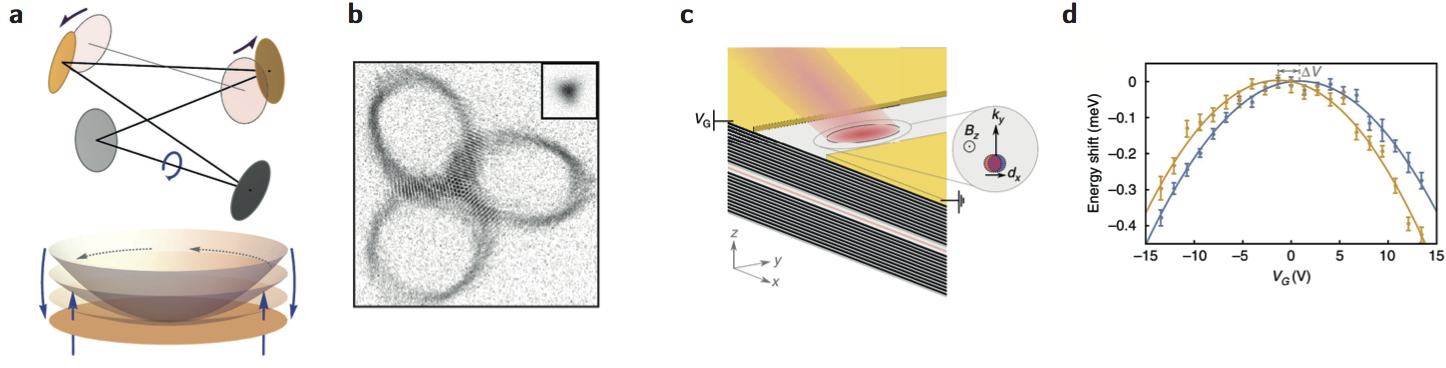}
\caption{\textit{\textsc{Synthetic magnetic fields for photons}} (a) Scheme of the non-planar optical cavity used to engineer optical Landau levels. The effective magnetic field is induced by the rotation of the cavity modes around the optical axis upon every round trip.
(b) Example of optical mode in the photonic Landau level, implying a large number of angular momentum modes (from \cite{schine2016synthetic}). 
(c) Scheme of the quantum well structure used to create cavity polariton gases with an artificial gauge field arising from the magnetoelectric Stark effect occuring under perpendicular electric and magnetic fields.
(d) Dispersion relation of the polariton modes as a function of the gate potential $V_G$, taken for two values of momenta $k_y$. The shift between the two parabolas corresponds to an effective vector potential controlled by the external magnetic field (from \cite{lim2017electrically}).  
\label{Fig_magnetic_field_light}}
\end{figure}

Photons trapped in a multimode optical cavity behave as a two-dimensional system of bosonic massive particles \cite{carusotto:2013}. The realization of Bose-Einstein condensation of photons in such physical systems \cite{klaers2010bose} brings novel perspectives in quantum simulation with optical photons. For instance, by using  a non-planar optical resonator, the optical mode can undergo a rotation on each round trip, leading to an effective magnetic field \cite{sommer2016engineering,schine2016synthetic} (see figures \ref{Fig_magnetic_field_light}a,b). Furthermore the motion of photons can be restricted to a cone, which allows exploring magnetic field effects in an effectively curved space \cite{schine2016synthetic}. The dynamics of resonator optical modes can be understood using the Floquet formalism described in section \ref{section_floquet}, where the time periodicity corresponds to the cavity round trip duration; in this prescription, the effective magnetic field can be readily estimated by calculating the effective Hamiltonian (see Ref.~\cite{sommer2016engineering,schine2016synthetic} and also Section \ref{sect:floquetuniform} below). Other proposals for creating artificial magnetic fields involve anamorphic optical elements \cite{longhi2015synthetic} or the combination of optical non-linearity with pump  fields
carrying orbital angular momentum \cite{westerberg2016synthetic}. The realization of photonic analogs of fractional quantum Hall states will require combining these techniques with the implementation of strong effective photon-photon interactions, e.g. using Rydberg electromagnetically induced transparency  \cite{pritchard2010cooperative,gorshkov2011photon,peyronel2012quantum,firstenberg2013attractive}.

In the last decade, the progress in the control over exciton-polariton gases led to the observation of their Bose-Einstein condensation \cite{kasprzak2006bose,balili2007bose,deng2002condensation}, opening new perspectives in the study of  bosonic superfluidity \cite{deng2010exciton,carusotto:2013,byrnes2014exciton}. Exciton-polaritons consist of the superposition of an exciton and a photon located inside semiconductor microcavities. They behave as bosonic quasi-particles that can arrange under appropriate pumping into a Bose-Einstein condensate  \cite{kasprzak2006bose,balili2007bose,deng2002condensation} exhibiting superfluid behavior \cite{amo2009superfluidity,amo2009collective} and solitonic/vortex excitations \cite{lagoudakis2008quantized,lagoudakis2009observation,sanvitto2010persistent,roumpos2011single,amo2011polariton,nardin2011hydrodynamic}. Several methods were developed to provide static or dynamically controlled in-plane potentials \cite{byrnes2014exciton}. Recently, the perspectives of polariton fluids for quantum simulation were enlarged by the realization of artificial gauge fields induced by the magnetoelectric Stark effect occuring under perpendicular electric and magnetic fields \cite{lim2017electrically} (see figures \ref{Fig_magnetic_field_light}c,d). Other types of hybrid light-matter quantum systems, such as stationnary light polaritons \cite{zimmer2008dark,bajcsy2003stationary}, could also be coupled to artificial gauge fields \cite{otterbach2010effective}.

\subsection{A Floquet realization of uniform magnetic fields}
\label{sect:floquetuniform}

Let us conclude this section by analyzing how effective magnetic fields naturally emerge through a Floquet-engineering approach (see Section~\ref{section_floquet}). Our discussion is based on a direct application of Eq.~\eqref{eq_expansion}, in the case where a quantum system is time-modulated according to a repeated four-step sequence of the form
\begin{equation}
\{  H_0+ A,  H_0+  B,  H_0- A, H_0-  B \}.\label{alpha_sequence}
\end{equation}
Here, the Hamiltonian $ H_0$ describes the system in the absence of the modulation, and $ A$ and $ B$ are arbitrary (Hermitian) operators. Following Ref.~\cite{Goldman:PRX2014}, the effective Hamiltonian $ H_{\text{eff}}$ of such a ``quantum walk" can be approximated by
\begin{align}
& H_{\text{eff}}\!=\!  H_0 \!+\! \frac{i \pi}{8   \omega} [A,  B] \!+\! \frac{\pi^2}{48  \omega ^2} \! \left (  [[  A,  H_0],  A] \!+\!  [[  B,  H_0], B] \right ) \!+\! \mathcal{O} (1/ \omega ^3),\label{four-step}
\end{align}
in the high-frequency limit. Based on this result, one can readily identify specific protocols leading to effective magnetic fields; as suggested in Ref.~\cite{Goldman:PRX2014}, a possible scheme is provided by the following set of operators: $ H_0=( p_x^2+  p_y^2)/2m$, $ A=( p_x^2-  p_y^2)/2m$ and $ B= \kappa x  y$. Inserting these operators in Eq.~\eqref{four-step} indeed results in the effective Hamiltonian 
\begin{align}
&H_{\text{eff}} = \frac{1}{2m} \left ( \left (  p_x - \mathcal A_x \right)^2 + \left (  p_y - \mathcal A_y  \right) ^2   \right ) + \frac{1}{2} m \omega_h^2 ( x^2 +  y^2) \notag \\
&\bs{\mathcal{A}} = (-m \Omega  y,  m \Omega  x) , \quad \Omega=\frac{\pi \kappa}{8 m \omega} , \quad \omega_h=\sqrt{\frac{5}{3}} \Omega,\label{bulk_eff_ham}
\end{align}
which describes the motion of a charged particle in the 2D plane, in the presence of a perpendicular magnetic field $\bs B = 2 m \Omega  \bs 1_z=\frac{\pi \kappa}{4 \omega} \bs 1_z$. Note that this scheme simultaneously realizes a trapping potential in Eq.~\eqref{bulk_eff_ham}. Physically, the four-step sequence introduced above  reads 
\begin{equation}
\left \{   \frac{p^2_x}{m} \, ,  \,  \frac{p^2_x+ p^2_y}{2m} + \kappa  x y  \, ,  \, \frac{p^2_y}{m}  \, ,  \, \frac{p^2_x+p^2_y}{2m} - \kappa x y  \right \}, \label{bulk_sorensen}
\end{equation}
which corresponds to restricting the direction of motion and applying pulsed quadrupole fields in a sequential manner. We point out that a similar scheme was originally proposed in an optical-lattice context~\cite{Sorensen:2005PRL}, to generate fractional Chern insulators~\cite{Parameswaran:2013fractional}. Besides, schemes based on pulsed magnetic fields were also proposed to generate spin-orbit coupling~\cite{Xu:2013PRA,Anderson:2013PRL,Goldman:PRX2014}, as was recently implemented in Ref.~\cite{Luo2016Scientific}.

\section{Artificial magnetic fields in lattices}
\label{}

\subsection{Physical realizations of uniform magnetic fluxes and the Harper-Hofstadter model}

\subsubsection{The Harper-Hofstadter model: From the butterfly to topology}\label{Sect:Hof}

In the continuum, the energy spectrum of a 2D electron gas in a magnetic field consists of highly degenerate Landau levels. At the same time electrons moving in a periodic potential develop a quantized energy spectrum consisting of discrete Bloch bands. As a consequence, the interplay between the two characteristic length scales, the lattice constant $a$ of the periodic potential and the magnetic length $l_B=\sqrt{\hbar/eB}$, leads to the emergence of a complex fractal energy spectrum know as Hofstadter's butterfly \cite{Harper:1955bj,Azbel:1964tk,Hofstadter:1976PRB}; see Fig.~\ref{Fig:Butterfly}b. It has been predicted more than 60 years ago~\cite{Hofstadter:1976PRB} but was only realized very recently, as we now explain. The challenge of exploring this ``Hofstadter butterfly physics" experimentally lies in the reconciliation of the two lengthscales. To have the magnetic length on the same order as the lattice constant, which is typically a few \aa ngstr\"oms large, unfeasibly large magnetic fields of several thousands Tesla would be required. One possibility to overcome this limitation is to lithographically design artificial superlattices with lattice constants up to $100\,$nm, a technique that has proven to be technically challenging mainly because of disorder \cite{Gerhardts:1991fl,Nakamura:1998cw,Albrecht:1999hv,Schlsser:1996ct,Albrecht:2001kz,Geisler:2004ei,Melinte:2004gb}. A related strategy is based on heterostructures, consisting of atomically thin materials, which can be used to engineer artificial superlattice structures; this can effectively increase the lattice unit cell, and therefore the lattice constant, depending on the relative alignment between the different layers. Three independent groups have succeeded to study the electronic properties of graphene placed on a substrate of hexagonal boron nitride \cite{Dean:2013bv,Ponomarenko:2013hl,Hunt:2013ef}, which exhibits a so-called moir\'e pattern; the latter acts as a superlattice with a lattice constant on the order of $10\,$nm and facilitated first studies of the fractal structure of the Hofstadter's butterfly in electronic systems. 

More generally, the Harper-Hofstadter model [Eq.~\eqref{TB_Hofstadter}] describes the dynamics of a charged particle in a 2D space-periodic system subjected to a constant external magnetic field (see Sect.~\ref{sect:hofstadter}). The model explicitly breaks time-reversal symmetry, through the Peierls phase factors in Eq.~\eqref{TB_Hofstadter}, which leads to the appearance of topologically non-trivial energy bands~\cite{Thouless:1982PRL,Bernevig:2013book}. The important parameter characterizing the electronic properties is the number of magnetic flux quanta per unit cell $\alpha$. The corresponding fractal single-particle energy spectrum (Fig.~\ref{Fig:Butterfly}b) emerges due to the broken translational symmetry of the lattice in the presence of the vector potential. The magnetic field effectively enlarges the unit cell \cite{Hofstadter:1976PRB,Chang:1996gv} and one can show that for rational values $\alpha=p/q$, with $p$ and $q$ integers, the effective unit cell is $q$ times larger and the lowest tight-binding band of the periodic potential splits into $q$ subbands. 

The general concept of geometric phases introduced in Sect.~\ref{sect:IntroGeoPhase} naturally translates to crystalline potentials \cite{Xiao:2010RMP}. Using Bloch's theorem, the Hamiltonian can be written in the momentum representation ${H}(\mathbf{k})$, whose eigenvectors are the cell-periodic Bloch functions $|u(\mathbf{k})\rangle_{\mu}$, where $\mu$ is the band index and where the parameter space is determined by the quasi-momentum $\bs k$ defined within the first Brillouin zone (FBZ). Adiabatic motion within a given Bloch band $E_{\mu}$ is associated with a geometric phase, which is characterized by the Berry curvature $\mathbf{\Omega^{\mu}_{\text{geom}}(\mathbf{k})}$; see Eq.~\eqref{Berry_curvature} and Ref.~\cite{Xiao:2010RMP}. As previously discussed, this geometric phase can be seen as an Aharonov-Bohm phase associated with the fictitious ``magnetic" field $\Omega^{\mu}(\mathbf{k})$ defined in quasi-momentum space. Beyond these geometric effects, energy bands also display topological properties, which are characterized by a topological invariant known as the Chern number~\cite{Thouless:1982PRL,Bernevig:2013book}
\begin{equation}
\nu_{\mu} = \frac{1}{2\pi}\int_{\text{FBZ}} \text{d}^2k\ \mathbf{\Omega^{\mu}_{\text{geom}}(\mathbf{k})}.
\label{eq:defChern}
\end{equation}
The latter can be interpreted as the number of monopole charges associated with the fictitious field $\Omega^{\mu}(\mathbf{k})$ within the FBZ~\cite{Wu:1975PRD}. Hence, the Chern number $\nu_\mu$ only takes  integer values and the energy band $E_{\mu}$ is topologically non-trivial if $\nu_\mu \neq 0$. This situation can give rise to intriguing physical effects. The most prominent example is the quantization of the Hall conductance in the integer quantum Hall effect \cite{Klitzing:1980kw,Klitzing:1986RMP,Thouless:1982PRL}, $\sigma_H \!=\! \frac{e^2}{h} \sum_{E_{\mu}<E_F} \nu_{\mu}$, where the summation runs over all occupied energy bands below the Fermi energy $E_{\mu}<E_F$. Indeed, one can show that the number of chiral edge modes, which contribute to transport in quantum-Hall experiments, is directly linked to the Chern numbers characterizing the occupied single-particle energy bands; this is the so-called ``bulk-edge correspondence"~\cite{Hatsugai:1993fc,Hatsugai:1993kd,Qi:2006co}. In fact, as will be illustrated below, the appearance of these chiral edge modes offers alternative experimental possibilities to reveal and study the topological properties of a system. 

Even though it seems natural to study the physics of the Hofstadter model with charged particles (e.g.~electrons) in square lattices subjected to a magnetic field, there has been a number of proposals and experiments for charge-neutral particles in different physical settings. In the following of this Section, we briefly introduce different platforms, where the physics of the Harper-Hofstadter model has been studied or is within reach. These include photonics \cite{Lu:2014NP,Lu:2016ib}, phonons \cite{Huber:2016tg} and cold atoms \cite{Goldman:2014RPP,Goldman:2016fa}. In some of these platforms, first steps have been undertaken to realize topological insulators, which are topological states that do not break time-reversal symmetry~\cite{Hasan:2010RMP}; the simplest realization of such topological insulators are two independent copies of a quantum-Hall system, whose magnetic fields point in opposite directions~\cite{Bernevig:2006PRL,Goldman:2010PRL}.

\subsubsection{Microwave (MW) and radiofrequency (RF) circuits}

\begin{figure}
\includegraphics{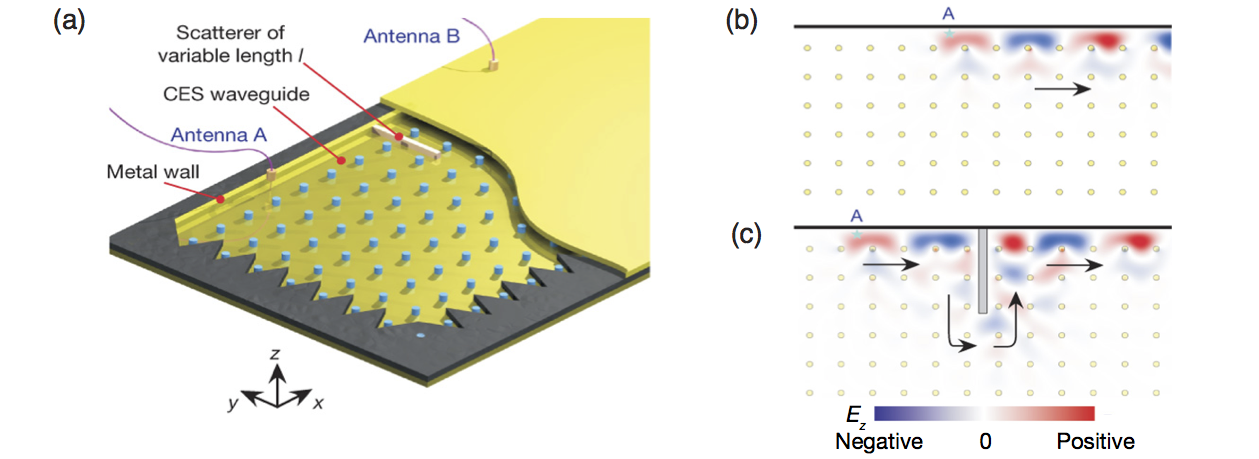}
\centering
%\vspace{0.4cm} 
\caption{\textit{\textsc{Topological microwave waveguide supporting chiral edge states} 
(a) Illustration of the 2D magneto-optic photonic crystal that consists of an array of ferrite rods in air (blue). The two parallel copper plates (yellow) confine the radiation in the $z$-direction. Chiral edge states (CES) exist at the interface between the topologically trivial metal wall and the photonic crystal. The two antennas (orange) are used for emission and detection of the MW radiation.
(b) Calculated field distribution $E_z$ of the CES. The signal propagates along the direction of the black arrow although the antenna (star) emits omnidirectional. 
(c) In the presence of a large metal obstacle the signal is expected to propagate around the obstacle because backscattering is completely suppressed. (Figure from Ref.~\cite{Wang:2009jo})}
\label{Fig:metamaterials}}
\end{figure}

Inspired by the topological interpretation of the integer quantum Hall effect~\cite{Thouless:1982PRL}, it was realized that similar phenomena could be realized and studied in photonic systems. The key idea was built on periodic metamaterials that explicitly include time-reversal symmetry breaking elements. First, this was proposed via the use of Faraday-effect media \cite{Haldane:2008cc,Raghu:2008do}. In these structures, the emergent photonic bands are characterized by non-zero Chern numbers, as in the electronic case; similarly, at the interface between two materials with different topological orders, unidirectional (chiral) edge modes are expected to appear~\cite{Hasan:2010RMP}. In their original proposal, Haldane and Raghu considered a hexagonal array of dielectric rods, where an additional Faraday term leads to a splitting of the Dirac points, hence creating non-degenerate bands with non-zero Chern numbers \cite{Haldane:2008cc}. Soon after, several experiments reported on the observation of chiral edge states (CES) in gyromagnetic photonic crystals based on square lattice geometries \cite{Wang:2008ie,Wang:2009jo,Fu:2010kg,Fu:2011fk}, as illustrated in Fig.~\ref{Fig:metamaterials}. CESs are remarkably robust to disorder, which may be introduced during the fabrication process. Therefore, topological systems are very promising candidates for technological developments such as the engineering of extremely robust waveguides (Fig.~\ref{Fig:metamaterials}c). Recently, these ideas have been extended in order to enable the realization of topological photonic bands with Chern numbers that are larger than one $|\nu|>1$ \cite{Skirlo:2014gx}; this permits the generation of multi-mode channels, as recently demonstrated by Skirlo \textit{et al.} \cite{Skirlo:2015gv}. 

\noindent It is well-known that characteristics of the 2D Harper-Hofstadter model can also be studied in suitably tailored 1D geometries; formally, this results from the dimensional reduction of the 2D Harper-Hofstadter model, which leads to the 1D Harper equation~\cite{Hofstadter:1976PRB}. This idea has indeed been used in pioneering experiments to study the fractal structure of the Hofstadter butterfly in a 1D geometry, where the transmission of a microwave signal through a waveguide with a variable periodic arrangement of scatterers was measured \cite{Kuhl:1998ev}. One can show that the transmission of the signal through an array of scatterers is formally described by the Harper equation~\cite{Harper:1955bj}.

An elegant way to implement the Harper-Hofstadter model in a RF-network was demonstrated in Ref.~\cite{Ningyuan:2015bx}. In this setup each lattice site consists of two inductors $A=(1,0)$ and $B=(0,1)$; the state on each lattice site is defined as the voltages across the inductors $(V_A,V_B)$. Neighboring lattice sites are capacitively coupled via permuted couplings to introduce a phase shift of $\pi/2$ for each round trip around one unit cell, which effectively implements the Harper-Hofstadter model for $\alpha=1/4$ (see Sect.~\ref{sect:hofstadter}). The two inductors further encode two different spin states, which results in an effective model that is time-reversal symmetric and supports edge modes with opposite chirality for the two spin states. Through site- and spin-resolved measurements the authors were able to study the dynamics of edge states and reveal the bulk energy bandgap~\cite{Ningyuan:2015bx}. 

\begin{figure}
\includegraphics{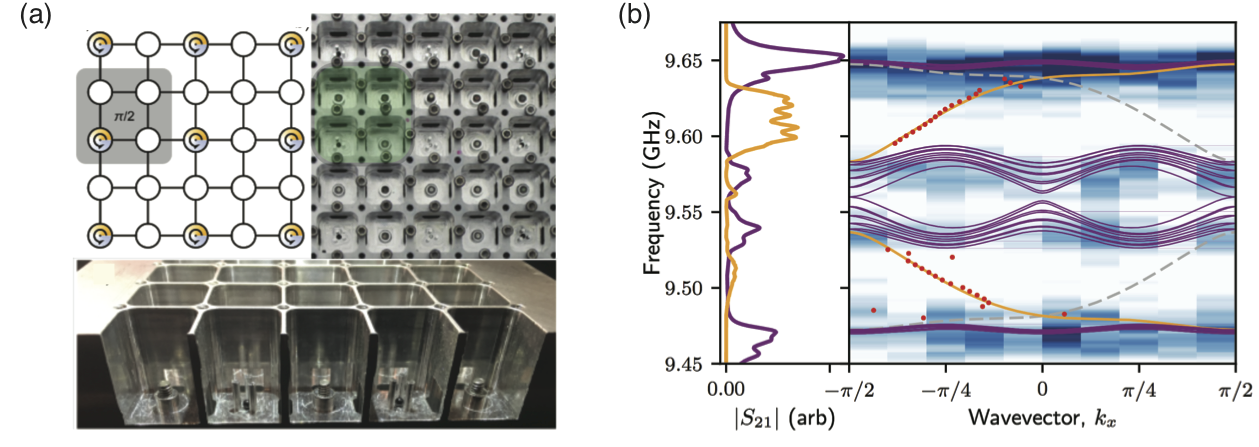}
\centering
%\vspace{0.4cm} 
\caption{\textit{\textsc{Microwave cavity array.} 
(a) Schematic drawing and photograph of the microwave cavity array simulating the Hofstadter model with flux $\alpha=1/4$. (top-left panel) The white circles represent cavities that do not introduce a phase shift, while the colored ones with an arrow introduce a phase shift of $\pi/2$ per lattice unit cell. (top-right panel) Top-view of a $5\times 5$ lattice with a lattice constant of $1.96\,$cm. (bottom panel) Side-view of the cavity array. The 2nd and 4th cavities are phase-shifting cavities. (b) Transmission spectrum (left) and energy bands (right) of the Hofstadter model for $\alpha=1/4$. The bulk response is shown in purple and blue, while the one from the edges is shown in orange and red. The four bulk bands originate from the increased magnetic Brillouin zone in the presence of the flux. (from Ref.~\cite{Owens:2017tw})}
\label{Fig:MWcavities}}
\end{figure}

Alternatively, time-reversal symmetry can be broken without the need of non-reciprocal elements, but using Floquet engineering (Sect.~\ref{section_floquet}). The basic theoretical concept relies on modulation-induced interband photonic transitions. In the work by Fang \textit{et al.} \cite{Fang:2012js}, the authors consider two photonic modes with frequencies $\omega_A$ and $\omega_B$ that are spatially separated and arranged in a square lattice geometry with sublattices $A$ and $B$. The key ingredient is a time-dependent coupling between them, which can be described by the following Hamiltonian 
\begin{equation}
{H}(t) = \omega_A \sum_i {a}^{\dagger}_i {a}_i^{\phantom{\dagger}} + \omega_B \sum_j {b}^{\dagger}_j {b}_j^{\phantom{\dagger}} + \sum_{<ij>} V \cos (\omega t +\phi_{ij}) ({a}^{\dagger}_i {b}_j^{\phantom{\dagger}}+{b}^{\dagger}_j {a}^{\phantom{\dagger}}_i),
\label{eq:redrivingresonators}
\end{equation}
\noindent where ${a}^{\dagger}_i$ and ${b}^{\dagger}_j$ are the creation operators in the two sublattices on site $i$ and $j$ respectively. The coupling strength $V$ is resonantly modulated with frequency $\omega = \omega_A-\omega_B$ and $\phi_{ij}$ is the phase of the modulation. In the rotating-wave approximation, namely following the formalism discussed in Sect.~\ref{section_floquet}, the evolution of the system can be described in the high-frequency limit by an effective time-independent Hamiltonian ${H}_{\text{eff}}$ with coupling matrix elements $V \text{e}^{i \phi_{ij}}/2$; therefore, this enables a direct engineering of effective Aharonov-Bohm phases in real space. Conceptually this technique is very general and can be realized in various photonic systems, ranging from the RF to the visible spectrum. In fact, there are different physical settings that make use of the same ``resonant-shaking" technique in order to engineer artificial magnetic fields, for instance, with cold atoms \cite{Goldman:2015kc}, ions \cite{Bermudez:2011PRL} or superconducting circuits \cite{Roushan:2016iu}. A first experiment on the engineering of artificial Aharonov-Bohm phases in the RF regime, based on this method, was reported in Ref.~\cite{Fang:2013ch}.

In photonic setups, a major challenge consists in mediating interactions between photons. Recently, a promising platform was developed based on microwave cavities where interactions can be introduced via Josephson junctions, which might enable future studies of topological many-body physics \cite{Owens:2017tw,Wallraff:2004dy,Anderson:2016kf}. The platform consists of an array of 3D microwave cavities, as illustrated in Fig.~\ref{Fig:MWcavities}a, which are evanescently coupled. In this scheme a time-reversal symmetry breaking flux is engineered on-site by manipulating the mode structure of individual cavities \cite{Anderson:2016kf}. This is in contrast with respect to other implementations, which are mainly based on the engineering of Peierls phases (see Sect.~\ref{sect:hofstadter}). Here, one cavity within each unit cell is engineered in such a way that a single mode with definite angular momentum is energetically isolated. By evanescently coupling the cavities, the angular momentum passively introduces an artificial flux in the photonic lattice (Fig.~\ref{Fig:MWcavities}a). This technique has been demonstrated for the Hofstadter model with flux $\alpha=1/4$ \cite{Owens:2017tw}. Using spectroscopic measurements, the authors revealed four bulk topological energy bands and topologically protected edge states (Fig.~\ref{Fig:MWcavities}b). The large potential of this platform relies in the fact that the cavities can be coupled to superconducting qubits in order to introduce strong photon-photon interactions \cite{Anderson:2016kf}. This could offer a promising platform to study fractional Chern insulators~\cite{Parameswaran:2013fractional}.

\subsubsection{Ultracold atoms in optical square lattices}
\label{sect:atomslattice}
\begin{figure}
\includegraphics[width=\textwidth]{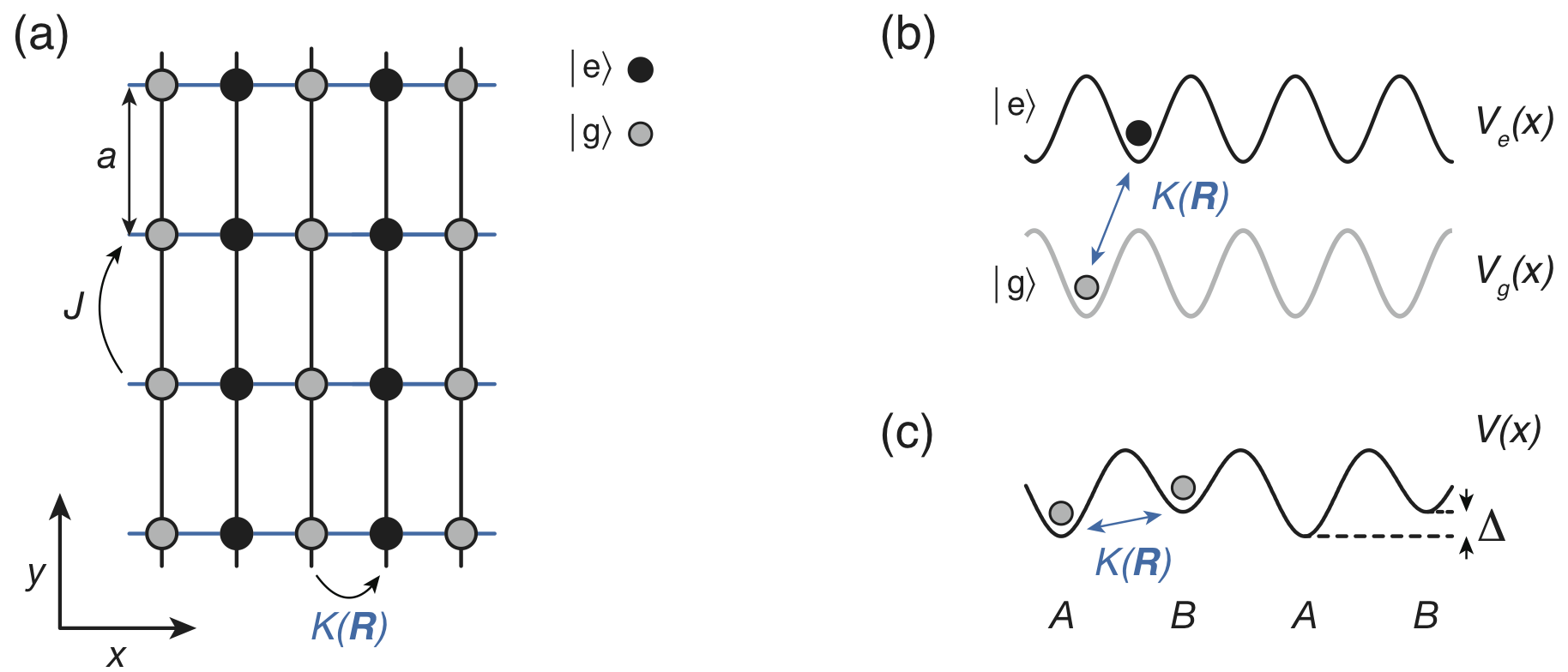}
\centering
%\vspace{0.4cm} 
\caption{\textit{\textsc{Laser-induced tunneling in an optical lattice.} 
(a) Schematic drawing of a 2D spin-dependent optical lattice; $|g\rangle$ atoms are trapped between $|e\rangle$. Along $y$ tunneling is determined by the hopping matrix element $J$ and along $x$ it is induced via additional laser beams. It is generally complex $K(\mathbf{R})$ and depends on the position $\mathbf{R}$ in the lattice.
(b) Illustration of Raman-assisted tunneling, where two internal states $|g\rangle$ and $|e\rangle$ are trapped in different lattice potentials $V_{g,e}(x)$ and are coupled with near-resonant lasers to realize $K(\mathbf{R})$.
(c) Laser-assisted tunneling in a superlattice potential with sublattices $A$ and $B$. Tunneling is inhibited by a potential energy offset $\Delta$ and can be resonantly restored by far-detuned laser beams to engineer complex hoppings $K(\mathbf{R})$.}
\label{Fig:JakschZoller}}
\end{figure}

In cold atoms, artificial magnetic fields have  been first realized within harmonic traps. As explained in Sect.~\ref{sect:rotation}, by setting a cold gas into rotation one can exploit the formal equivalence between the Coriolis force and the Lorentz force, which has been demonstrated successfully through the observation of quantized vortices in a BEC \cite{Madison:2000fg,AboShaeer:2001go,Schweikhard:2004hk,Bretin:2004jj}. These ideas have been extended to rotating optical lattices \cite{Tung:2006fa,Hemmerich:2007bc,Williams:2010cb,Sachdeva:2010cd,Gemelke:2010uqa} and to even more exotic settings; e.g.~a rotating BEC was predicted to induce an effective magnetic field for impurity atoms trapped in an optical lattice~\cite{Klein:2009kn}. 

Later, alternative methods have been developed on possible ways to generate magnetic field effects for neutral particles in periodic potentials~\cite{Jaksch:2003gd,Ruostekoski:2002fs,Mueller:2004PRA,Sorensen:2005PRL,Gerbier:2010ho}. For cold atoms, tunneling in an optical lattice is typically determined by a hopping matrix element $J$, which is real [Eq.~\eqref{TB_effective}]. The proposal by Jaksch and Zoller~\cite{Jaksch:2003gd} considers a 2D optical lattice, where tunneling along one axis ($x$-axis) is inhibited by the use of a state-dependent optical lattice; moreover, atoms in different internal states $| g \rangle$ and $| e \rangle$ are trapped in different columns of the optical lattice; see Fig.~\ref{Fig:JakschZoller}a. Tunneling can then be resonantly restored using Raman lasers, that couple the two internal states and introduces a spatially dependent optical coupling (Fig.~\ref{Fig:JakschZoller}b). The latter is in general complex and can be written in the following form
\begin{equation}
K(\mathbf{R})_{| g \rangle \rightarrow | e \rangle} = \int \text{d} \mathbf{r}\ w^*(\mathbf{r}-\mathbf{R})\  \Omega_R\ \text{e}^{i \mathbf{k}\cdot \mathbf{r}} \ w(\mathbf{r}-\mathbf{R}-\mathbf{d}_x) = K_0 \text{e}^{i \mathbf{k}\cdot \mathbf{R}}.\label{eq:RamanCoupling}
\end{equation}
Here $\mathbf{R}$ is the position in the lattice, $\Omega_R$ is the amplitude of the coupling field, $w(\mathbf{r})$ is the localized Wannier function on site $\mathbf{R}$ and $\mathbf{d}_x$ is the lattice unit vector along the $x$-axis. The phase terms in the induced hopping matrix elements lead to a phase accumulation when a particle hops around a close loop that leads to an effective magnetic field of the following strength $\alpha = k_y a / (2\pi)$, where $k_y$ is the projection of the Raman wavevector along the perpendicular direction. Since the induced tunneling is accompanied with a spin-flip or change of the internal state, the technique is also known as \textit{Raman-assisted tunneling}. Without additional potentials this scheme would lead to the generation of a staggered effective magnetic field because $K(\mathbf{R})_{| g \rangle \rightarrow | e \rangle}=K^*(\mathbf{R})_{| e \rangle \rightarrow | g \rangle}$ and therefore it would result in a magnetic field with zero mean. In order to rectify the field an additional linear potential \cite{Jaksch:2003gd} or a superlattice potential \cite{Gerbier:2010ho} could be used. Both proposal thereby rely on the idea that an additional potential is needed in order to split the degeneracy in the resonance condition for the induced tunnel couplings $| g \rangle \rightarrow | e \rangle$ and $| e \rangle \rightarrow | g \rangle$ so as to control the sign of the induced phase.

Alternatively the scheme can be implemented with a single internal state if potential energy offsets between neighboring sites $\Delta_{\mathbf{R}}$ in the lattice are introduced (Fig.~\ref{Fig:JakschZoller}c). If this energy offset is large compared to the tunneling $\Delta_{\mathbf{R}} \gg J$ the dynamics along the corresponding lattice axis will be frozen. Resonant tunneling can then be restored with an additional pair of laser beams whose frequency difference matches the energy offset between neighboring sites \cite{Kolovsky:2011EPL,Creffield:2013gp}. The resulting flux is then computed in a similar manner $\alpha=\delta k_y a / (2\pi)$, where $\delta k$ is the wavevector difference between the two driving laser beams. In this scheme the relevant energy scales can be on the same order of magnitude $\Delta \sim J$ and it may be advantageous to describe the explicitly time-dependent system in the Floquet formalism (see Sect.~\ref{section_floquet} and Ref.~\cite{Goldman:2015kc}). Since only one internal state of the atom is involved in this scheme it is typically referred to as \textit{photon-assisted} or \textit{laser-assisted tunneling}. Depending on the spatial profile of the additional potential energy used to inhibit tunneling different flux patterns can be realized. An alternating energy offset $\Delta_{\mathbf{R}}=(-1)^m \Delta$, where $m$ labels the lattice site $\mathbf{R}=m \mathbf{d}_x$ (Fig.~\ref{Fig:JakschZoller}c), leads to a staggered flux pattern with zero mean \cite{Aidelsburger:2011hl}. Instead a linear potential gradient $\Delta_{\mathbf{R}}=m \Delta$ results in homogeneous effective magnetic fields \cite{Aidelsburger:2013PRL,Miyake:2013PRL,Atala:2014uo,Kennedy:2015dw,Tai:2016arxiv}. This led to the observation of chiral Meissner-like currents in bosonic flux ladders \cite{Atala:2014uo}, but also to the first cold-atom measurement of the topologically-invariant Chern number~\cite{Aidelsburger:2014NatPhys}, which was achieved by loading bosonic atoms into the lowest Hofstadter band for $\alpha\!=\!1/4$; this topological transport experiment was performed by measuring the center-of-mass displacement of the atomic cloud in response to an applied force~\cite{Dauphin:2013PRL,Price:2016PRB,Dauphin:20172DMat}. In both schemes the strength of the effective magnetic fields depends only on the term $k_y a$, which can be tuned easily by changing the angle between the Raman laser and the underlying lattice potential or by changing the ratio between the wavevector of the Raman laser and the lattice constant. Therefore it is fully tunable by changing the geometry of the configuration. Similar theoretical proposals have been put forward for trapped ions, where the complex coupling is engineered via induced coupling of phonon modes \cite{Bermudez:2011PRL,Bermudez:2012gl}.

The topological properties of the 2D Harper-Hofstadter model can be further studied in a dynamical way, through the implementation of topological pumps in 1D~\cite{Thouless:1983PRB,Niu:1984JPA,Niu:1990PRL}. These ideas have been successfully realized in photonic systems  \cite{Kraus:2012kz,Verbin:2015gd,Wimmer:2017jw} and ultracold atoms \cite{Lohse:2016jua,Nakajima:2016vb,Lu:2016fl,Schweizer:2016ju}, and recently enabled studies of 4D quantum Hall physics based on 2D topological pumping \cite{Lohse:2017wx,Zilberberg:2017ww}.

\subsubsection{Ring-resonator arrays\label{section_ring_resonators}}

\begin{figure}
\includegraphics[width=\textwidth]{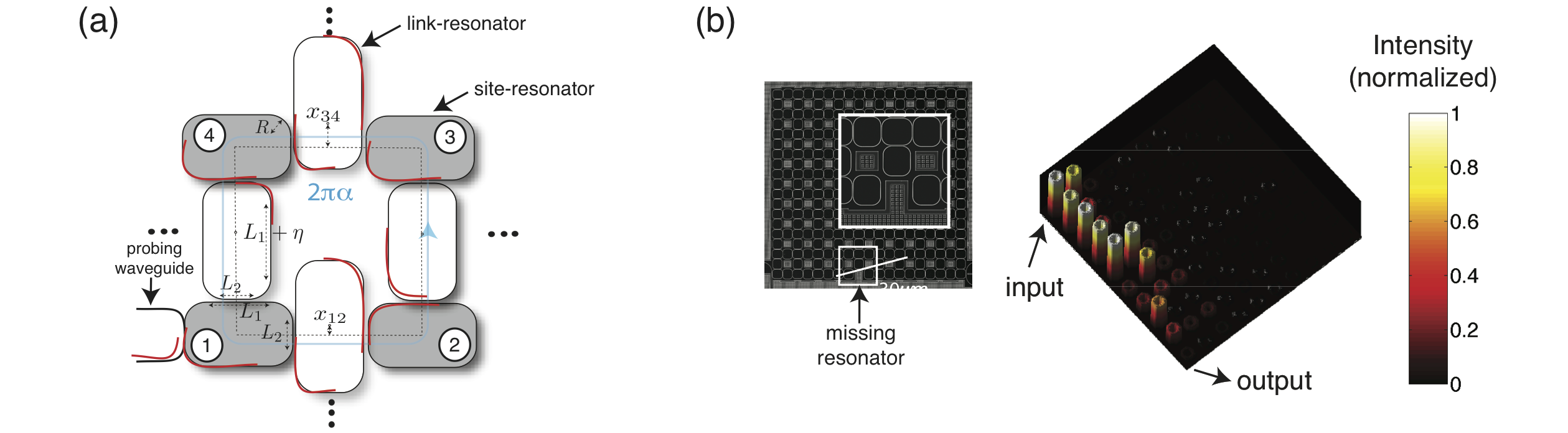}
\centering
%\vspace{0.4cm} 
\caption{\textit{\textsc{Effective magnetic fields in ring resonator arrays.} 
(a) Schematic drawing of the resonator array consisting of Link and Site resonators, which are resonant at different frequencies due to the length difference $2\eta$. Two Link resonators (1,2) and (3,4) are placed asymmetrically, i.e. $|x_{34}-x_{12}|>0$, which leads to an overall phase shift $2\pi\alpha$ for a photon hopping around the plaquette (blue arrow).
(b) Topological protection of the edge state. A resonator was intentionally removed from the array (left). The light is propagating around the missing resonator or defect, which indicates topological protection of the edge state.
(Image from Ref. \cite{Hafezi:2013jg})}
\label{Fig:RingResonator}}
\end{figure}

The Harper-Hofstadter model can also be  realized in photonic systems that consist of a network of ring resonators \cite{Liang:2013bu}. In this setting, one lattice site consists of a ring resonator and the spin is encoded in the propagation direction in the ring. Using waveguides these ring resonators can be coupled to neighboring resonators as discussed in Ref.~\cite{Hafezi:2011dt}. Restricting the discussion to a single spin component and neglecting spin mixing, the network can be described with a transfer matrix formalism which relates the wave amplitudes between neighboring resonators. The couplings are treated in a more abstract way through reflection and transmission coefficients, which is more general than a tight-binding description and works also in the strong-coupling regime. Interestingly, when applying this formalism one finds that the network hosts a topological insulating phase even without the usual addition of a tight-binding flux. While in the tight-binding description this zero-field case corresponds to a topologically trivial phase \cite{Hafezi:2011dt}, there exists a topological insulator in the strong-coupling regime \cite{Liang:2013bu}. A nice advantage of the transfer matrix formalism is that it is relatively easy to include gain and loss in the system. This topological regime of a network without additional tight-binding flux was explored recently in a MW network \cite{Hu:2015cz} via the implementation of a topological pump. 

An experimental implementation of the ring resonator network with tunable tight-binding flux was reported by Hafezi \textit{et al.} \cite{Hafezi:2013jg} based on silicon technology. In this setup the effective magnetic field is realized via asymmetric placements of site and link resonators such that a photon that is propagating clockwise around one plaquette will pick up a phase $2\pi \alpha$ or  $-2\pi\alpha$ if it is propagating anticlockwise. A schematic drawing of such a resonator plaquette is shown in Fig.~\ref{Fig:RingResonator}a. The system can be formally described by photons hopping on a lattice with non-zero Peierls phases, as in Eq.~(\ref{TB_Hofstadter}). The phase a photon acquires is determined by the optical length, for instance, $\phi_{12}=4\pi n x_{12}/\lambda$, where $x_{12}$ is the relative shift of the link resonator, $\lambda$ is the wavelength and $n$ the index of refraction. Since two of the link resonators are placed asymmetrically the photons acquire a non-zero phase shift when hopping around the plaquette, where $\alpha=2n(x_{34}-x_{12})/\lambda$, which can be interpreted as an effective magnetic flux piercing the unit cell. Hence, the spectrum is equivalent to that of the Hofstadter model and the system supports topological edge states, which have been observed experimentally \cite{Hafezi:2013jg} as shown in Fig.~\ref{Fig:RingResonator}b. Moreover this setting facilitated the first Chern-number measurement in a truly 2D photonic system, exploiting the bulk-edge correspondence~\cite{Mittal:2016cy}.

\subsubsection{Mechanical systems}
At first sight, there is no obvious connection between mechanical systems and topological quantum matter. However, the universality of geometrical and topological concepts revealed new possibilities in designing mechanical systems that are well suited to study certain aspects of topological insulators~\cite{Wang:2015NJP,Huber:2016tg,Susstrunk:2016dfa}. The field of topological mechanics has experienced an impressively fast development during a short period of time. Within the scope of this review, we present a brief intuitive link between classical mechanics and topological electronic systems followed by a list of important advances in the field. The easiest way to understand the connection is to map the mechanical problem onto an equation that formally resembles the Schr\"odinger equation. A set of coupled undamped linear mechanical oscillators can be described by the equations of motion
\begin{equation}
\ddot{x}_i(t) = \sum_{j=1}^{N}[ -D_{ij} x_j(t) + \Gamma_{ij} \dot{x}_j(t) ],
\label{eq:Eom}
\end{equation}

\noindent where $t$ denotes time, $x_i(t)$ describes the displacement of the $i$-th oscillator, $i \in \mathbb{N}$ and $D_{ij}$ are related to spring constants coupling different degrees of freedom \cite{Susstrunk:2015uo,Nash:2015eua}. The terms $\Gamma_{ij}$ arise from velocity-dependent forces and formally make the connection to the Lorentz force acting on charged particles in a magnetic field. These terms can be exploited to engineer models that break time-reversal symmetry (e.g.~the Harper-Hofstadter model), in metamaterials with non-reciprocal elements. Equation~(\ref{eq:Eom}) can be recast into a Hermitian eigenvalue problem for the frequencies $\omega$
\begin{equation}
i \frac{\partial}{\partial t} 
\begin{pmatrix}
\sqrt{D}^T x\\
i \dot{x}
\end{pmatrix}
=
\begin{pmatrix}
0 & \sqrt{D}^T\\
\sqrt{D} & i\Gamma
\end{pmatrix}
\begin{pmatrix}
\sqrt{D}^T x\\
i \dot{x}
\end{pmatrix},
\end{equation}
which resembles the Schr\"odinger equation~\cite{Huber:2016tg}. There are different routes to engineer topological mechanical systems. The first one builds on the intrinsic particle-hole symmetry of the system \cite{Kane:2014if}. For each frequency $\omega$, there exists also a solution at $-\omega$ and topological modes can appear close to zero frequency. Kane and Lubensky \cite{Kane:2014if} studied the appearance of topological boundary modes in mechanical systems near collapse, for instance granular materials close to the jamming transition \cite{Lubensky:2015jd,Sussman:2016er}. These are systems close to collapse, where the number of degrees of freedom is equal to the number of constraints. Interestingly, there exist zero-frequency modes, where parts of the system can move freely. Topological modes in isostatic lattices have been observed in deformed kagome lattices \cite{Paulose:2015dc} and in a 1D mechanical system of rigid links and rotors \cite{Chen:2014ec}, which resembles the topological Su-Schrieffer-Heeger (SSH) model \cite{Su:1979cb} and might open new possibilities to study effects beyond electronic systems \cite{Vitelli:2014up}. A different kind of zero-frequency modes has been observed as localized buckling regions under external stress \cite{Paulose:2015hd}. Indeed, topological concepts may lead to novel design mechanisms for tailored mechanical properties of materials \cite{Chen:2016bk}. Recently, these ideas have been extended to static non-reciprocal metamaterials \cite{Coulais:2017ck} and transformable topological metamaterials \cite{Rocklin:2017cd}. 

A second class of topological metamaterials displays topological properties at finite frequencies. This was recently demonstrated in a mechanical system of dimers, each consisting of two rigidly connected brass balls, which are coupled by springs \cite{Prodan:2017bv}. A classification scheme for topological phononic crystals is presented in Ref.~\cite{Susstrunk:2016dfa}. All the above examples are discrete realizations of topological metamaterials. For practical applications it will be interesting to extend these studies to continuous media. First steps along these lines have been demonstrated in a 1D continuous periodic acoustic system \cite{Xiao:2015ig} and in 2D based on an array of metallic rods \cite{He:2016cra}. In general the construction of finite-frequency topological modes is challenging because the conceptually simplest way requires time-reversal symmetry breaking terms, which are not readily available in acoustic systems. However, if such terms could be realized, for instance, with acoustic structures that contain a circulating fluid \cite{Fleury:2014wo,Yang:2015hq}, unidirectional topological edge states are expected to exist \cite{Prodan:2009ee}. The realization of these ideas is very challenging experimentally. One the other hand working directly with topological time-reversal symmetric systems is not straightforward because phononic systems lack a complete classification. Recently, there has been progress regarding both possibilities. It has been demonstrated that a system of coupled gyroscopes in a honeycomb configuration is intrinsically time-reversal symmetry broken and supports unidirectional topological edge states \cite{Nash:2015eua,Wang:2015jv}. Models associated with topological Bloch bands have been studied with an array of coupled mechanical pendula~\cite{Susstrunk:2015uo} and in a honeycomb lattice of metallic cylinders \cite{He:2016cra}; see also the proposal~\cite{Wang:2015NJP}. 

\begin{figure}
\includegraphics[width=\textwidth]{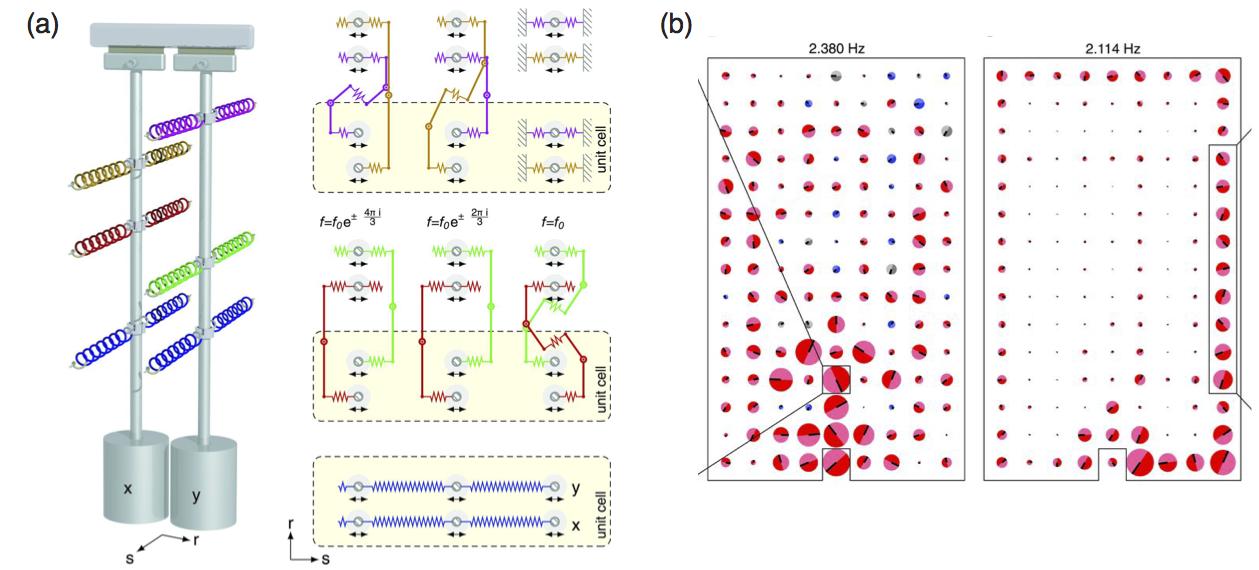}
\centering
%\vspace{0.4cm} 
\caption{\textit{\textsc{Array of coupled pendula exhibiting topological helical edge modes.} 
(a) One lattice site consists of two pendula $x$ and $y$ (left). Along the $r$-axis the pendula are coupled in a non-trivial manner via lever arms, whose number determines the sign of the coupling: cross-coupling between $x$ and $y$ (magenta and brown); $x-x$ and $y-y$ couplings (red and green). Along $s$ the pendula are simply coupled by the blue springs. For $|\alpha|=1/3$ the unit cell consists of three lattice sites. Here $f$ denote the hopping matrix elements, with amplitude $f_0$~\cite{Susstrunk:2015uo}.
(b) Observed steady states at a bulk ($2.380\,Hz$) and edge excitation frequency ($2.114\,$Hz). The system is excited with left circular polarization at the site that is excluded from the bottom row. The size of the circle illustrates the strength of the measured deflection. (Image from Ref.~\cite{Susstrunk:2015uo})}
\label{Fig:Hofstadter_Pendula}}
\end{figure}

Specifically, the system of Ref.~\cite{Susstrunk:2015uo}, which is illustrated in Fig.~\ref{Fig:Hofstadter_Pendula}a realizes two independent copies of the Harper-Hofstadter model with $\alpha=\pm1/3$, where the sign depends on a pseudospin (see below). The system consists of a 2D array of coupled pendula, where each site itself is composed of two pendula $x$ and $y$ (Fig.~\ref{Fig:Hofstadter_Pendula}a). The pseudo-spin is encoded in left and right circularly polarized motion, which is determined by the relative phase between the oscillation of $x$ and $y$ pendula. In analogy to the Harper-Hofstadter model expressed in the Landau gauge (Eq.\ref{TB_Hofstadter}) one may understand the basic working principle of the coupled pendula as follows. Along the $s$-axis the pendula are simply coupled by the blue springs, which corresponds to the axis with real hopping matrix elements in the Harper-Hofstadter model. The non-trivial direction with complex hopping matrix elements ($r$-axis) requires $x-x$, $y-y$ as well as $x-y$ cross-couplings and gives rise to the Peierls phase factors in the Harper-Hofstadter model [Eq.~\eqref{TB_Hofstadter}]. For the configuration realized in Ref.~\cite{Susstrunk:2015uo}, the system is characterized by three well-separated bulk bands and helical edge modes in the band gaps. This can be seen in the response of the system, when it is excited at frequencies that correspond to bulk or edge frequency modes respectively (Fig.~\ref{Fig:Hofstadter_Pendula}b). The topological robustness of the observed phononic helical edge states paves the way for future applications, for instance, the realization of phonon waveguides \cite{Schuetz:2015dx}.

\begin{figure}
\includegraphics[width=\textwidth]{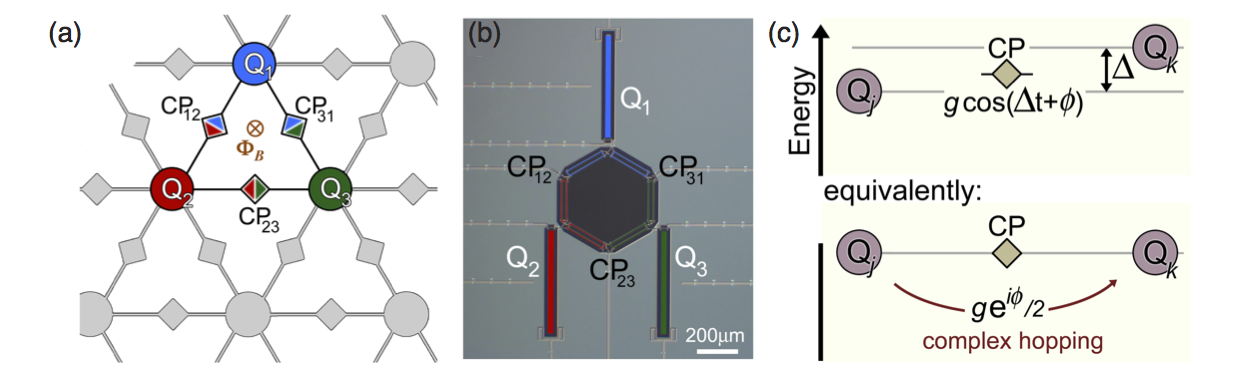}
\centering
%\vspace{0.4cm} 
\caption{\textit{\textsc{Effective magnetic fields in superconducting circuits} 
(a) Illustration of a lattice of superconducting qubits $Q_j$, where $j$ labels the lattice site. The unit cell consists of three coupled qubits. 
(b) Optical image of a superconducting circuit with three qubits, which are coupled via adjustable couplers $CP_{jk}$. (c) Illustration of the modulation induced complex coupling between neighboring qubits. The frequency difference between the two qubits is $\Delta$. Via periodic modulation with frequency $\Delta$, phase $\phi$ and amplitude $g$ and effective complex coupling $g\text{e}^{i\phi}/2$ can be engineered. (Image from Ref.~\cite{Roushan:2016iu})}
\label{Fig:SuperconductingCircuits}}
\end{figure}

\subsubsection{Superconducting circuits}

Superconducting circuits promise large progress in simulating interacting quantum systems with photons. In these settings interactions between photons arise due to the nonlinearity of the qubits. There have been a number of proposals for realizing topological systems with superconducting circuits \cite{Wang:2016gv}, however, it took until the year 2016 to successfully demonstrate this strategy experimentally, using a system made of three coupled qubits \cite{Roushan:2016iu}. Interestingly, despite the differences between the physical platforms discussed so far, the conceptual ideas are very similar. In this setting each lattice site consists of a single qubit and the hopping rate of MW photons between neighboring qubits can be adjusted via the properties of an additional coupling loop (Fig.~\ref{Fig:SuperconductingCircuits}). If the on-site energies between two lattice sites differ by an amount $\Delta$, a sinusoidal modulation of the tunneling term with frequency $\Delta$ (here: $\hbar=1$), phase $\phi$ and amplitude $g$ can restore resonant coupling between the sites. Using the formalism introduced in Sect.~\ref{section_floquet}, the dynamics can be described in terms of an effective Hamiltonian, which shares similarities with the Harper-Hofstadter model:~the engineered hopping matrix elements become complex (Fig.~\ref{Fig:SuperconductingCircuits} c), as briefly discussed below Eq.~(\ref{eq:redrivingresonators}), and the photon's wavefunction picks up a phase; as explained in previous chapters, this acquired phase is analogous to the Peierls phases inherent to the Harper-Hofstadter model [Eq.~\eqref{TB_Hofstadter}]. 

Besides, in the experiment of Ref.~\cite{Roushan:2014ch}, a single qubit was used to simulate the topological properties of the Haldane model (see the following section Sect.~\ref{sect:haldane}) via a formal but simple mapping. 

%Building upon these techniques,  electronic band-structures with a larger number of bands could be studied, in principle, if one scales the system to a larger number of qubits. 

Very recently, Ref.~\cite{Roushan:2017arXiv} reported on the realization of the 1D Harper Hamiltonian (i.e. the dimensional reduction of the 2D Harper-Hofstadter Hamiltonian) using nine superconducting qubits; this platform was used to reveal signatures of the Hofstadter butterfly through time-domain spectroscopy, but also to study many-body localization using two interacting photons.

\subsubsection{Synthetic dimensions: cold atoms and photonics}
\begin{figure}
\includegraphics[width=\textwidth]{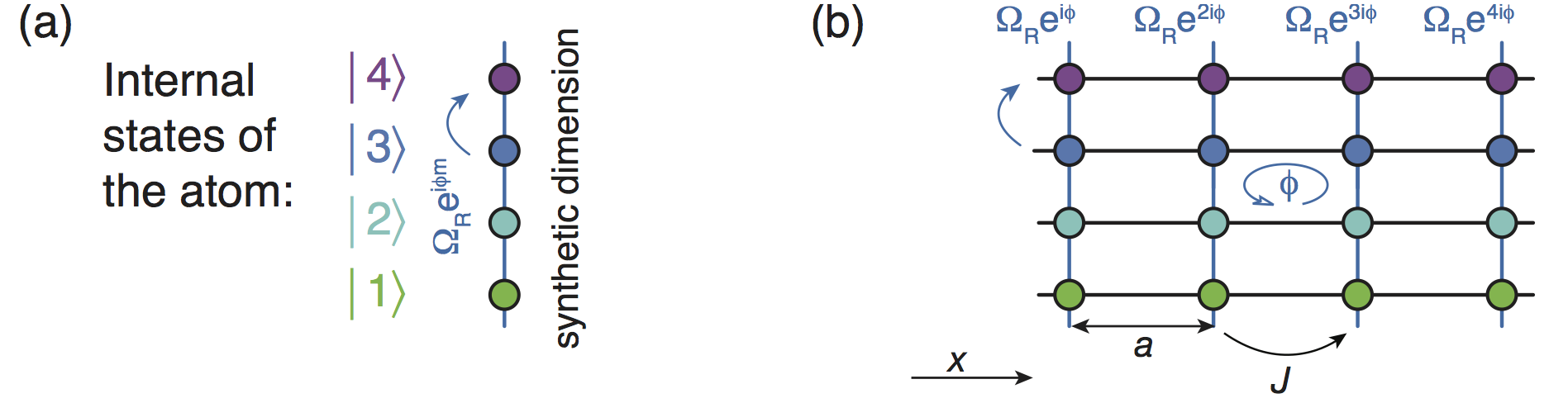}
\centering
%\vspace{0.4cm} 
\caption{\textit{\textsc{Synthetic Dimensions}
(a) Schematic illustration of a synthetic dimension, which consists of different internal states of an atom, here illustrated with different colors. Coupling between these states can be realized, for instance, with Raman beams, which are naturally complex $\Omega_R \text{e}^{i\phi m}$ and whose phase depends on position, here labeled by the index $m$.
(b) 2D synthetic lattice, where one dimension is a normal optical lattice in real space, where the sites are labeled by the index $m$ and the second dimension is realized with internal states as in (a). This setting mimics an effective magnetic field with flux $\alpha=\phi/(2\pi)$.}
\label{Fig:SynthDim}}
\end{figure}

A conceptually different approach has been put forward in the cold-atoms and photonics communities, which is based on so-called \textit{synthetic dimensions}. Here the spatial direction of a physical lattice is replaced by more abstract degrees of freedom, for instance by coupling different internal states of an atom (Fig.~\ref{Fig:SynthDim}a). The first proposal considered the use of additional degrees of freedom to realize 4D quantum systems, either via spin-dependent lattices or an on-site dressed lattice~\cite{Boada:2012jma}. The latter scheme was worked out in more detail in Ref.~\cite{Celi:2014dg}, and was eventually realized almost simultaneously by two experimental groups with bosonic~\cite{Stuhl:2015cb} and fermionic~\cite{Mancini:2015fb} atoms; these works were then followed by realizations based on optical clock transitions \cite{Cooper:2015cb,Livi:2016cn,Kolkowitz:2017iv}. 

As illustrated in Fig.~\ref{Fig:SynthDim}b, the system proposed in Ref.~\cite{Celi:2014dg} consists of a 1D lattice with ordinary tunnel coupling between neighboring sites along the spatial dimension $x$. The synthetic direction consists of different internal states of the atom, for instance, the Zeeman levels in the lower hyperfine manifold. These states can be interpreted as synthetic lattice sites. Initially the states are separated in energy by applying a magnetic field $E_{Z} = g_F \mu_B B$, where $g_F$ is the Land\'{e} $g$ factor, $\mu_B$ is the Bohr magneton and $B$ is the strength of the magnetic field. Coupling between the states is then introduced using a pair of Raman lasers whose frequency difference $\hbar \omega = E_{Z}$ is resonant with the Zeeman splitting. The Raman transition imparts the momentum $2 \mathbf{k}_R \cdot {\mathbf{1}}_x= 2 k_R^x$ onto the atoms along the lattice direction, where $k_R$ is the wavevector of the Raman beams and ${\mathbf{1}}_x$ is the unit vector along the lattice axis. The system can be described by the following Hamiltonian
\begin{equation}
{H} = \sum_{n,m} \left(-J {a}^{\dagger}_{n,m+1}{a}_{n,m} + \Omega_R \text{e}^{i \phi m} {a}^{\dagger}_{n+1,m}{a}_{n,m}\right) + \text{h.c.},
\end{equation}
here $m$ labels the lattice sites, $n$ the internal states, $\Omega_R$ is the Raman coupling and ${a}^{\dagger}_{n,m}$ is the creation operator in the extended lattice. The Hamiltonian is characterized by the Peierls phases $\phi=2 k^x_R a$ that give rise to a magnetic flux $\alpha = k_R^x a / \pi$ per plaquette, thus enabling a realization of the Hofstadter model using synthetic dimensions. Synthetic dimensions can be further realized by coupling momentum states with resonant two-photon Bragg transitions \cite{Gadway:2015gh,Meier:2016fj,Meier:2016bo,An:2017gf,An:2017tz} or by coupling different harmonic oscillator states \cite{Price:2017gm}. The former has been realized experimentally to study the SSH \cite{Meier:2016bo} and the Hofstadter model \cite{An:2017gf}. The concept of synthetic dimensions was also introduced for photonic systems using orbital angular momentum states \cite{Luo:2015fa}, higher modes in an optical ring-resonator array \cite{Ozawa:2016et,Yuan:2016ui} and optomechanical systems \cite{Marquardt:2015bt}.

\subsection{Physical realizations of local magnetic fluxes, the Haldane model and Chern insulators}
\label{sect:haldane}

In the previous Section, we presented possible realizations of the Harper-Hofstadter model~\cite{Hofstadter:1976PRB}, which describes the motion of a charged particle on a square lattice penetrated by a uniform magnetic flux. As we discussed, the corresponding band structure exhibits Bloch bands with non-trivial topology (Chern numbers~\cite{Thouless:1982PRL,Kohmoto:1989PRB}). In this sense, the Harper-Hofstadter model is the simplest lattice model that captures the quantum Hall effect~\cite{Thouless:1982PRL,Kohmoto:1989PRB}, as discovered in two-dimensional electronic systems subjected to large magnetic fields~\cite{Klitzing:1986RMP}.

As realized by Haldane~\cite{Haldane:1988PRL}, Bloch bands with non-zero Chern numbers can also be found in lattice systems that do not feature \emph{uniform} magnetic fields; filling such ``Chern bands" with electrons leads to a quantized Hall conductivity~\cite{Thouless:1982PRL}:~this is the ``quantum anomalous Hall effect" (\emph{anomalous} since it is not due to an external and uniform magnetic field~\cite{Nagaosa:2010RMP}). In order to understand the origin of these anomalous Chern bands, let us analyze the general definition of the Chern number of a given band, as in Eq.~(\ref{eq:defChern}). If a system is invariant under time-reversal, then one can verify that the Berry curvature satisfies the symmetry $\mathbf{\Omega^{\mu}_{\text{geom}}(-\mathbf{k})}=-\mathbf{\Omega^{\mu}_{\text{geom}}(\mathbf{k})}$, which then indicates that all Chern numbers~\eqref{eq:defChern} are necessarily trivial, $\nu_{\mu}\!=\!0$; see Ref~\cite{Bernevig:2013book}. As a corollary, Chern bands can be found in models that explicitly break time-reversal symmetry. This is the case for the Harper-Hofstadter model, through the presence of the Peierls phases (associated with the uniform magnetic flux); see Section~\ref{Sect:Hof}. However, time-reversal symmetry can be broken by other means in a lattice model, for instance, by inserting local fluxes~\cite{Haldane:1988PRL}, as we explain in the next paragraph.

\subsubsection{Introducing the model: from local fluxes to Chern bands}\label{sect:Haldane_local_chern}

\begin{figure}
\includegraphics[width=\textwidth]{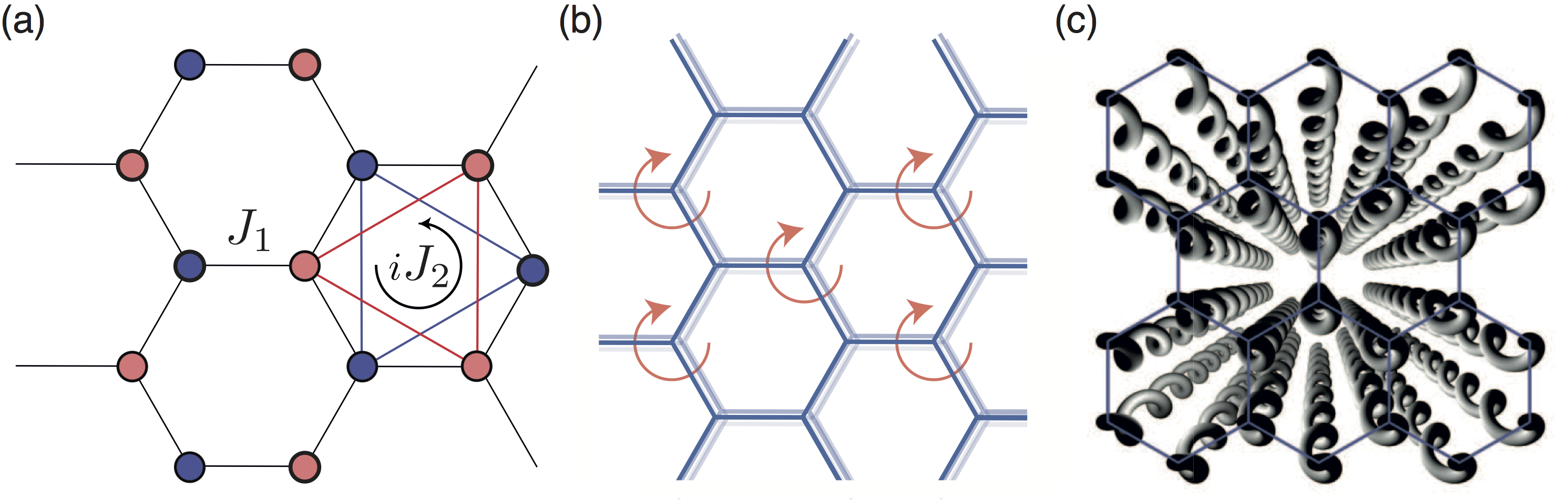}
\centering
%\vspace{0.4cm} 
\caption{\textit{\textsc{The Haldane model and its realizations} (a) Illustration of the Haldane model; the orientation-dependent second-neighbor hopping processes are represented in a unit cell. Note that the complex nature of the related tunneling matrix elements induces local (staggered) fluxes within sub-regions of the honeycomb unit cell~\cite{Haldane:1988PRL}. (b) The Haldane model can be engineered by circularly shaking a honeycomb-lattice structure~\cite{Oka:09PRB,Kitagawa:2010PRB,Rechtsman:2013Nature,Jotzu:2014Nat}. (c) Photonic realization of the Haldane model, using helical waveguides; image from Ref. \cite{Rechtsman:2013Nature}.}
\label{Fig:Haldane}}
\end{figure}

Historically, it is Haldane who first constructed a simple 2-band lattice model that exhibits Chern bands~\cite{Haldane:1988PRL}. This model is defined on a 2D honeycomb lattice with nearest-neighbor tunneling, i.e.~the tight-binding model usually used to describe graphene~\cite{Neto:2009RMP}. The corresponding spectrum exhibits two bands, which touch at singular points in the Brillouin zones (the well-known Dirac points of graphene). The latter singularities are due to the invariance of the model under time-reversal and inversion symmetries, hence, breaking one of these symmetries can open a gap in the spectrum~\cite{Haldane:1988PRL}. Haldane proposed to break time-reversal by introducing \emph{local and staggered magnetic fluxes} within each unit cell of the honeycomb lattice (the total flux penetrating each honeycomb cell being zero); formally, this is realized by adding \emph{second-neighbor} tunneling terms to the graphene tight-binding model, with complex phase factors (i.e.~Peierls phase factors). In its simplest form, the Haldane-model Hamiltonian can be written as
\begin{equation}
H = - J_1 \sum_{\langle j, k \rangle} \vert j \rangle \langle k \vert + J_{2} \sum_{\langle \langle m, n \rangle \rangle}  i^{\circlearrowright} \vert m \rangle \langle n \vert  ,  \label{Haldane_ham}
\end{equation}
where the first term describes the standard nearest-neighbor hopping term of the graphene-tight-binding model, and where the second term corresponds to the second-nearest-neighbor hopping term introduced by Haldane to break time-reversal symmetry; this term has complex matrix elements $i^{\circlearrowright}\!=\!\pm i$, whose sign  depends on the orientation of the hopping; see Fig.~\ref{Fig:Haldane} (a). One readily verifies that this second term indeed generates (staggered) local fluxes within each honeycomb unit cell, and that this opens a gap at the Dirac points, leading to two isolated bands with non-zero Chern numbers; see Ref.~\cite{Haldane:1988PRL} for details.

In fact, given a two-band lattice model, it is easy to estimate whether the latter can exhibit Chern bands with non-zero Chern numbers. This is due to an elegant relation that connects the Chern number in Eq.~\eqref{eq:defChern} to a simple (integer) winding number, which can be directly evaluated from the Hamiltonian~\cite{Yakovenko:1990PRL,Qi:2006PRB}, as we now recall. For a two-band model, the Hamiltonian can be expressed in momentum-representation as
\begin{equation}
 H (\bs k) = \varepsilon (\bs k) 1_{2\times 2} + d_x (\bs k)  \sigma_x+ d_y (\bs k)  \sigma_y+ d_z (\bs k)  \sigma_z ,\label{2_band_ham}
\end{equation}
where $\bs d (\bs k)$ defines a vector field over the Brillouin zone, and where $ \sigma_{x,y,z}$ are Pauli matrices. Applying the expression for the Chern number~\eqref{eq:defChern} to this 2-band case [Eq.~\eqref{2_band_ham}] leads to the compact expression~[see Ref.~\cite{Goldman:2013NJP} for a simple derivation]
\begin{equation}
\nu_{\text{ch}}=(1/4 \pi) \int_{\text{FBZ}} \bs{n}(\bs k)\cdot \left ( \partial_{k_x}\bs n(\bs k) \times \partial_{k_y}\bs n(\bs k) \right )   \text{d}^2k ,\label{winding}
\end{equation}
where $\bs n(\bs k)\!=\!\bs d (\bs k)/\vert \bs d (\bs k)\vert$ is a map from the FBZ (i.e.~a torus $\mathbb{T}^2$) to the sphere ($S^2$). In this picture, the Chern number in Eq.~\eqref{winding} simply counts the number of times the map $\bs n(\bs k): \mathbb{T}^2 \rightarrow S^2$ covers the sphere, which indeed defines a topological winding number~\cite{Yakovenko:1990PRL,Qi:2006PRB}. A variety of two-band models exhibiting non-zero winding numbers have been proposed,~\cite{Qi:2006PRB,Qi:2011RMP}, including the Haldane model~\cite{Haldane:1988PRL} and lattices subjected to Rashba spin-orbit coupling~\cite{Qiao:2010PRB,Beugeling:2012PRB,Qiao:2012PRB}. In the literature, such 2-band systems realizing non-trivial winding numbers (Chern bands) are loosely referred to as ``Chern insulators". A solid-state realization of these Chern insulators has been reported in 2013 in Ref.~\cite{Chang:2013Science}.

Besides, if a Hamiltonian $H (\bs k)$ associated with non-zero winding numbers [Eq.~\eqref{winding}] is realized as an effective Hamiltonian of a periodically-driven lattice system [Section~\ref{section_floquet}], one generally uses the terminology ``Floquet Chern insulators"~\cite{Oka:09PRB,Kitagawa:2010PRB,Lindner:2011NatPhys,Cayssol:2013PS}. This Floquet-engineering approach to realize Chern bands was experimentally demonstrated in 2013 in photonics, using helical optical wave-guides~\cite{Rechtsman:2013Nature}, and in 2014 in ultracold atoms trapped in shaken optical lattices~\cite{Jotzu:2014Nat}, as we further discuss below.

\subsubsection{Graphene with circularly polarized light}

In solid-state physics, a simple route to ``Floquet Chern insulators" consists in subjecting a material to circularly-polarized light~\cite{Oka:09PRB,Kitagawa:2011PRB,Lindner:2011NatPhys}; see Ref.~\cite{Cayssol:2013PS} for a review. In the case of irradiated graphene, this naturally leads to an effective Haldane-like Hamiltonian [Eq.~\eqref{Haldane_ham}]. To see this, recall that the effect of circularly-polarized light on graphene can be modeled by the modified graphene-tight-binding Hamiltonian
\begin{equation}
 H = - J_1 \sum_{\langle j, k \rangle} \vert j \rangle \langle k \vert e^{i A_{jk}(t)},  \qquad A_{jk}(t)= \bs{A} (t) \cdot (\bs r_k - \bs r_j) , \label{Haldane_graphene_irrad}
\end{equation}
which describes hopping between the nearest-neighboring sites of the honeycomb lattice; here $ \bs r_j$ is the position of the $j$th site, the electron charge is $e\!=\!1$, and the time-dependent gauge potential is given by $ \bs{A}(t)\!=\! E \left (\sin (\omega t) \bs{1}_x+ \cos (\omega t) \bs{1}_y \right )$; see Section~\ref{scope_section}. Note that the electric field associated with the irradiation is given by $\bs E (t)\!=\!\partial_t \bs{A}$ and that it enters the graphene-tight-binding Hamiltonian through the Peierls substitution [Eq.~\eqref{TB_Peierls}]. 

The time-dependent Hamiltonian in Eq.~\eqref{Haldane_graphene_irrad} can be treated using the effective-Hamiltonian approach introduced in Section~\ref{section_floquet}. In particular, in the high-frequency limit ($\omega \gg J_1$), a simple calculation~\cite{Kitagawa:2011PRB,Eckardt:2015NJP} shows that the effective Hamiltonian $H_{\text{eff}}$, as obtained from Eq.~\eqref{eq_expansion}, reproduces the Haldane Hamiltonian in Eq.~\eqref{Haldane_ham} up to second-order corrections $\mathcal{O} (1/\omega^2)$. Specifically, in this Floquet-engineering context, the tunneling matrix elements $J_1$ and $i^{\circlearrowright} J_2$ of the original Haldane model~\eqref{Haldane_ham} are explicitly given by the effective tunneling matrix elements 
\begin{equation}
J_1^{\text{eff}}\!=\! J_1\mathcal{J}_0(E) , \qquad J_2^{\text{eff}}\!=\!i^{\circlearrowright} [\sqrt{3}(J_1)^2/\omega] \left [\mathcal{J}_1(E)\right ]^2,\label{effective_tunneling}
\end{equation}
which parametrically depend on the driving field strength $E$ and frequency $\omega$; here $\mathcal{J}_{0,1}$ denote Bessel functions of the first kind. Note that the time-reversal-breaking nature of this Floquet-engineered Haldane model is due to the chirality of the driving field (whose circular polarization indeed imposes a privileged orientation to the system); indeed, reversing the orientation of the circular field changes the sign of the local fluxes penetrating the honeycomb lattice ($i^{\circlearrowright}\rightarrow - i^{\circlearrowright}$). In particular, it is this feature of the driven model that leads to (effective) Bloch bands with non-zero Chern numbers~\cite{Oka:09PRB,Kitagawa:2011PRB,Lindner:2011NatPhys}.

\subsubsection{Helical photonic waveguides\label{section_photonic_waveguides}}

Interestingly, Floquet Chern insulators were first realized in the field of photonics and reported in Ref.~\cite{Rechtsman:2013Nature}. Here, the strategy was to design a photonic crystal that exactly reproduces the irradiated-graphene Hamiltonian in Eq.~\eqref{Haldane_graphene_irrad}, by exploiting the paraxial propagation of light in 2D arrays of photonic waveguides (as created by femto-second-laser-writing techniques~\cite{Szameit:2010JPB}). Specifically, this method builds on the similarity between the Schr\"odinger equation and the optical paraxial Helmholtz equation for the evolution of the electrical field envelope~\cite{Szameit:2010JPB}
\begin{equation}
i \partial_z \mathcal{E}(\bs x)= - \frac{1}{2k_0} \nabla^2 \mathcal{E}(\bs x) - k_0 \frac{\delta n (\bs x)}{n_0} \mathcal{E}(\bs x),
\end{equation}
where $\mathcal{E}(\bs x)$ is the electric field envelope, where $k_0$ (resp.~$n_0$) is the wavenumber (resp.~bare refractive index) in the medium, and where $\delta n (\bs x)$ describes local deviations from the bulk refractive index $n_0$; see Ref.~\cite{Szameit:2010JPB,Mukherjee:2016thesis} for details. Note that here, the propagating direction $z$ plays the role of a fictitious ``time" direction. Using femto-second-laser-writing techniques, the effective potential $\delta n (\bs x)$ can be tuned so as to create arrays of waveguides in the $x-y$ plane, which are directed along $z$~\cite{Rechtsman:2013Nature,Mukherjee:2016thesis}.  In the regime of evanescently-coupled waveguides, the amplitude in the $j$-th wave-guide satisfies a ``tight-binding" Schr\"odinger equation~\cite{Szameit:2010JPB,Mukherjee:2016thesis}
\begin{equation}
i \partial_z \mathcal{E}_j(z)=\sum_{\langle k \rangle} C_{jk} \mathcal{E}_k(z) , \label{TB_optical}
\end{equation}
where the sum is taken over nearest-neighboring sites and where $C_{jk}\approx C $ is some coupling constant. Hence, in the case of a honeycomb-shaped array of waveguides, Eq.~\eqref{TB_optical} is equivalent to the equations of motion of an electron propagating in graphene (in the tight-binding approximation); see Ref.~\cite{Plotnik:2014NatMat}.

Inspired by Ref.~\cite{Haldane:2008cc,Oka:09PRB,Kitagawa:2011PRB,Lindner:2011NatPhys}, Rechtsman \emph{et al.} designed a method to turn their photonic graphene-like crystal into a Floquet Chern insulator~\cite{Rechtsman:2013Nature}. To understand this method, let us return to the time-dependent Hamiltonian in Eq.~\eqref{Haldane_graphene_irrad}; performing a gauge transformation, this Hamiltonian can be re-written in a more suggestive form,
\begin{equation}
H = - J_1 \sum_{\langle j, k \rangle} \vert j \rangle \langle k \vert  - \sum_{j} \bs E (t)\cdot \bs{r}_j \vert j \rangle \langle j \vert, \label{Haldane_graphene_irrad_bis}
\end{equation}
where $ \bs{E}(t)\!=\! E\omega \left (\cos (\omega t) \bs{1}_x - \sin (\omega t) \bs{1}_y \right )$ is the electric field associated with the gauge potential $\bs A (t)$ in Eq.~\eqref{Haldane_graphene_irrad}. The form~\eqref{Haldane_graphene_irrad_bis} allows for a simple interpretation:~the particle hops on a 2D honeycomb lattice that is circularly shaken, with a shaking amplitude $E \omega$ and a frequency $\omega$; indeed, the second term in Eq.~\eqref{Haldane_graphene_irrad_bis} reflects the (time-dependent) inertial force felt by the particle in the frame that moves with the circularly-driven lattice. As a technical note, we point out that the effective Hamiltonian associated with the time-dependent Hamiltonian in Eq.~\eqref{Haldane_graphene_irrad_bis} is also equivalent to the Haldane model [Eq.~\eqref{effective_tunneling}], up to the above gauge transformation.

The latter observation has an important corollary for honeycomb-shaped photonic crystals discussed above:~``shaking" a honeycomb array of photonic waveguides, circularly along the ``time" direction ($z$),  effectively produces a Floquet Chern insulator for light; this assumes that the shaking operates in the high-frequency regime, $\omega\!\gg\!J_1$, where $J_1$ is the nearest-neighbor coupling constant for evanescently-coupled waveguides. This was realized in the experiment of Ref.~\cite{Rechtsman:2013Nature}, where \emph{helical waveguides} were finely designed, using the femto-second-laser-writing technique~\cite{Szameit:2010JPB}. The underlying topological band structure (i.e.~the effective bands with non-zero Chern numbers) was then probed through the observation of chiral edge modes, at the boundaries of the 2D photonic crystal~\cite{Rechtsman:2013Nature}. Similar photonic crystals were recently produced to engineer ``anomalous Floquet topological bands" for light~\cite{Mukherjee:2017NatCom,Maczewsky:2017NatCom}, whose topological classification goes beyond that of static systems~\cite{Kitagawa:2010PRB,Rudner:2013,Nathan:2015}. 

\subsubsection{The Haldane model with ultracold atoms}

The previous discussion highlighted the important fact that Floquet Chern insulators can be realized by shaking a 2D honeycomb lattice in a circular manner [Eq.~\eqref{Haldane_graphene_irrad_bis}]; indeed, we saw that the underlying dynamics are well captured by an effective Hamiltonian that is reminiscent of the Haldane model~\eqref{Haldane_ham}. This strategy was first demonstrated for light propagation in Ref.~\cite{Rechtsman:2013Nature}, but was then also implemented in the context of ultracold matter. In Ref.~\cite{Jotzu:2014Nat}, Jotzu \emph{et al.} reported on the first realization of a circularly-shaken honeycomb optical lattice producing Chern bands for ultracold atoms, as we now explain. Before doing so, let us emphasize that cold-atom experiments based on shaken triangular optical lattices were pioneered in Hamburg~\cite{Struck:2011Science,Struck:2013NP}, where the generation of local fluxes was exploited to study frustrated magnetism~\cite{Eckardt:2010EPL}.

The experiment of Ref.~\cite{Jotzu:2014Nat} first consists in loading a Fermi gas of ultracold $^{40}$K atoms into the lowest band of a honeycomb optical lattice; see Ref.~\cite{Tarruell:2012Nature} for the laser configuration that produces this tunable optical potential. Then, a circular shaking of the lattice is produced through piezo-electric actuators, which sinusoidally modulate two mirrors that are responsible for the creation of the (initially static) optical lattice. The time-dependent Hamiltonian that describes the motion of atoms within this shaken optical lattice is then precisely of the form~\eqref{Haldane_graphene_irrad_bis}, in the frame that moves with the lattice (see also Refs.~\cite{Zheng:2014PRA,Eckardt:2015NJP}). Hence, in the high-frequency regime, this cold-atom system realizes the Haldane model (up to corrections discussed above). The non-trivial Berry curvature associated with the resulting (effective) Bloch bands  was then demonstrated by evaluating the anomalous velocity~\cite{Karplus:1954PR,Xiao:2010RMP} of the cloud in response to an external and static force~\cite{Jotzu:2014Nat}. 

More recently, a similar cold-atom setting was realized in Hamburg to investigate topological signatures of quench dynamics in Chern bands~\cite{Flaschner:2016arxiv,Tarnowski:2017arxiv}. As was predicted by Wang et al.~\cite{Wang:2017PRL}, quenching a two-band system from a trivial to a non-trivial region of its topological phase diagram generates complex dynamics, which involve the creation and annihilation of vortices in momentum-space; the topological nature of the post-quenched Hamiltonian can then be determined by measuring a winding number associated with the trajectories of the vortices. Such a phenomenon was observed and reported in Ref.~\cite{Tarnowski:2017arxiv}, where the trajectories of the emerging momentum-space vortices were fully reconstructed through state tomography~\cite{Flaschner:2016Science}, demonstrating clear evidence for the dynamical generation of non-trivial winding numbers.

\subsection{Gauge fields from topological defects and strain}
\label{sect:strain}

\subsubsection{Graphene}
Since the first successful electrical measurements on a single layer of carbon, i.e. graphene \cite{Novoselov:2004it,Berger:2004jh,Novoselov:2005es}, there has been a growing interest of related physics in particular because it has a deep relation to quantum electrodynamics \cite{Geim:2007hy,Katsnelson:2007cl,DasSarma:2007vi,Neto:2009RMP,Geim:2009id}. Here we want to briefly review the connection between the physics of graphene on the one hand and gauge fields on the other. There are several different scenarios that give rise to perturbations that are mathematically equivalent to gauge fields \cite{Vozmediano:2010dl}. Generally speaking it appears due to a change in the hopping between the two sublattices. On the one hand, they may occur in the presence of topological defects or dislocations, which change the structure of the lattice itself. These types of gauge fields do not depend on any material parameters but only on general topological features of the real space potential. The second type arises through modifications of the hopping amplitudes, which might be caused by changes in the distance between the orbitals or a change of symmetry. This situation indeed depends on the microscopic details of the material.

\paragraph{Topological defects}
Using high-resolution Transmission Electron Microscopy topological defects consisting of pentagon-heptagon pairs have been directly observed in single wall carbon nanotubes \cite{Suenaga:2007bm}. These are known as Stone-Wales defects \cite{Stone:1986cj}. An intuitive way to understand the emergence of gauge fields due to dislocations in the lattice structure may be achieved in the following way: In the presence of a dislocation an electron that encircles the defect will experience a phase mismatch, which can be interpreted as an effective magnetic field that is present at the position of the defect. The general strategy in describing the defects consists in determining the geometric phase around a defect, which in turn determines the gauge field \cite{Lammert:2004fr,Furtado:2008br,Osipov:2005uo}. In fact, following this approach, one realizes that the Stone-Wales defects are rather associated with gauge fields that emerge due to deviations from a perfectly flat graphene sheet, such as ripples \cite{deJuan:2007cy}. In a general situation, it is difficult to disentangle the different effects that lead to an effective gauge field.

\paragraph{Strain Field}
Gauge fields may also arise in defect-free graphene due to elastic deformations in the presence of inhomogeneous external stress. This was already realized more than 20 years ago by Kane and Mele \cite{Kane:1997fj} but the relation between a particular designed strain and the corresponding pseudo-magnetic field is not straightforward and motived multiple studies, that were using strain-engineering to generate homogeneous magnetic fields \cite{Katsnelson:2007gm,Guinea:2008iv,Guinea:2010fl,Guinea:2010ff}. In general the coupling matrix elements in the tight-binding model of a deformed graphene lattice are all nonequivalent and one can show that the Dirac Hamiltonian, that describes the low-energy dynamics of the system, with non equal hopping parameters results in the following effective Hamiltonian:

\begin{equation}
H=-i\hbar v_{F} \vec{\sigma} \cdot \left( \vec{\nabla} - i \vec{A} \right) , \qquad A_x = \frac{\sqrt{3}}{2}(t_3 - t_2), \quad A_y = \frac{1}{2}(t_2 + t_3 - 2t_1) ,
\end{equation}

\noindent where the effective gauge field $\vec{A}$ depends on the values of the different tunnel couplings $t_i, i\in \{1,2,3\}$ \cite{Katsnelson:2007gm}. Such systems can support topological edge states even in the absence of time-reversal symmetry breaking. Indeed strain-induced effective magnetic fields larger than 300$\,$Tesla have been observed in graphene nanobubbles, where the appearance of Landau levels has been observed in the electronic spectrum using scanning tunneling microscopy \cite{Levy:2010hl}. The existence of helical edge state in strained honeycomb lattices depends strongly on the type of strain and on the termination of the edges. A systematic study has been presented recently \cite{Salerno:2017bg}, which also applies to artificial graphene systems \cite{Polini:2013ib}, some of them are briefly discussed below. In particular using strain engineering large pseudo-magnetic fields have been demonstrated in molecular graphene that was assembled using individual CO molecules, which are places on a Cu surface \cite{Gomes:2012et}. 

Theoretically, the effects of strain-induced electromagnetic fields can also be  analyzed through a semiclassical (wave-packet) approach, which takes the geometric effects captured by the Berry curvature into account~\cite{Roy:2017arxiv}. Such an approach was recently exploited to identify schemes realizing strain-induced gauge fields in cold-atom realizations of 3D Weyl semimetals~\cite{Roy:2017arxiv}; see also Ref.~\cite{Cortijo:2015PRL} on the generalization of strain-induced gauge fields in three-dimensional systems.

\begin{figure}
\includegraphics[width=\textwidth]{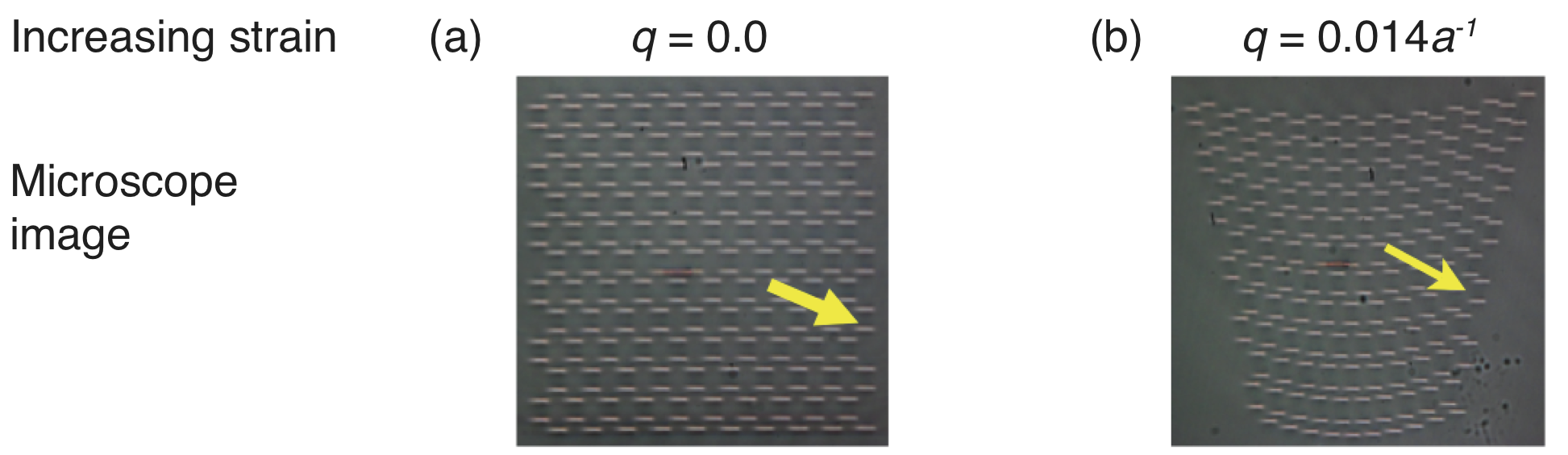}
\centering
%\vspace{0.4cm} 
\caption{\textit{\textsc{Strain-induced gauge fields} 
Microscope images of the coupled waveguide array for increasing strain. (a) Regular lattice without strain. (b) Strained lattice with $q=0.014a^{-1}$, where $a$ is the lattice constant. (Image from Ref. \cite{Rechtsman:2013bp})}
\label{Fig:strainphotonics}}
\end{figure}

\subsubsection{Photonics}
It has been proposed and demonstrated that strain-induced gauge fields can also be engineered in coupled waveguide arrays, hence leading to the appearance of photonic Landau levels in the spectrum \cite{Rechtsman:2013bp,Schomerus:2013by}. The system consists of an array of evanescently coupled waveguides (see also Sect.~\ref{section_photonic_waveguides}), which are arranged in a honeycomb configuration in the $x$-$y$ plane and are invariant along the propagation direction. The coupling strength $c(r)$ between the modes depends on the distance between neighboring waveguides $c(r)=c_0 e^{-(r-a)/l_0}$, here $a$ is the lattice constant ($a=14\mu$m in Ref.~\cite{Rechtsman:2013bp}), $c_0$ is the coupling constant at distance $r=a$ and $l_0$ characterizes the decay of the coupling strength with distance. For a uniform lattice $c(r)$ would be constant, while in the presence of strain $c(r)$ becomes non-uniform and can give rise to a pseudomagnetic field. As already presented in Eq.~\eqref{TB_optical}, the photonic graphene lattice can be described in a tight-binding picture. For a fixed frequency the band-structure resembles that of graphene, i.e. it displays two bands which are connected by two inequivalent Dirac-cones in the Brillouin zone. In order to generate the strain-induced pseudo-magnetic field, waveguides are created at transverse positions that correspond to a certain configuration of $c(r)$, as displayed in Fig.~\ref{Fig:strainphotonics}. In the case of Ref.~\cite{Rechtsman:2013bp} this should correspond to fields on the order of 5500$\,$T. The nature of the gauge field was then studied via the dynamics of edge modes, which can be excited with a tightly focused laser beam.

\subsubsection{Cold atoms}
To the best of our knowledge there are no cold-atom realizations of strain-induced gauge fields. The reason being that strain is not easily generated in optical lattice potentials. One possibility for an implementation similar to the one discussed above, relies on the generation of spatially inhomogeneous tunnel couplings \cite{Tian:2015wv}. In a hexagonal optical lattice the standard laser-beam configuration consists of three beams that are intersecting at $120^{\circ}$. Now one can show that by simply displacing the beams a pseudo magnetic field can be generated that is nearly homogeneous in real space and leads to the appearance of Landau levels in the spectrum. This proposal is particularly appealing because it solely relies on static optical lattices and might be a way to overcome the heating rates observed in other cold-atom realizations that implement artificial magnetic fields based on Floquet engineering or Raman transitions (see Sect.~\ref{sect:atomslattice} for Refs). Simple schemes to create strain-induced gauge fields in optical-lattice realizations of 3D Weyl semimetals were also proposed in Ref.~\cite{Roy:2017arxiv}.

\subsubsection{Acoustics}
Very recently, the concepts of strain-engineering have been translated to lattice vibrations in mechanical metamaterials \cite{Abbaszadeh:2016tj} and to 2D acoustic honeycomb lattices \cite{Yang:2017gu}. The first setting consists of a honeycomb lattice of nodes connected by rods, where uniform gauge fields can be engineered for acoustic vibrations by smoothly varying, for instance, the thickness of the rods, which influences the effective spring constant \cite{Abbaszadeh:2016tj}. The second proposal considers a honeycomb lattice of acoustic resonators that are coupled with thin cylindrical waveguides \cite{Yang:2017gu}. The coupling strength between the resonators decays exponentially with the distance between them and, similarly to the photonic system described above, strain can be generated by displacing the resonators from their original position. This leads to the appearance of Landau levels in the spectrum.

\section{Spin-orbit coupling}\label{sect:SOC}
Spin-orbit coupling, referring to the interaction between the motion of a particle and its internal spin, plays a prominent role in emerging fields of condensed matter physics \cite{Winkler:2003SPR}, from the observation of the spin Hall effect \cite{kato_observation_2004,konig_quantum_2007} to the development of spintronic semiconductor devices \cite{zutic_spintronics:_2004,chappert_emergence_2007} and topological insulators \cite{kane_z_2005,bernevig_quantum_2006,hsieh_topological_2008}. A spin-orbit coupling naturally occurs for electrons moving in ionic lattice structures with proper symmetry properties. It can also be engineered in quantum well structures, as well as in two-dimensional atomic crystal layers such as graphene. 
The concept of spin-orbit coupling was also extended to various analog physical systems, from ultracold atomic gases to photonic and mechanical systems, as the interaction between the motion of a particle or a wave propagation mode and an internal degree of freedom.

\subsection{Spin-orbit coupling in condensed matter systems}

Spin-orbit coupling usually refers to the interaction between the motion and the spin of an electron orbiting around an atomic core, that gives rise in particular to the fine-structure splitting of electronic quantum levels. In a non-relativistic picture of the Dirac equation, the spin-orbit coupling is described by a term
\[
H_{\mathrm{SO}}=\frac{\hbar}{4m_{\mathrm{e}}^2c^2}\bm{\sigma}\cdot[\boldsymbol{\nabla} V\times\mathbf{p}],
\]
where $V$ is the Coulomb potential of the atomic core, $\bm{\sigma}$ is the vector of Pauli matrices, $m_{\mathrm{e}}$ is the electron mass and $\mathbf{p}$ refers to its momentum.

\begin{figure}
\includegraphics[width=\linewidth]{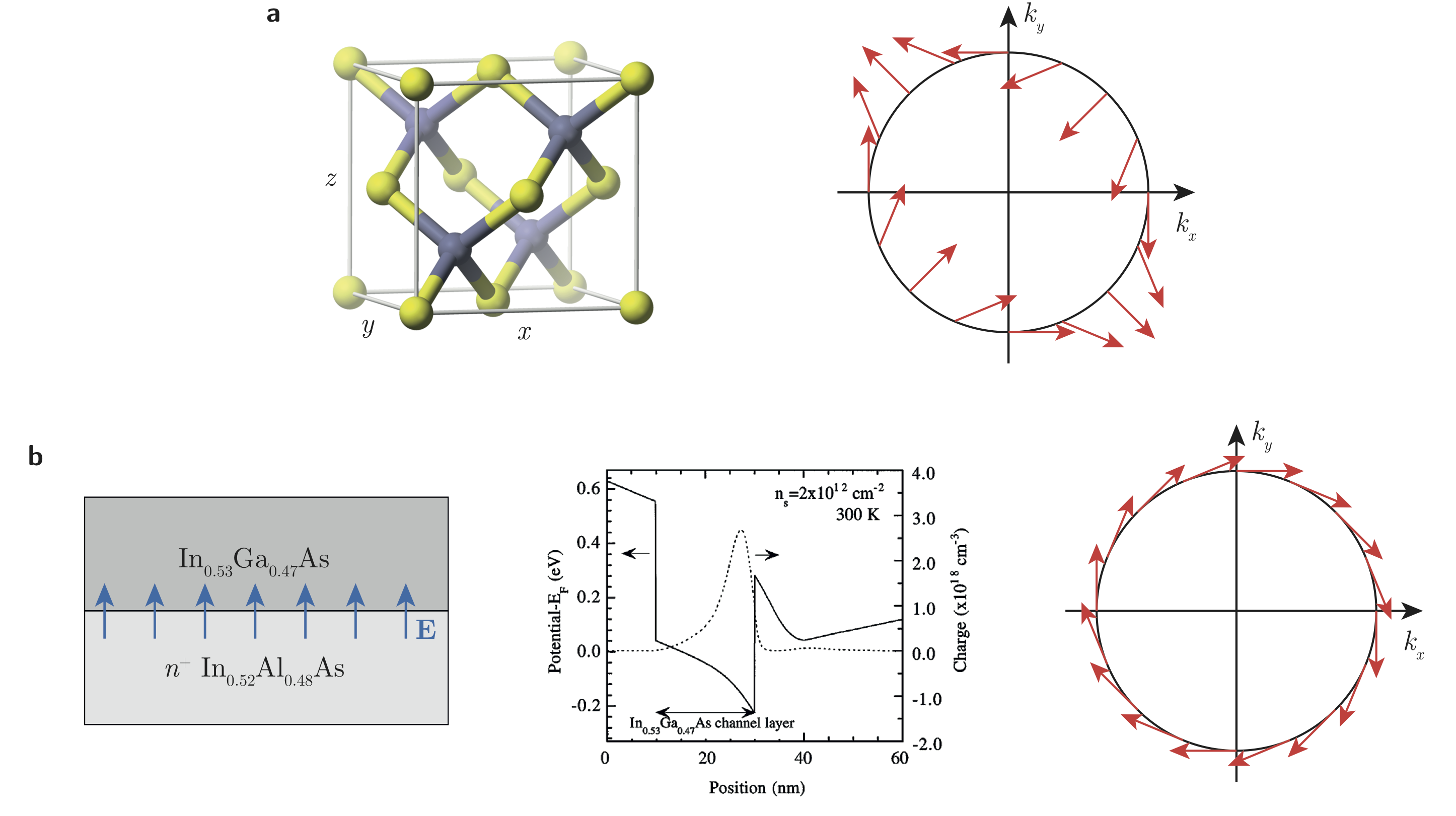}
\caption{\textit{\textsc{Spin-orbit coupling in solids}} (a) Scheme of the zinc-blende crystallographic structure corresponding to several semi-conductors such as GaAs, GaP, InAs. The lack of inversion symmetry leads to a Dresselhaus spin-orbit coupling, with a spin polarization at the Fermi surface corresponding to the red arrows.  (b) Simplified scheme of a In$_{0.53}$Ga$_{0.47}$As/In$_{0.52}$Al$_{0.48}$As heterostructure. At the interface between the two materials one expects an electric field (blue arrows), as calculated in the conduction band diagram \cite{Nitta:1997PRL}. The resulting Rashba spin-orbit coupling leads to a spin polarization at the Fermi surface corresponding to the red arrows.\label{Fig_SO_Dresselhaus_Rashba}}
\end{figure}

By analogy, a spin-orbit coupling occurs for electrons in solid-state materials, where the Coulomb potential takes its origin in the electric field of the ionic crystal \cite{Elliott:1954PR,Winkler:2003SPR}. The crystal structure of mostly used semiconductors, such as diamond-like Si and Ge, obeys spatial inversion symmetry. This symmetry, combined with time-reversal invariance at zero magnetic field, guarantees a double spin degeneracy of the Bloch bands \cite{Kittel:1963WI}. This degeneracy is lifted, even at zero magnetic field, for bulk 3D crystals lacking inversion symmetry  \cite{dresselhaus_spin-orbit_1955}, such as semiconductors with a zinc blende structure, e.g. GaAs and InSb (see figure \ref{Fig_SO_Dresselhaus_Rashba}a).  Such band structures can be understood as a result of an effective spin-orbit coupling. Considering for simplicity electron motion restricted to $x$ and $y$ directions, the spin-orbit coupling can be written in the `Dresselhaus' form $H_{\mathrm{D}}=-\beta(\sigma_x q_y+\sigma_y q_x)$. A different type of spin-orbit coupling can arise as a result of strong electron confinement, e.g. at the surface of semi-conductors \cite{ohkawa1974quantized} or  in two-dimensional electron gases confined in quantum well structures \cite{Rashba:1960PSS,yu_a_bychkov_and_e_i_rashba_oscillatory_1984,Manchon:2015NM}. There, in a simplified picture, the materials from both sides of the electronic plane produce a perpendicular electric field, leading to a `Rashba'-type spin-orbit coupling $H_{\mathrm{R}}=\alpha(\sigma_x q_y-\sigma_y q_x)$ (see figure \ref{Fig_SO_Dresselhaus_Rashba}b). A more precise description of spin-orbit coupling in quantum wells requires taking into account the band structures of the different layer materials, see for example reference \cite{Winkler:2003SPR}.  Altogether, the generic form of the spin-orbit coupling occurring in 2D electron systems is given by the sum of Rashba and Dresselhaus couplings, as
\[
H_{\mathrm{SO}}=\alpha(\sigma_x q_y-\sigma_y q_x)-\beta(\sigma_x q_y+\sigma_y q_x),\label{eq_SO}
\]
where the weights $\alpha$ and $\beta$ can be independently measured \cite{Ganichev:2004PRL,Meier:2007NP}, and their ratio can be tuned using additional gate control over the electron confinement \cite{Nitta:1997PRL,Grundler:2000PRL} and density \cite{Heida:1998PRB,Matsuyama:2000PRB}. Mastering of the spin-orbit coupling structure is a prerequisite in order to precisely control spin dynamics/transport and implement semiconductor spintronics devices, such as a spin field-effect-transistor \cite{Datta:1990APL,Schliemann:2003PRL,Zutic:2004RMP}.

Spin-orbit coupling can also be engineered in two-dimensional materials, including graphene, graphene-like materials such as silicene and germanene, or other monolayer materials such as  transition-metal dichalcogenides \cite{butler_progress_2013,xu_graphene-like_2013,ren_topological_2016}. In graphene, the intrinsic spin-orbit coupling due to Coulomb interaction with the ionic crystal is too small to be accessible experimentally \cite{kane_quantum_2005}. Several schemes were developed to artificially produce a coupling between the electron motion and its spin, or the effective spin 1/2 formed by the valley degree of freedom. The malleability of graphene sheets allows for the application of strong stress and deformations, leading to important modifications of the band structure \cite{pereira_tight-binding_2009,amorim_novel_2016} and to a quantum spin Hall insulator phase \cite{Guinea:2010fl,Levy:2010hl}. Other techniques are based on depositing adatoms on the graphene layer \cite{weeks_engineering_2011,balakrishnan_colossal_2013,balakrishnan_giant_2014}, on the hybridization with a different atomic layer  \cite{marchenko_giant_2012,calleja_spatial_2015},  the application of an external magnetic field \cite{young_tunable_2014}, the proximity effect with semiconducting materials \cite{avsar_spin-orbit_2014}. 

Graphene-like materials, such as silicene \cite{liu_quantum_2011,ni_tunable_2011,vogt_silicene:_2012}, germanene \cite{davila_germanene:_2014} or stanene \cite{xu_large-gap_2013,zhu_epitaxial_2015} are expected to exhibit intrinsic spin-orbit coupling of much larger strength than for graphene, which could allow direct observation of the quantum spin Hall effect. More complex two-dimensional materials, such as transition-metal dichalcogenides \cite{wang_electronics_2012}, are also promising alternatives to explore spin-orbit coupling effects and create topological phases of matter \cite{cazalilla_quantum_2014,qian_quantum_2014,ren_topological_2016}.

\subsection{Spin-orbit coupling in cold atomic gases}
In cold atom systems, a coupling between the motion of an atom and its spin can be simulated e.g. by coupling spin states with laser light. Its consequences on the single-atom dynamics was extensively studied in experiments. In many-body quantum gases, the interplay between spin-orbit coupling, interactions between atoms and their quantum statistics leads to novel states of matter with strong fundamental interest~\cite{Zhai:2015RPP}.

\subsubsection{Simulating a spin-orbit coupling for atoms}
\begin{figure}
\includegraphics[width=\linewidth]{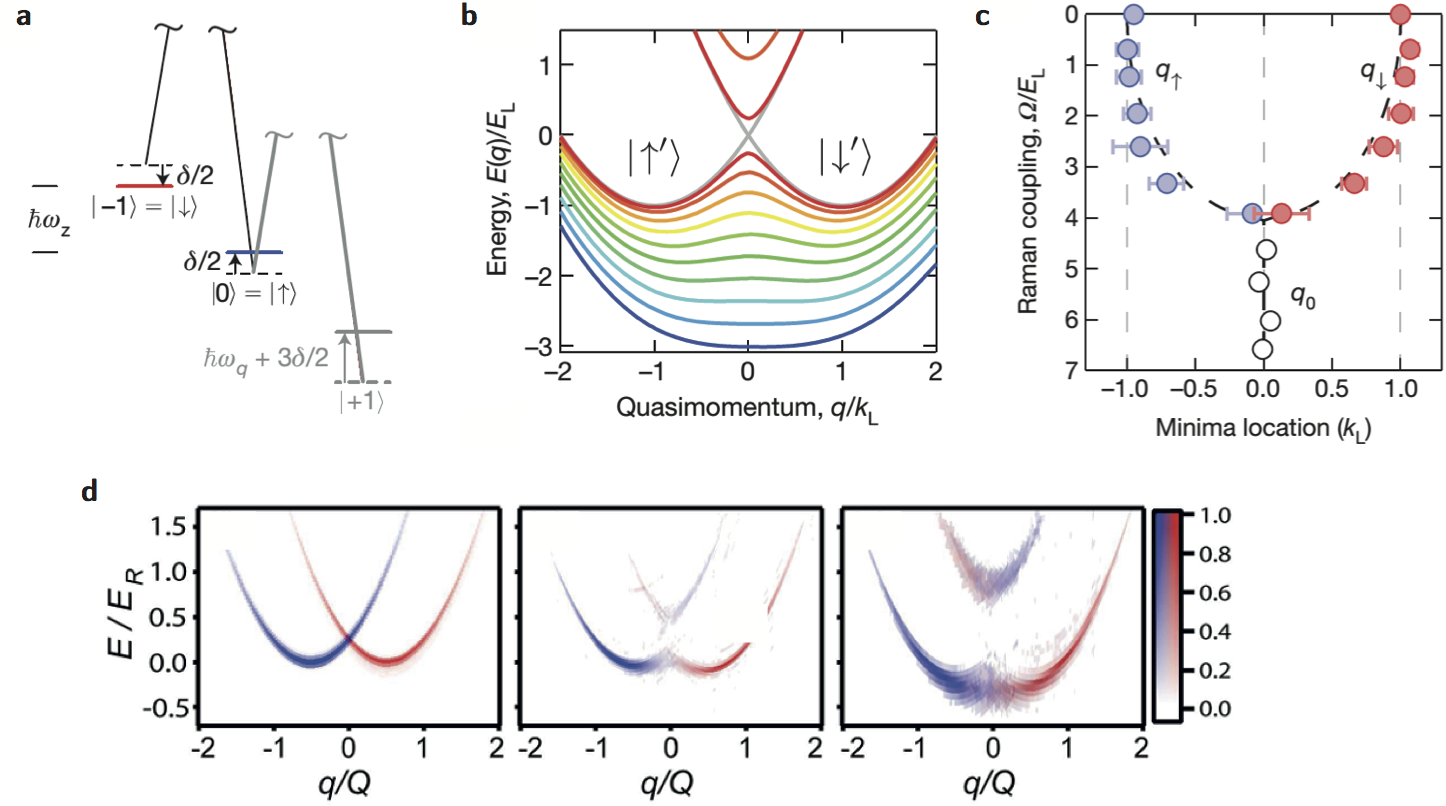}
\caption{\textit{\textsc{Spin-orbit coupling in cold atoms}} (a) Scheme of atom-laser couplings used to simulate a 1D spin-orbit coupling. 
(b) Dispersion relation for a spin-orbit coupled atom, calculated for various values of the effective Zeeman field amplitude $\Omega$. 
(c) Location of the minima of the dispersion relation measured from the momentum distribution of spin-orbit coupled Bose-Einstein condensates (from \cite{Lin:2011Nat}).
(d) Dispersion relation and spin polarization measured using a spin-injection spectroscopy technique (from \cite{cheuk2012spin}). 
\label{Fig_SO_1D}}
\end{figure}

As proposed in \cite{Liu:2009PRL} and first implemented in \cite{Lin:2011Nat}, a spin-orbit interaction can be produced by coherently coupling two spin states of an atom with a resonant two-photon optical transition (see figure \ref{Fig_SO_1D}a). The modeling of the atom dynamics is identical to the one introduced for  the simulation of an orbital magnetic field via laser dressing (see Section \ref{section_laser_dressing} and Section~\ref{non_Abelian_section}). Assuming a resonant coupling $\delta=0$, equations (\ref{eq_laser_dressinga}) and (\ref{eq_laser_dressingb}) can be recast as
\begin{equation}\label{eq_SO_1D}
H=\sum_{\mathbf{q}}\frac{\hbar^2q^2}{2m}-2\lambda\, q_x\sigma_z+\frac{1}{2}\hbar\Omega\sigma_x,\quad\lambda=\frac{\hbar^2 K}{2m},
\end{equation}
where we dropped the irrelevant constant $\hbar^2K^2/2m$ for simplicity, and we assumed $\mathbf{K}=K{\mathbf{1}}_x$. This single-particle Hamiltonian corresponds to a spin-orbit coupling (\ref{eq_SO}) with equal Rashba and Dresselhaus amplitudes $\alpha=\beta=\lambda$ (after redefining the spin directions $x\rightarrow y\rightarrow z\rightarrow x$), and a Zeeman field of amplitude $\propto\Omega$ oriented along $x$. 
%The resulting dispersion relation,  plotted in figure\,\ref{fig_SO}, will be commented in section \ref{section_SO_single}. 
Such a spin-orbit coupling has been realized for bosonic $^{87}$Rb atoms \cite{Lin:2011Nat}, as well as for fermionic $^6$Li, $^{40}$K, $^{161}$Dy and $^{173}$Yb atoms \cite{cheuk2012spin,wang2012spin,burdick2016long,song_spin-orbit-coupled_2016}. Additional control over the spin-orbit amplitude can be provided by a suitable time-modulation of the Raman laser coupling \cite{jimenez2015tunable}.

A generic issue encountered with the laser dressing method is the residual incoherent Rayleigh scattering of the dressing laser light by the atoms, which leads to heating of the atomic samples. In the case of alkali atoms, different Zeeman sublevels share the same electronic orbital. As light fields only couple different electronic orbitals, the light-spin coupling only occurs indirectly, via the intrinsic electronic spin-orbit coupling in the excited electronic state -- the one leading to fine-structure splitting. As a result, large heating rates are observed for atoms with small fine-structure splitting, such as Li or K \cite{cheuk2012spin,wang2012spin}; using alkali atoms with large fine-structure splitting, such as Rb of Cs \cite{Lin:2011Nat}, or atoms with more complex electronic structure, such as Lanthanides \cite{nascimbene_realizing_2013,cui_synthetic_2013,burdick2016long}, is required to reach low-temperature phases of matter.
An interesting alternative consists in encoding a pseudo-spin in external degrees of freedom, e.g. between the two lowest orbitals of a double-well potential \cite{hugel_chiral_2014,Atala:2014uo,sun_tunneling-assisted_2015}. Ultracold atoms also grant the ability to explore spin-orbit coupling involving more than two spin levels \cite{wang2010spin,lan2014raman,natu2015striped}, as recently demonstrated in $F=1$ Rb BECs \cite{campbell2016magnetic}.

Extending the laser dressing scheme to a two-dimensional spin-orbit coupling, e.g. of the Rashba form, requires implementing more complex laser configurations, as proposed in \cite{unanyan1999laser,ruseckas2005non,juzeliunas2010generalized,campbell2011realistic} and realized with ultracold Fermi gases \cite{huang_experimental_2016,meng_experimental_2016} and Bose-Einstein condensates \cite{wu_realization_2016}. We emphasize that the spin-orbit coupling configuration of Ref.~\cite{wu_realization_2016}, which was implemented in an optical lattice, led to topological Bloch bands with non-zero Chern numbers (see Section~\ref{sect:Haldane_local_chern}); these non-trivial topological invariants were revealed through spin-polarization measurements, as originally proposed in Ref.~\cite{Liu:2013PRL}. Three-dimensional spin-orbit coupling $\propto\sigma_xq_x+\sigma_yq_y+\sigma_zq_z$ could be implemented using similar techniques, which would lead to the occurrence of 3D topological states \cite{Bermudez:2010PRL,anderson_synthetic_2012,mazza_optical-lattice-based_2012,dubvcek2015weyl}.

Another approach to create artificial spin-orbit coupling is to periodically drive the atoms with suitable magnetic field gradient pulses, as proposed in \cite{Xu:2013PRA,Anderson:2013PRL,struck2014spin,Goldman:PRX2014} and implemented in \cite{Luo2016Scientific}. Spin-orbit coupling can also be simulated using a combination of laser couplings and magnetic field gradients \cite{kennedy_spin-orbit_2013,Aidelsburger:2013PRL}. These methods avoid the heating from spontaneous emission discussed for the laser dressing scheme. However, the time modulation introduces another source of incoherent relaxation, involving the absorption of energy quanta at the modulation frequency \cite{choudhury2015transverse,bilitewski2015scattering,weinberg2015multiphoton,lellouch2017parametric,eckardt2017colloquium}.

A further approach is to use a clock transition to coherently couple two different electronic states of very long lifetimes \cite{wall_synthetic_2016,Livi:2016cn,Kolkowitz:2017iv}, which naturally leads to long coherence times compared to the laser dressing scheme with alkali atoms.

\subsubsection{Dynamics of spin-orbit coupled atoms\label{section_SO_single}}
We first consider the case of the 1D spin-orbit coupling corresponding to equation (\ref{eq_SO_1D}). The dispersion relation associated with the motion along $x$ -- directly measured in reference \cite{cheuk2012spin}  (see figure \ref{Fig_SO_1D}d) -- is represented in figure  \ref{Fig_SO_1D}b. It corresponds to parabolic dispersion relations shifted in opposite momentum directions for each spin due to the spin-orbit coupling. The Zeeman coupling further leads to an avoided crossing around $q_x=0$. The most important features of this dispersion relation are (i) the single-particle degeneracy occurring for $\lambda K>\hbar\Omega/4$ \cite{Lin:2011Nat}  (see figure \ref{Fig_SO_1D}c), (ii) the spin-momentum locking for an atom adiabatically following the lowest energy branch  (see figure \ref{Fig_SO_1D}d).
The structure of the dispersion relation was probed via studies of collective oscillation modes in trapped Bose-Einstein condensates \cite{Zhang:2012PRL}. Zitterbewegung oscillations between the two energy branches were observed, following a sudden quench of the spin-orbit coupling \cite{qu_observation_2013,leblanc_direct_2013}. Using a spatial variation of the Raman coupling lasers, it was possible to demonstrate a spin Hall effect, corresponding to spin-dependent effective Lorentz forces, leading to the realization of an atomtronic spin field-effect transistor \cite{vaishnav_spin_2008,Datta:1990APL,Beeler:2013Nat}. Landau-Zener transitions across the avoided crossing at zero momentum were investigated in \cite{olson_tunable_2014}, providing insight on the robustness of spin-momentum locking. The lack of Galilean invariance in spin-orbit coupled systems was probed via the response of an atomic gas to a moving optical lattice \cite{zhu_exotic_2012,hamner_spin-orbit-coupled_2015}.

For the 2D Rashba spin-orbit coupling, the dispersion relation is expected to feature a largely degenerate ground state,  occurring on a circle in the $(q_x,q_y)$ plane. The low-energy dispersion relation was investigated with Bose-Einstein condensates in the presence of a Rashba-type spin-orbit coupling induced by an optical Raman  lattice \cite{wu_realization_2016}. In this scheme, higher-order spin-orbit coupling terms break the continuous symmetry, leaving a four-fold degeneracy. The dispersion relation also exhibits a Dirac point that was probed using radio frequency spectroscopy in reference \cite{huang_experimental_2016}. 

\subsubsection{Interactions between spin-orbit coupled atoms and quantum phases of matter}
Spin-orbit coupling also leads to a profound modification of two-body interaction and pairing properties in ultracold atomic samples \cite{zhang_p_2008,vyasanakere2011bound,jiang2011rashba,zhou2011topological}. 

\begin{figure}
\includegraphics[width=\linewidth]{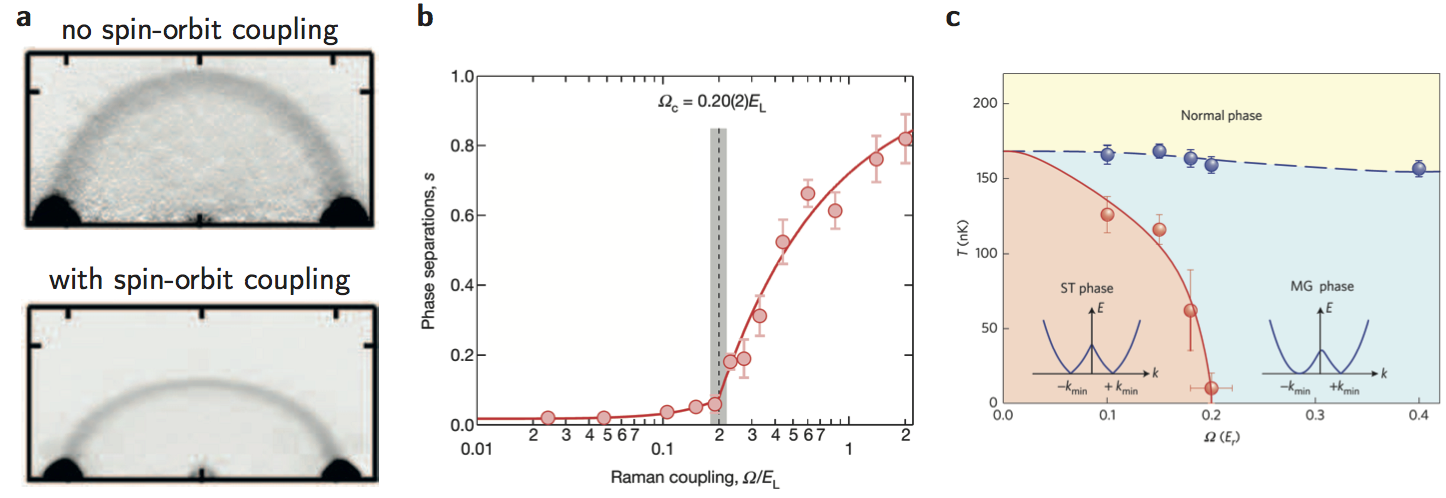}
\caption{\textit{\textsc{Phases of spin-orbit coupled atomic gases}} (a) Consequence of spin-orbit coupling on the scattering between atoms, which no longer corresponds to an isotropic $s$-wave scattering (from \cite{williams_synthetic_2012}).
(b)Low-temperature phase diagram of spin-orbit coupled Bose-Einstein condensates. For weak effective Zeeman fields $\Omega$ the two spin states are mixed, while they separate for $\hbar\Omega>0.2\,E_{\mathrm{r}}$ (from \cite{Lin:2011Nat}). 
(c) Finite-temperature phase diagram measured from the magnetization of spin-orbit coupled Bose-Einstein condensates (from \cite{Ji:2014NP}).
\label{Fig_SO_BEC}}
\end{figure}

In particular, it leads to scattering between atoms in higher-order partial waves \cite{williams_synthetic_2012} (see figure \ref{Fig_SO_BEC}a), and to interactions between spin-polarized fermions occupying the lowest energy branch of the dispersion relation \cite{williams_raman-induced_2013,fu_production_2014}. Spin-orbit coupling also modifies the binding of atom pairs into molecular states \cite{fu_radio-frequency_2013,fu_production_2014}. 

The specific dispersion relation and the modification of interaction properties lead to a rich landscape of quantum phases of matter expected for spin-orbit coupled gases in the quantum degenerate regime~\cite{wang2010spin,ho_bose-einstein_2011,li_quantum_2012}. Most of the experimental studies investigated the phase diagram of Bose-Einstein condensates under a one-dimensional spin-orbit coupling. There, the competition between the two single-particle ground states of opposite momenta leads to condensation in either a single momentum state or in a quantum superposition between two momenta, depending on the strength of the spin-orbit coupling, the interaction properties of the atom, and the sample temperature (see figure \ref{Fig_SO_BEC}b,c) \cite{Lin:2011Nat,Ji:2014NP,hamner_dicke-type_2014,ji2015softening,campbell2016magnetic,li2017stripe}. 

Original phases of matter could be realized in future experiments with spin-orbit coupled gases, including topological superfluids (with the associated Majorana bound states) \cite{kitaev_unpaired_2001,zhang_p_2008,sato2009non,jiang_majorana_2011,zhu2011probing} and Weyl semi-metals \cite{wan2011topological,jiang_tunable_2012,dubvcek2015weyl,zhang_simulating_2015,he_realization_2016}. We refer the interested reader to several existing reviews on the subject \cite{zhai_spin-orbit_2012,zhou_unconventional_2013,galitski_spinorbit_2013,Goldman:2014RPP,Zhai:2015RPP}.

\subsection{Spin-orbit coupling in photonic systems}

\subsubsection{Spin-orbit coupling of light}
A light beam conveys  both spin and orbital angular momentum, encoded in its polarization and spatial mode, respectively. While these two degrees of freedom remain decoupled for plane waves, an effective spin-orbit coupling arises in inhomogeneous media, of particular importance in the case of materials structured at sub-wavelength scales. 

Spin-orbit coupling already affects the propagation of paraxial light beams in inhomogeneous media \cite{liberman_spin-orbit_1992,bliokh_modified_2004,bliokh_spin-orbit_2015}. While in geometrical optics light propagation does not depend on polarization, its first correction introduces polarization-dependent effects. In a semi-classical framework, the trajectory of light, described by its mean momentum $\mathbf{k}(s)$, position $\mathbf{r}(s)$ and polarization $\mathbf{e}(s)$, where $s$ is the trajectory arc length, is governed by the set of equations
\begin{align}
\frac{\mathrm{d}\mathbf{\hat{k}}}{\mathrm{d}s}&=\nabla \log n-(\mathbf{\hat{k}}\cdot\nabla \log n)\mathbf{\hat{k}},\label{eq_geom_k}\\
\frac{\mathrm{d}\mathbf{r}}{\mathrm{d}s}&=\mathbf{\hat{k}}-\frac{\sigma}{k}\,\mathbf{\hat{k}}\times\nabla \log n,\label{eq_geom_r}\\
\frac{\mathrm{d}\mathbf{e}}{\mathrm{d}s}&=-(\mathbf{\hat{e}}\cdot\nabla \log n)\mathbf{\hat{k}},\label{eq_geom_e}
\end{align}
where $\mathbf{\hat{k}}=\mathbf{k}/k$ and $\sigma=\mathrm{i}(\mathbf{e}\times\mathbf{e}^*)\cdot\mathbf{\hat{k}}$ refers to the polarization helicity \cite{liberman_spin-orbit_1992}. This light beam dynamics features a spin-orbit interaction that manifests in two ways. First, equation (\ref{eq_geom_e}) describes the Rytov-Vladimirsky rotation of the light polarization for a bent light beam required to ensure the transverse nature of electromagnetic waves \cite{rytov_transition_1938,vladimirsky1940plane}. This effect can be clearly illustrated from the optical activity of a helical optical fiber \cite{chiao_manifestations_1986,tomita1986observation}. Second, equation (\ref{eq_geom_r}) includes the so-called optical Magnus effect, corresponding to a spin-dependent deflexion of the light beam \cite{liberman_spin-orbit_1992,bliokh_modified_2004}. Further spin-dependent beam displacements are predicted for light beams passing through or reflected by an interface between two homogeneous media \cite{onoda_hall_2004,bliokh2006conservation}, a particular case being the Goos-H\"anchen and Fedorov-Imbert effects for total internal reflexion \cite{goos1947neuer,fedorov1955k,imbert1972calculation,bliokh2013goos}.

\begin{figure}
\includegraphics[width=\linewidth]{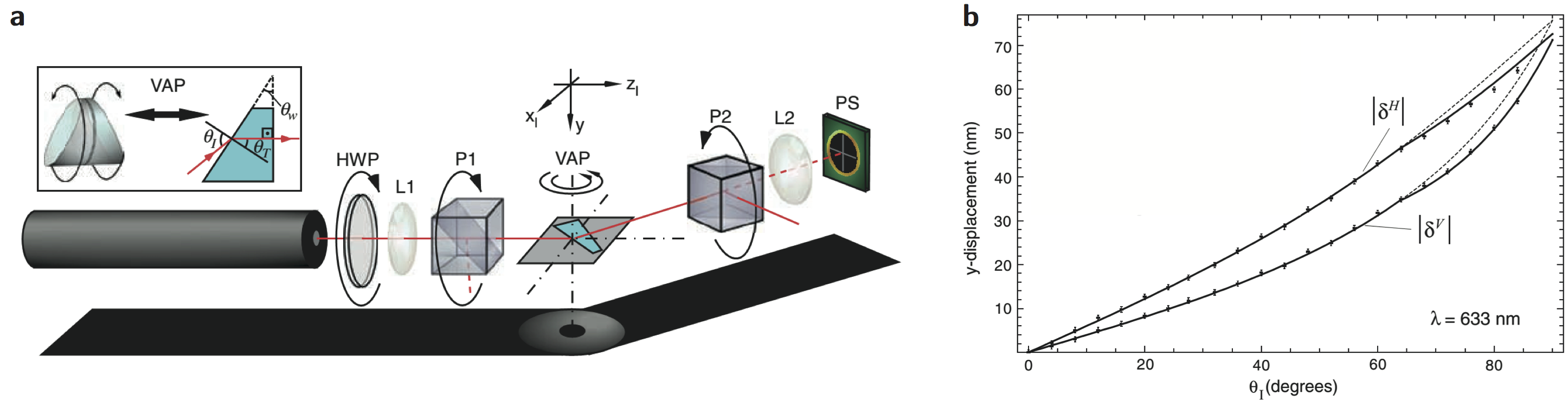}
\caption{\textit{\textsc{Spin-orbit coupling for light}} (a) Scheme of the experiment demonstrating the spin-Hall effect of light passing through the air-glass interfaces of a prism (b) Measurement of the light beam displacement as a function of the prism angle, for horizontal and vertical incident polarizations  (from \cite{hosten_observation_2008}).\label{Fig_SO_light}}
\end{figure}

Such a spin-orbit interaction of light typically produces very small beam displacements for index variations occuring on lengthscale above the light wavelength. Spin-dependent sub-wavelength shifts were observed using light beams passing through or reflected by an air-glass interface \cite{pillon_experimental_2004,hosten_observation_2008,qin_measurement_2009} (see figure \ref{Fig_SO_light}) or propagating in a helical mode at the grazing angle inside a glass cylinder \cite{bliokh_geometrodynamics_2008}, as well as other types of interfaces \cite{qin_spin_2010,menard_ultrafast_2010,hermosa_spin_2011,zhou_experimental_2012,zhou_identifying_2012,yin_photonic_2013}. 

A strong enhancement of spin-orbit coupling effects can be achieved using tighly focused light or using light field structures at the sub-wavelength scale, including nano-optic, photonic or plasmonic systems \cite{bliokh_spin-orbit_2015}. Light beams strongly focused by a microscope objective or scattered by small objects exhibit a spontaneous conversion of spin to orbital angular momentum \cite{zhao_spin--orbital_2007,haefner_spin_2009,rodriguez-herrera_optical_2010,marrucci2011spin,li_spin-enabled_2013}, as well as large spin-dependent beam shifts and deflections \cite{baranova_transverse_1994,bliokh2008coriolis,gorodetski_observation_2008,shitrit_optical_2011,gorodetski_weak_2012,ling2014realization,kruk_spin-polarized_2014}. Spin-orbit effects are also magnified inside anisotropic structures such as metasurfaces \cite{gorodetski_observation_2008,li_spin-enabled_2013,yin_photonic_2013,lin_polarization-controlled_2013,shitrit_spin-optical_2013,meinzer_plasmonic_2014,oconnor_spin-orbit_2014,liu_photonic_2017} or liquid crystals \cite{marrucci2006optical,nagali_quantum_2009,brasselet_optical_2009,brasselet_electrically_2011}.

\subsubsection{Spin-orbit coupling of exciton-polariton systems\label{section_polariton}}
Spin-orbit coupling play an important role for light confined in optical cavities or waveguides, due to the splitting of TE- and TM-polarization modes. While in vacuum all polarizations are degenerate, the TE and TM modes experience different phase shifts upon internal reflection, leading to a lift of degeneracy. The splitting of the polarization components can be written as
\[
H(\mathbf{k})=H_0(\mathbf{k})I+\boldsymbol{\Omega(\mathbf{k})}\cdot\boldsymbol{\sigma},
\]
where $\mathbf{k}$ is the in-plane momentum, $I$ and $\boldsymbol{\sigma}$ are the identity and vector of Pauli matrices associated to the light polarization. The variation of the splitting $\Omega$ with momentum comes from the dependence of the reflection phase shifts on the incoming angle \cite{born1980principles}. It can be viewed as an effective momentum-dependent Zeeman field, leading to an optical spin Hall effect. 

\begin{figure}
\includegraphics[width=\linewidth]{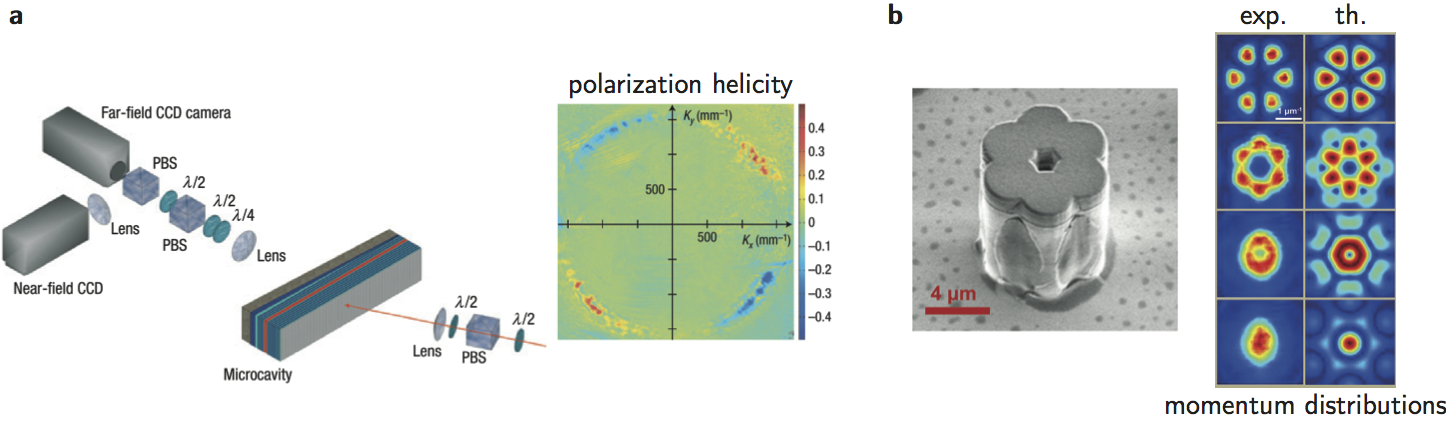}
\caption{\textit{\textsc{Spin-orbit coupling in polaritons}} (a) Observation of the optical spin Hall effect in microcavity exciton-polariton systems. Due to a splitting of different microcavity polarization modes, a coupling between the pump laser and the polariton spin occurs, leading to a correlation between the emitted light direction (in the $k_x-k_y$ plane) and polarization helicity (adapted from \cite{leyder_observation_2007}) (b) Observation of spin-orbit coupling in a chain of six coupled micropillars. The spin-orbit coupling, arising from the polarization-dependent confinement and tunneling of photons between adjacent micropillars, leads to polariton eignstates with characteristic polarization patterns (adapted from \cite{sala2015spin}).\label{Fig_SO_polaritons}}
\end{figure}

In exciton-polariton systems, the pseudospin is inherited from both the spin of the quantum well exciton and the cavity photon, and its dynamics is governed by the polarization of the exciting light. In addition to the TE-TM splitting, more complex mechanisms such as electron and hole spin-relaxation significantly contribute to the effective spin-orbit coupling \cite{dyakonov_current-induced_1971,pikus_exchange_1971,maialle_exciton_1993,kavokin_optical_2005,leyder_observation_2007,shelykh2009polariton}. In such systems,   a polariton spin current could be generated under linearly polarized light excitation, resulting in the development of circular polarizations in different directions, and a spatial spin separation  \cite{leyder_observation_2007} (see figure \ref{Fig_SO_polaritons}a).

A higher degree of control over the polariton spin is achieved using in-plane micro-structures, such as a chain or a lattice of micro-pillars \cite{bloch_strong_1998}. The tunneling of photons between two adjacent micro-pillars depends on the relative angle between the link between pillars and the photon polarization. Combined with  the TE-TM splitting discussed above, it leads to the emergence of a spin-orbit coupling for polaritons  \cite{nalitov_spin-orbit_2015}, as demonstrated in a chain of six coupled pillars \cite{sala2015spin} (see figure \ref{Fig_SO_polaritons}b). Extending to two-dimensional lattice structures \cite{jacqmin_direct_2014} would grant access e.g. to a Rashba spin-orbit coupling \cite{nalitov_spin-orbit_2015}. Non-abelian gauge fields could also be produced using incident light under oblique incidence on the polariton plane \cite{tercas_non-abelian_2014}.

\subsubsection{Photonic topological insulators}
The structure of quantum Bloch states of electrons in ion crystals, including the ones forming topological insulator phases, can be mimicked using light modes in photonic crystals \cite{haldane2008possible,Wang:2009jo,Lu:2014NP,sun2017two}. The organization of  light modes  can be controlled by tailoring the crystal symmetry and the elementary cell properties; in particular an effective spin can be defined either  from the light polarization or the structure of the lattice elementary cell. The coupling between this spin and the light propagation modes may lead to an arrangement of modes analog to  a topological insulator.

Arrays of coupled ring resonators \cite{Hafezi:2011dt,Hafezi:2013jg,gao_probing_2016} and evanescently coupled waveguides \cite{Rechtsman:2013Nature,Maczewsky:2017NatCom,Mukherjee:2017NatCom} can be engineered to exhibit a structure of modes analog to a topological insulator. These systems can be viewed as two copies of quantum Hall systems, where the motion of each spin evolves under an effective magnetic field. We described their physical behavior in sections \ref{section_ring_resonators} and \ref{section_photonic_waveguides}.

\begin{figure}
\includegraphics[width=\linewidth]{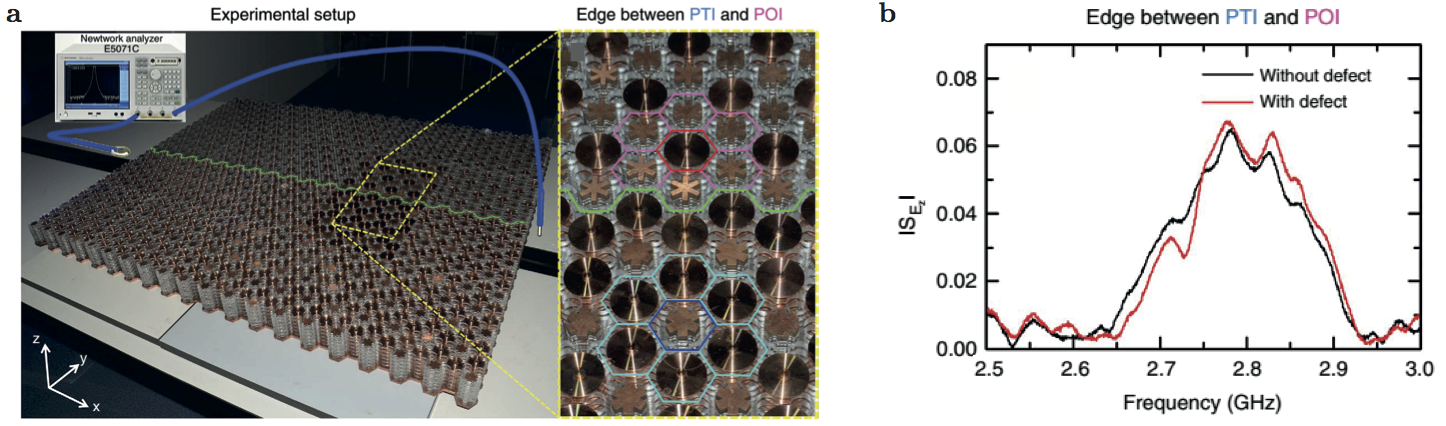}
\caption{\textit{\textsc{Photonic topological insulators}} (a) Anisotropic metamaterial crystal allowing to engineer photonic topological insulators. The structure of electromagnetic propagation modes can form photonic ordinary or topological insulators (POI and PTI, respectively), depending on the geometry of the unit cell.  (b) Topologically protected edge transport was probed at the interface between POI and PTI regions (from \cite{chen_experimental_2014}).\label{Fig_photonic_metamaterial}}
\end{figure}

Bianisotropic metamaterials were also used to engineer photonic topological insulators \cite{khanikaev_photonic_2013}. These systems are characterized by tailored periodic modulations of the permeability, permitivity and bianisotropic coupling between the electric and magnetic fields. Such a modulation induces a spin-orbit coupling on the effective spin 1/2 defined from the transverse electric/magnetic modes, resulting in topologically protected spin-polarized edge transport \cite{chen_experimental_2014,cheng_robust_2016,slobozhanyuk_experimental_2016} (see figure \ref{Fig_photonic_metamaterial}). 

We also mention that tailored photonic crystal structures allows exploring other types of topological systems, for example with dispersion bands exhibiting topological Weyl points \cite{lu2013weyl,lu_experimental_2015,chen_photonic_2016}.

\subsection{Spin-orbit coupling in mechanical systems}

Topological effects can also be induced in classical mechanical systems \cite{Susstrunk:2015uo,Yang:2015hq,khanikaev_topologically_2015,kariyado_manipulation_2015,Nash:2015eua,Wang:2015jv}, based on the recent development of various types of acoustic meta-materials  \cite{ma2016acoustic}. Similarly to their electronic counterpart, topological acoustic systems exhibit topological surface modes whose propagation is protected from environment perturbations, which could be used in various applications such as acoustic cloaking or vibration isolation. 

In these systems a spin-orbit coupling can be engineered by encoding an effective spin in two acoustic modes brought close to degeneracy. The experimental systems studied so far consist in discrete lattices of simple elementary mechanical systems, such as pendula \cite{Susstrunk:2015uo,salerno_spinorbit_2017} or gyroscopes \cite{Nash:2015eua}, as well as acoustic crystals, made of a continuous fluid flowing in an engineered lattice geometry   \cite{He:2016cra,yu_surface_2016,ye2017observation,wei2017experimental,xia2017topological}.

\begin{figure}
\includegraphics[width=\linewidth]{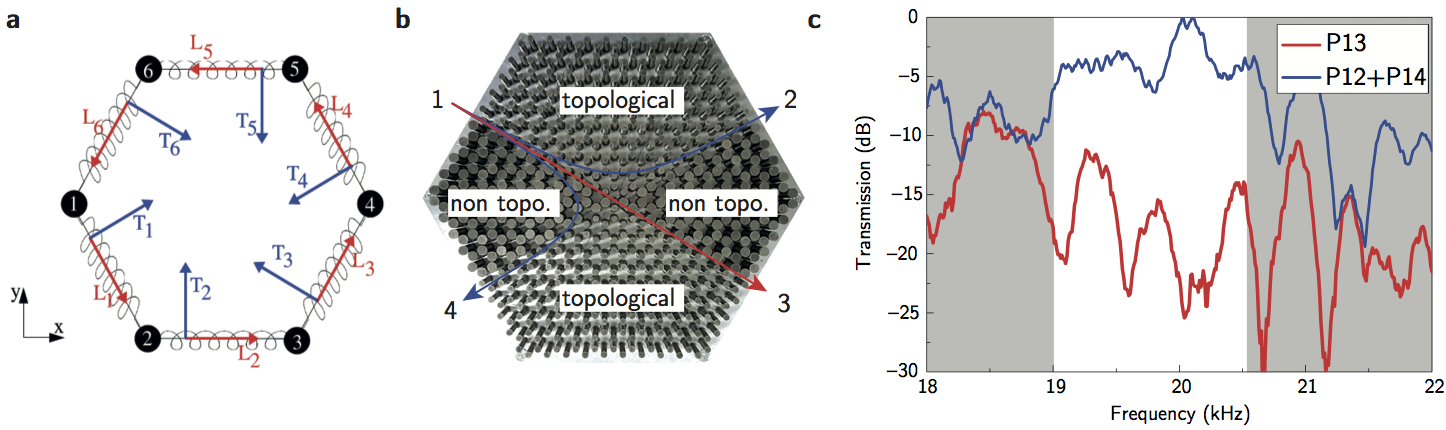}
\caption{\textit{\textsc{Spin-orbit coupling and topology in mechanical systems}} (a) Scheme of the chain of coupled pendula. An effective spin 1/2 can be encoded in  the transverse (blue arrow) and longitudinal (red arrow) oscillation modes. An effective spin-orbit coupling naturally occurs, being induced by the difference in the spring couplings for the two modes (from \cite{salerno_spinorbit_2017}) (b) Picture of the acoustic lattice structure, made of a hexagonal array of steel rods. Two different values of the rod radii are used in different regions, leading to areas with different topological character. Edge sound sound waves are emitted from the edge 1 and may exit from 2, 3 and 4. (c) Evidence for the spin-Hall effect, via the observation of edge propagation along the 12 and 14 paths only, the 13 path corresponding to a different spin orientation (from \cite{He:2016cra}).\label{Fig_SO_sound}}
\end{figure}

We illustrate the simulation of a spin-orbit coupling in discrete lattice systems with the example of a chain of coupled pendula, as recently realized in~\cite{salerno_spinorbit_2017}. A set of six pendula is arranged in a hexagonal geometry as shown in figure \ref{Fig_SO_sound}a. Each pendulum is connected to its neighbours using  springs, and may evolve in the $xy$ plane. An effective spin $1/2$ can be defined from the transverse and longitudinal oscillation modes. The spin-orbit coupling naturally arises from the different spring coupling amplitudes associated with the two modes. This system consitutes the mechanical counterpart of the micropillar chain hosting polariton modes with an effective spin-orbit coupling (see section \ref{section_polariton}). 

Acoustic crystals also feature spin-orbit coupling effects. We discuss here the experiment described in \cite{He:2016cra}, which consists in a study of sound propagation across an array of solid rods arranged in a hexagonal lattice geometry  (see figure \ref{Fig_SO_sound}b). The sound dispersion features a topological transition when varying the rod filling, with a acoustic topological insulator phase for small rod radii. This behavior can be related to an effective spin-orbit coupling occuring between different dispersion bands. In a structure with regions of different rod fillings corresponding to different topological phases, it was possible to exhibit edge sound modes with spin-dependent propagation, analogous to the quantum spin Hall effect  (see figure \ref{Fig_SO_sound}c).

\section{Conclusions and outlook}

This review proposed a general overview on the current efforts and progress related to the realization and tunability of a wide range of gauge potentials, which effectively reproduce the effects of electromagnetic fields or spin-orbit couplings in solids and engineered systems; while the quantum simulation of condensed-matter phenomena typically require quantum-engineered systems (e.g. ultracold atomic gases, ion traps, superconducting devices), we have also put the emphasis on classical realizations of gauge fields, such as those created through arrays of photonic wave-guides and mechanical oscillators. 

Today, applications of engineered gauge fields mainly concern solid-state-oriented phenomena, such as topological states of matter (topological insulators and superconductors), topological defects (e.g. vortex physics), and frustrated magnetism. However, it is worth pointing out that such engineered gauge fields also open the door for the exploration of physical effects that stem from other fields of research, such as high-energy physics (e.g.~particle physics and cosmology); examples include the study of axion electrodynamics~\cite{Wilczek:1987PRL}, the physics of Weyl fermions~\cite{Bernevig:2015NP} and their related quantum anomalies~\cite{bertlmann2000anomalies}, Majorana fermions~\cite{Elliott:2015RMP}, and non-Abelian monopoles~\cite{nakahara:2003book}, just to name a few. It is worth pointing out that all these exotic concepts and phenomena are in fact strongly related to the physics of topological quantum matter~\cite{Hasan:2010RMP,Qi:2011RMP}. While many of these effects can therefore be approached through the engineering of static, external and classical gauge fields (i.e. classical gauge potentials), great efforts are currently devoted to the realization of \emph{dynamical gauge fields} in quantum-engineered systems~\cite{Wiese:2013Ann,Zohar:2015Rep,Dalmonte:2016hd}. This outstanding goal would open the possibility of simulating lattice gauge theories and studying elementary aspects of QCD in a highly controllable environment; see Ref.~\cite{Martinez:2016Nature} for a recent experimental realization of such dynamical gauge fields and Refs.~\cite{Wiese:2013Ann,Zohar:2015Rep,Dalmonte:2016hd} for reviews on this exciting topic.

% \appendix

% \section{}
% \label{}

% The Acknowledgements are also a un-numbered section
\section*{Acknowledgements}
% Acknowledgements text here
NG is supported by the ERC Starting Grant TopoCold and by the FRS-FNRS (Belgium). SN acknowledges support from European Research Council (Synergy grant UQUAM) and the Idex PSL Research University (ANR-10-IDEX-0001-02
PSL\raisebox{0.5mm}{\small{$\bigstar$}}). MA is supported by the Deutsche Forschungsgemeinschaft (FOR2414), the European Commission (Synergy grant UQUAM), the Nanosystems Initiative Munich and received funding from the European Union's Horizon 2020 research and innovation programme under the Marie Sklodowska-Curie grant agreement No. 703926.

%%%%%%%%%%%%%%%%%%%%%%%%%%%%%%%%%%%

%\bibliographystyle{plain} 
%\bibliographystyle{ieeetr} 

%\bibliography{ReviewGaugeFields}

%\begin{thebibliography}{00}
%% please try to use the bibitem system -
%%   the references should be in order of citation in the text


%\bibitem{label}
%\end{thebibliography}

\end{document}